\documentclass[12pt]{article}

\usepackage{authblk}
\usepackage{amstext}
\usepackage{amssymb}
\usepackage{mathtools}
\usepackage{amsthm}
\usepackage{bibentry}
\usepackage{comment}
\usepackage{array}
\usepackage{booktabs}
\usepackage{multirow}
\usepackage{xfrac}
\usepackage{subfig}
\usepackage{graphicx}
\usepackage{tabularx}
\usepackage{float}
\usepackage[compact]{titlesec}
\usepackage{makecell}
\usepackage{algorithm}
\usepackage{algorithmic}

\newcommand{\blind}{0}
\usepackage{natbib}
\usepackage{caption}
\usepackage{enumerate}
\usepackage{enumitem}

\usepackage{amsmath}
\usepackage{mathrsfs}
\usepackage{bm}
\usepackage{setspace}
\usepackage{color,xcolor}                          

\usepackage[colorlinks=true,allcolors=blue]{hyperref}
\usepackage[capitalize,noabbrev]{cleveref}
\crefformat{equation}{#2\textup{(#1)}#3}
\DeclareMathOperator*{\argmin}{arg\,min}

\newtheorem{theorem}{Theorem}
\newtheorem{proposition}{Proposition}
\newtheorem{corollary}{Corollary}
\newtheorem{lemma}{Lemma}
\newtheorem{assumption}{Assumption}

\newtheorem{remark}{Remark}

\usepackage[toc,page,header]{appendix}
\usepackage{minitoc}
\noptcrule 

\crefname{assumption}{Assumption}{Assumptions}

\DeclareMathOperator{\diag}{diag}
\newcommand{\mat}[1]{{\bm{#1}}}
\newcommand{\dd}{\,\text{d}}
\newcommand{\lWt}[1][]{\overline{\bm W}_{t#1}}
\newcommand{\lwt}[1][]{\overline{\bm w}_{t#1}}
\newcommand{\lXt}[1][]{\overline{\bm X}_{t#1}}
\newcommand{\lYt}[1][]{\overline{\bm Y}_{t#1}}
\newcommand{\lHt}[1][]{\overline{H}_{t#1}}
\newcommand{\lct}[2]{\overline{#1}_{#2}}

\newcommand{\commentL}[1]{{\color{red} \{LZ: #1\}}}

\newcommand{\projm}{(\mat{Z}_t^{\top}\mat{Z}_t)^{-1}\mat{Z}_t^{\top}}
\newcommand{\projmk}{(\mat{Z}_{t}^{\top} \mat{Z}_{t})^{-1} \mat{Z}_{t,k}^{\top}}
\newcommand{\solvem}{(\mat{Z}_t^{\top}\mat{Z}_t)^{-1}\mat{Z}_t^{\top}}
\newcommand{\dto}{\overset{d}{\to}}
\newcommand{\pto}{\overset{p}{\to}}
\newcommand{\shiftmean}{\frac{1}{T-M+1}\sum_{t=M}^T}
\DeclareMathOperator*{\plim}{plim}
\newcommand{\ttheta}{\bm \theta}
\newcommand{\ggamma}{\bm \gamma}
\newcommand{\eeta}{\bm \eta}
\newcommand{\bbeta}{\bm \beta}
\newcommand{\mmu}{\bm \mu}

\newcommand{\aat}{\bm A_t}
\newcommand{\bbetat}{\bbeta_t^\ast}
\newcommand{\basef}[1][_l]{z#1}

\newcommand{\ft}{\mathcal{F}_{t-1}}

\newcommand{\partialgamma}[1][_j]{\frac{\partial}{\partial\gamma#1}}
\newcommand{\partialtheta}[1][_j]{\frac{\partial}{\partial\theta#1}}
\newcommand{\partialggamma}[1][^\top]{\frac{\partial}{\partial\ggamma#1}}

\newcommand{\shiftfrac}{\frac{1}{\sqrt{T-M+1}}\sum_{t=M}^T}

\newcommand{\Var}{\mathbb{V}}
\newcommand{\E}{\mathbb{E}}
\renewcommand{\P}{\mathbb{P}}

\usepackage{xparse}
\NewDocumentCommand{\esteq}{ O{} O{} O{\ttheta}}{s #2(\lHt[-1],W_t,Y_t;#3#1)}
\NewDocumentCommand{\sesteq}{ O{} O{} O{\ttheta} }{s #2(\lct{h}{t-1},w_t,y_t;#3#1)}
\NewDocumentCommand{\scorefunction}{ O{} O{} }{\psi #2(W_t,\lHt[-1];\ggamma#1)}
\NewDocumentCommand{\sscorefunction}{ O{} O{} O{t} }{\psi #2(w_{#3},\lct{h}{#3-1};\ggamma#1)}
\NewDocumentCommand{\effm}{ O{t} O{\bm} }{#2 R_{#1}}
\NewDocumentCommand{\seffm}{ O{\bm} }{#1 r}
\NewDocumentCommand{\tautr}{ O{h'} O{h''} O{^\ast_{t}}}{\tau^{\mathrm{Proj.}}_{t,#1,#2}(r; \bbeta#3)}

\NewDocumentCommand{\Nti}{ O{} O{\bm} O{it}}{N_{#3}(F_{#2{h}#1})}
\NewDocumentCommand{\tauti}{ O{\bm} }{\tau_{it}(F_{#1{h}'},F_{#1{h}''})}
\NewDocumentCommand{\catet}{ O{\bm} O{\bm} }{\tau_{t,#1{h'},#1{h''}}(#2{r})}
\NewDocumentCommand{\catetproj}{ O{\bm} O{\bm} O{}}{\tau^{\mathrm{Proj.}}_{t,#1{h'},#1{h''}}(#2{r};\bbeta_t#3)}
\NewDocumentCommand{\pseudoOutH}{ O{} O{t} O{\bm} O{\hat} O{}}{\widetilde{Y}^{H}_{#2}(F_{#3 h#1};#4\ggamma#5)}
\NewDocumentCommand{\pseudoOutI}{ O{} O{t} O{\bm} O{\hat}}{\widetilde{Y}^{I}_{#2}(F_{#3 h#1};#4\ggamma)}
\NewDocumentCommand{\pseudoEffH}{O{t} O{\bm} O{\hat}}{\widetilde{\tau}^{H}_{#1}(F_{#2 h'},F_{#2 h''};#3\ggamma)}
\NewDocumentCommand{\pseudoEffI}{O{t} O{\bm} O{\hat}}{\widetilde{\tau}^{I}_{#1}(F_{#2 h'},F_{#2 h''};#3\ggamma)}
\NewDocumentCommand{\pseudoEffIVec}{O{t,\bm h',\bm h''} O{\bm}}{\widetilde{#2\tau}^{I}_{#1}}
\NewDocumentCommand{\pseudoEffHVec}{O{t,\bm h',\bm h''} O{\bm}}{\widetilde{#2\tau}^{H}_{#1}}
\NewDocumentCommand{\tYt}{O{H}}{\widetilde {\bm Y_t}^{#1}}

\renewenvironment{proof}[1][\proofname]{{\noindent\bfseries #1. }}{\qed}

\begin{document}

\def\spacingset#1{\renewcommand{\baselinestretch}%
{#1}\small\normalsize} \spacingset{1}


\title{{\bf Estimating Heterogeneous Treatment Effects for Spatio-Temporal Causal Inference:\\ {\Large How Economic Assistance Moderates the Effects of Airstrikes on Insurgent Violence}}}

\if0\blind
{
  \title{\bf Estimating Heterogeneous Treatment Effects for Spatio-Temporal Causal Inference\thanks{We thank the anonymous reviewer from the Magaro Peer Pre-Review Program at the Institute for Quantitative Social Science, Harvard University, for their valuable feedback. 
This material is based upon work partially supported by the National Science Foundation under Grant No. 2124124, 2124463, and 2124323.
}}
  \author{Lingxiao Zhou\thanks{Department of Statistics, University of Florida} \quad
    Kosuke Imai\thanks{Department of Government and Department of Statistics, Harvard University} \quad
    Jason Lyall\thanks{Department of Government, Dartmouth College} \quad
    Georgia Papadogeorgou\thanks{Department of Statistics, University of Florida}}

} \fi

\maketitle

\begin{abstract}
Scholars from diverse fields increasingly rely on high-frequency spatio-temporal data. Yet, causal inference with these data remains challenging due to spatial spillover and temporal carryover effects. We develop methods to estimate heterogeneous treatment effects by allowing for arbitrary spatial and temporal causal dependencies. We focus on common settings where the treatment and outcomes are time-varying spatial point patterns and where moderators are either spatial or spatio-temporal variables. We define causal estimands based on stochastic interventions where researchers specify counterfactual distributions of treatment events. We propose the H\'ajek-type estimator of the conditional average treatment effect (CATE) as a function of spatio-temporal moderator variables, and establish its asymptotic normality as the number of time periods increases. We then introduce a statistical test of no heterogeneous treatment effects. Through simulations, we evaluate the finite-sample performance of the proposed CATE estimator and its inferential properties. Our motivating application examines the heterogeneous effects of US airstrikes on insurgent violence in Iraq. Drawing on declassified spatio-temporal data, we examine how prior aid distributions moderate airstrike effects. Contrary to expectations from counterinsurgency theories, we find that prior aid distribution, along with greater amounts of aid per capita, is associated with increased insurgent attacks following airstrikes.

\end{abstract}

\noindent%
\vfill

\newpage
\spacingset{1.9} 
\section{Introduction}\label{sec:intro}

Estimation of causal effects based on spatio-temporal data is important in many disciplines, including public health, ecology, and environmental and social sciences \citep[e.g.,][]{christiansen2022toward,wang2022causalgnn}. Recent examples include the effects of air pollution on child cognitive development \citep{wodtke_22}, and the use of Call Detail Records to track population mobility during war \citep{christia_22}.

Yet valid causal inference with spatio-temporal data remains a challenge due to spatial spillover and temporal carryover effects.  Indeed, much of the existing causal inference literature has not directly dealt with spatio-temporal data.
In particular, prior studies on estimating heterogeneous treatment effects have primarily considered i.i.d. settings \citep[e.g.,][]{imai2013estimating,wager2018estimation,fan2022estimation,kennedy2020towards}. While a few methods account for spatial spillover effects (but not temporal carryover effects), they typically impose a specific interference structure that may not hold in practice \citep[e.g.,][]{giffin2023generalized,tchetgen2021auto}. 

We develop a statistical methodology for analyzing treatment effect heterogeneity in spatio-temporal data with arbitrary spillover and carryover effects (\cref{sec:framework}).  In our setting, the treatment and outcome variables are time-varying spatial point patterns, while the moderator of interest is either spatial or spatio-temporal variables.
We begin by defining the conditional average treatment effect (CATE) in this setting. Our definition employs a stochastic intervention \citep{munoz2012population,haneuse2013estimation}, which specifies a counterfactual probability distribution of treatment assignment across space and time. This intervention can be specified to alter the spatial and temporal distribution of treatment assignment, its overall intensity, and other features. It can also be adaptive and depend on past data, including those affected by previous treatments.

We propose a two-step estimation procedure.  We first use the stabilized IPW weights to estimate the treatment effect for each time period and in each small geographical area, which we refer to as ``pixel.'' A pixel can be any geographical unit of choice, including a fine spatial grid cell and an administrative unit. In the second step, we characterize treatment effect heterogeneity by 
fitting pixel-on-pixel regression of the effect estimate on a low-dimensional function of moderators for each time period, and then aggregating these regression results over time. 

Leveraging martingale theory, we establish that the proposed CATE estimator is consistent and asymptotically normal, provided that the propensity score model is correctly specified. Unlike existing asymptotic results based on Lindeberg-type conditions and mixing or partial-independence assumptions \citep[e.g.,][]{andrews1994asymptotics,bradley2005basic}, our analysis works directly with the martingale structure induced by the sequential spatio-temporal treatment assignment process, avoiding assumptions on the decay of dependence across space or time.  While stabilized weights substantially improve statistical efficiency in spatio-temporal settings, existing asymptotic results do not apply to the stabilized IPW estimator. Our theoretical results overcome this limitation.

While our method builds on the spatio-temporal causal inference framework of \cite{papadogeorgou2022causal}, we overcome a key theoretical challenge that arises from the pixel-on-pixel regression within our martingale framework.  Namely, pooling data across all time periods, as done in the standard meta regression, results in a single projection that is only valid under correct model specification without effect heterogeneity across time. Violations of these assumptions lead to an inconsistent estimator. We instead develop an approach based on time-specific regressions, and establish valid estimation and inference without these additional assumptions.

Although our asymptotic variance is not identifiable, we derive an upper bound and propose its consistent estimator. 
This new estimator has a much improved finite-sample performance compared to a direct extension of the variance bound estimator developed for the 
IPW estimator in \cite{papadogeorgou2022causal}.
Simulation studies indicate that the empirical coverage of the resulting confidence intervals based on the variance bound is close to the nominal level, even in complex data-generating scenarios (\cref{sec:simulation}).
In addition, we show that the asymptotic variance of the proposed estimator with stabilized IPW weights is no greater when the propensity score is estimated than when it is known.  A similar result has been obtained for the standard IPW estimator in the i.i.d.  \citep{hirano2003efficient} and spatio-temporal settings \citep{papadogeorgou2022causal}.  We generalize these results to the stabilized IPW estimator in the spatio-temporal setting.

Furthermore, we develop an asymptotically valid statistical test of no heterogeneous treatment effect. 
Our test contrasts with the existing ones that are only applicable to i.i.d. data with no spatial or temporal interference \citep[e.g.,][]{crump2008nonparametric, sant2021nonparametric}. Lastly, we propose a sensitivity analysis framework to evaluate the robustness of our estimates to violations of key assumptions (\cref{sec:sensitivity}).


Our application comes from the Iraq War (2003-2011), where the United States (US) used both economic aid and airstrikes as part of the “hearts and minds’’ counterinsurgency campaign (\cref{sec:application}).
Drawing on newly declassified geolocated data, we examine how prior aid programs moderated the effects of US airstrikes on insurgent violence.
We find that districts receiving aid tend to experience increased insurgent attacks following intensified airstrikes, and these effects grow as aid per capita rises, even when the amounts are relatively small.
Our findings challenge the prevailing expectations that economic aid and airstrikes work together to reduce short-term insurgent attacks.

 
\paragraph*{Related Literature.}
The literature on spatio-temporal causal inference remains sparse due to the challenges of complex confounding and interference across both space and time \citep{wang2021causal}. Only a few studies consider causal heterogeneity in such settings. For example, \cite{christiansen2022toward} proposes a latent spatial model to adjust for confounders that vary across either time or space, but not both, while assuming a specific graphical structure. \cite{zhang2023spatiotemporal} introduces a spatially interrupted time-series design to estimate heterogeneous treatment effects, while \cite{pmlr-v213-jerzak23a} explores heterogeneity by incorporating image-based moderators to capture geographic variation in randomized controlled trials. In contrast, building on \cite{papadogeorgou2022causal}, we develop a methodology to estimate heterogeneous treatment effects while accounting for arbitrary spatial spillover and temporal carryover effects.  In addition, under our framework, moderators can be either spatial or spatio-temporal variables.


A greater number of studies focus on spatial causal inference problems but without the temporal dimension.
These works rely on design-based inference with randomized experiments \citep[e.g.,][]{wang2020design} or impose structures on the patterns of spatial spillover effects \citep[e.g.,][]{papadogeorgou2019causal, tchetgen2021auto, giffin2023generalized}.  
Unlike these studies, we leverage the temporal dimension to avoid structural assumptions about spillover and carryover effects under both experimental and observational settings.

Our work also contributes to the broader causal inference literature on CATE estimation by considering spatio-temporal settings. 
Much of the existing literature on CATE has focused on i.i.d. settings without interference \citep[e.g.,][]{imai2013estimating, abrevaya2015estimating, athey2019generalized, hahn2020bayesian, fan2022estimation}. Many existing approaches, such as the DR-learner \citep{kennedy2020towards} and FW-learner \citep{yang2023forster}, employ a two-step estimation procedure using pseudo-outcomes.  However, these methods assume i.i.d. data with unit-specific covariates.  
Existing approaches to CATE estimation for spatial data typically lack a temporal dimension, assume finitely many spatial units, and rely on structural assumptions about spillover effects, such as clustered network interference \citep[e.g.,][]{credit2023structured, bargagli2020heterogeneous}.



Our estimator is also related to several strands of statistical literature that combine noisy local estimators with smoothing or weight stabilization. In survey sampling, H\'ajek-type estimators normalize inverse-probability weights to reduce variance relative to Horvitz-Thompson estimators \citep{hajek1971comment,sarndal1992model}. While we employ a similar normalization strategy, our target is the treatment effect of a counterfactual intervention rather than a finite population mean as in survey sampling. Another connection is to the small-area estimation literature, where direct estimates for small geographic areas are combined with regression or smoothing models to improve precision \citep{fay1979estimates,ghosh1994small,rao2015small}. Our approach shares a similar two-step structure but differs in its goal. In small-area estimation the second stage is primarily used to stabilize noisy area-level estimates, whereas in our framework it is used to learn heterogeneous treatment effects as a function of moderators.


\section{The Proposed Methodology}\label{sec:framework}

In this section, we describe the proposed methodology for estimating the CATE based on spatio-temporal data.
We begin by defining the causal estimands and then introduce the new CATE estimator.
We then derive the asymptotic properties of the proposed estimator, develop an inferential procedure, and formalize a test for the null hypothesis of no heterogeneity.

\subsection{Setup}

We adopt the framework of \cite{papadogeorgou2022causal}. 
Suppose we observe the data over $T$ discrete time periods, $t\in\mathcal{T}=\{1,\dots,T\}$.
Let $\Omega$ denote the entire space under consideration that contains a potentially infinite number of possible treatment and outcome locations.  Let $W_t(s) \in \{1, 0\}$ indicate whether location $s \in \Omega$ receives the treatment at time $t$, whereas $W_t = W_t(\Omega) \in \mathcal{W}$ denotes the treatment point pattern at time $t$ with $\mathcal{W}$ representing the set of all possible point patterns, which is assumed to be identical over time. We also assume that the number of treated locations at each time period is finite.  
We use $\lWt = (W_1,\dots,W_t)$ to denote the collection of treatment history up to time $t,$ whose realization is $\overline{\bm w}_t=(w_1,\dots, w_t)$.

Let $Y_t(\lwt)$ represent the potential outcome for the space $\Omega$ at time $t\in\mathcal{T}$ for any given treatment sequence $\lwt \in \mathcal{W}^t = \mathcal{W}\times\dots\times \mathcal{W}$. We do not restrict $Y_t(\lwt)$ so that it can depend on the entire treatment path $\lwt$ in an arbitrary way.  This allows for unrestricted spatial spillover effects based on all treated locations at time $t$ as well as temporal carryover effects of past treatment patterns. We only observe the outcome point pattern at time $t$ corresponding to
the observed treatment sequence, 
$Y_t = Y_t(\lWt)$. We use $\lYt = \{Y_1,\dots, Y_t\}$ to represent the collection of observed outcomes up to and including time $t$.  Lastly, we use $\lct{\mathcal{Y}}{T} = \{Y_t(\lwt):\lwt\in\mathcal{W}^t,t\in\mathcal{T}\}$ to denote the collection of potential outcomes for all treatment sequences and all time periods.

In addition, $\bm X_t$ denotes the set of possibly time-varying confounders realized prior to $W_t$ but after $W_{t-1}$, where $\lXt = (\bm X_1, \dots , \bm X_t)$ is the set of observed covariates up to time $t$. Let $\lct{\mathcal{X}}{T} = \{\bm X_t(\lwt[-1]): \lwt[-1]\in\mathcal{W}^{t-1},t\in\mathcal{T}\}$ denote the set of potential covariate values under all treatment sequences and all time periods. Then, $\bm X_t = \bm X_t(\lWt[-1])$ represents the observed covariates.  
Lastly, we use $\lHt = \{\lWt,\lYt,\lXt[+1]\}$ to present the observed history preceding the treatment at time $t+1$.  Since our statistical inference is based on a single time series, the randomness comes only from the treatment assignment $W_t$ given the complete history that includes all observed and potential values $\lct{H}{t-1}^\ast = \{\lWt,\lct{\mathcal{Y}}{T},\lct{\mathcal{X}}{T}\}$.  

\subsection{Causal estimands}\label{subsec: estimand}

We consider a stochastic intervention $F_h$ that defines a counterfactual distribution over the treatment point pattern.
For simplicity, we begin by studying fixed stochastic interventions that do not depend on $\bm X_t$ and $Y_t$, both of which can be affected by the treatment. In \Cref{sec: adaptive}, we extend the proposed framework to adaptive interventions that may depend on the observed history.
For concreteness, we consider the Poisson process, which is specified by its intensity function $h: \Omega \to [0,\infty)$, though our methodology can accommodate alternative intervention distributions.

This setup contrasts with most existing work, which considers a finite set of units with discrete treatments in longitudinal studies \citep[e.g.,][]{munoz2012population,haneuse2013estimation}. In spatio-temporal settings, stochastic interventions are useful because deterministic intervention regimes may violate the positivity assumption. By considering stochastic interventions that remain close to the observed treatment assignment, we can formulate causal estimands that are likely to be identifiable even under arbitrary interference. To our knowledge, \cite{papadogeorgou2022causal} is the only prior work that specifies stochastic interventions for spatio-temporal point patterns, but this work neither considers adaptive interventions nor estimates heterogeneous effects.

We study how the causal effects of a change in the stochastic intervention vary across regions based on their characteristics.
These characteristics are potential moderators for the treatment effect and may be spatial or spatio-temporal variables.
Formally, we partition the space $\Omega$ into a total of $p$ regions or ``pixels'' in $\mathcal{S} = \{S_1, \dots, S_p\}$, where $\Omega = \bigcup_{i=1}^p S_i$ with $S_i \bigcap S_{i^\prime} = \varnothing$ for $i \ne i^\prime$. These pixels can be arbitrary geographical units.
For each pixel, we measure the moderator using either its original value (if constant within the pixel) or a summary statistic within the pixel (e.g., mean).
We use $\effm[it] \in \mathcal{R}$ to denote the value of the potential moderator for pixel $S_i$ at time $t$ where $\mathcal{R}$ is the support of the variable. 

The discretization of space for the moderator enables us to analyze effect heterogeneity at the pixel level without any restrictions on the structure of spatial spillover or temporal carryover effects. This discretization determines how spatial treatment effect heterogeneity is summarized and hence is part of the estimand definition rather than a numerical approximation.
In practice, the choice of pixels should be guided by the spatial scale desired for summarizing heterogeneity, the resolution of the data, and the spatial variability of the moderator.
If the treatment assignment and the moderator
vary substantially within each pixel, a finer partition may be more appropriate as the resulting estimand would not capture such within-pixel variation. 
In our application, we treat equally sized small geographic areas as pixels, while the moderator is the amount of humanitarian aid provided during the previous month in the district to which each pixel belongs.

\paragraph*{Intervention over a single time period.}
We define the expected number of outcome events within pixel $S_i$ at time $t$ under stochastic intervention $F_h$ as:
\begin{equation*}
  \Nti[][] \ = \ \int_{\mathcal{W}}N_{S_i}(Y_t(\lWt[-1],w_t)) \dd F_{h}(w_t)
\end{equation*}
where
$N_{S_i}(Y_t(\lWt[-1],w_t))$ is the number of outcome events that fall within pixel $S_i$ at time $t$ based on the treatment path $(\lWt[-1],w_t)$.  Then, the treatment effect of $F_{h'}$ versus $F_{h''}$ for time period $t$ and pixel $S_i$ is defined as the contrast in the expected number of outcome events between the two stochastic intervention regimes:
$\tauti[] \ = \ \Nti[''][]-\Nti['][]$.

We are interested in estimating how this pixel-level treatment effect varies as a function of the moderator $\effm[it]$.  We define the CATE at time $t$ when the moderator takes a specific value $\seffm \in \mathcal{R}$ as:
\begin{equation}\label{eq:catet}
\tau_{t}(\seffm; F_{h'},F_{h''}) \ = \ \frac{1}{\sum_{i=1}^p I(\effm[it] = \seffm)} \sum_{i=1}^p \tauti[] I(\effm[it] = \seffm),
\end{equation}
where $I$ represents the indicator function.
For notational simplicity, we write the CATE as $\catet[]$. For each time $t$, this quantity averages pixel-level effects over all pixels with moderator value $\seffm$, thereby integrating over the spatial variation conditional on $\effm[it]=\seffm$.
In our application, if we use the receipt of aid in the previous month as the dichotomous moderator, then $\tau_{t,h',h''}(1)$ represents the average effect on the expected number of insurgent attacks for a district that received aid, when changing the distribution of airstrikes from $F_{h'}$ to $F_{h''}$.

Since the moderator can take many values, we characterize the CATE by projecting the pixel-level treatment effect $\catet[]$ onto a space of $\effm[it]$ using an interpretable model $\catetproj[]$, which is indexed by a low-dimensional vector $\bbeta_t$. For example, one may choose a linear model with $\effm[it]$ as predictors and $\bbeta_t$ as their coefficients \citep[see][for similar estimands]{kennedy2019robust,van2011targeted,ding2019decomposing}.
Formally, we focus on the projection estimand:
\begin{equation*}
 \catetproj[][][^\ast] \quad \text{where} \quad \bbeta_t^\ast
  =\argmin_{\bbeta_t}\sum_{i=1}^p  \left[ \tau_{t,h',h''}(\bm R_{it})-\tau^{\mathrm{Proj.}}_{t,h',h''}(\bm R_{it};\bbeta_t) \right]^2.
\end{equation*}
This estimand represents the best approximation to the true CATE curve that can be achieved with the selected parametric model $\catetproj[]$.

Finally, we average the time-specific projected CATEs over all time periods to define the following overall projected CATE:
\begin{equation*}
  \tau_{h',h''}^{\mathrm{Proj.}}(\seffm;\bbeta^\ast_{1},\dots,\bbeta^\ast_{T}) = \frac{1}{T}\sum_{t=1}^T \tau^{\mathrm{Proj.}}_{t,h',h''}(r; \bbeta^\ast_{t}).
\end{equation*}
When the moderator takes a finite number of values, $\catet[]$ defined in Eqn~\cref{eq:catet} can be represented exactly based on a saturated parametric model, and the projection estimand coincides with the true CATE.
For a continuous moderator, projecting this estimand onto a lower-dimensional space provides an interpretable and stable summary of how the CATE varies with the moderator, while potentially smoothing over localized nonlinear patterns. 

Our inferential target is $\tau_{h',h''}^{\mathrm{Proj.}}(\seffm;\bbeta^\ast_{1},\dots,\bbeta^\ast_{T})$, i.e., the projection of the CATE onto the working model class. This projection should be interpreted as a summary of effect heterogeneity rather than a structural assumption on the true effect surface. Even if the working model is misspecified, our estimation and inference remain valid with respect to this projection parameter.

\paragraph*{Intervention over multiple time periods.}
We generalize our estimand to stochastic interventions defined over multiple consecutive time periods.
Consider a  stochastic intervention over $M$ time periods, denoted by $F_{\bm h} = F_{h_1}\times \dots\times F_{h_M}$,  where the treatment is assigned according to $F_{h_{1}}$ at time $t$, $F_{h_{2}}$ at time $t-1$, and so on until time $t - M + 1$.

Formally, we define the expected number of outcome events in pixel $S_i$ at time $t$ under this multi-period stochastic intervention as follows:
\begin{equation}\label{eq:Nti}
  \Nti
  = \int_{\mathcal{W}^M}N_{S_i}(Y_t(\lWt[-M],w_{t-M+1},\dots,w_t)) \dd F_{\bm h}(w_{t-M+1},\dots, w_t).
\end{equation}
As before, we define the CATE with respect to $F_{\bm h'}$ and $F_{\bm h''}$ for time $t$ as
\begin{equation*}
  \catet  \ = \ \frac{1}{\sum_{i=1}^p I(\effm[i,t-M+1] = \seffm)} \sum_{i=1}^p \tauti I(\effm[i,t-M+1] = \seffm),
\end{equation*}
where $\tauti = \Nti['']-\Nti[']$ is the pixel-specific effect and $\effm[t-M+1]$ denotes the moderator that immediately precedes the intervention time periods, avoiding post-treatment bias.  

Finally, the projection estimand for this multi-period case is defined as \begin{align*}
 \tau_{t,\bm h',\bm h''}^{\mathrm{Proj.}}(\bm R_{i,t-M+1};\bbeta_t^\ast) \quad \text{where} \quad \bbeta_t^\ast
  &=\argmin_{\bbeta_t}\sum_{i=1}^p \left(\tau_{t,\bm h',\bm h''}(\bm R_{i,t-M+1})-\tau_{t,\bm h',\bm h''}^{\mathrm{Proj.}}(\bm R_{i,t-M+1};\bbeta_t)\right)^2.
\end{align*}
Averaging this quantity over time, we have the overall CATE, which is defined as
\begin{equation*}
  \tau^{\mathrm{Proj.}}_{\bm h',\bm h''}(\seffm;\bbeta^\ast_{M},\dots,\bbeta^\ast_{T}) = \frac{1}{T-M+1}\sum_{t=M}^T \tau^{\mathrm{Proj.}}_{t, \bm h'\bm h''}(\seffm;\beta^\ast_t).
\end{equation*}

\paragraph*{Working model.}
When the moderator is continuous or takes many distinct values, the working model $\catetproj$ represents an approximation to the true CATE curve.  Since this model does not have to be correctly specified, model specification can depend on one's analysis goal.  Here, for simplicity and interpretability, we consider the following model for each time period:
\begin{equation}\label{eq:model}
\catetproj =\sum_{l = 1}^{L} \beta_{t,l} \basef(\seffm)
=\bm z(\seffm)^\top\bbeta_t, 
\end{equation}
where $\bm z(\seffm) = (\basef[_1](\seffm),\dots,\basef[_L](\seffm))^\top$ are, for example, the basis functions of splines.

Although we focus on a working model that is linear in the pre-specified $\basef(\seffm)$ functions, more flexible approaches could be used in the second stage (e.g., series estimators \citep{kennedy2020towards} and kernel smoothing) so that we can better capture nonlinear or locally varying patterns in the CATE. While these approaches are well developed for i.i.d. data, extending them to spatio-temporal settings with arbitrary interference is not trivial. In particular, they require the selection of tuning parameters (e.g., the number of basis functions or bandwidth), which cannot be easily done via cross-validation in the presence of spatial and temporal dependence. Moreover, these methods may exhibit greater variability in finite samples. We leave these extensions to future work.


\subsection{Assumptions}\label{subsec:assumption}

We require two assumptions about the treatment assignment mechanism. Here, we focus on the scenario, in which the intervention distribution is identical and independent over $M$ time periods, $F_{\bm h} = F_{h}\times\dots\times F_{h}$.  It is straightforward to extend our theory to different and/or dependent intervention distributions across time periods (see \Cref{a:sec:extension}). Adaptive interventions that depend on the observed history are discussed in \cref{sec: adaptive}. 
\begin{assumption}
\label{assump: unconfoundedness}
  $f(W_t\mid\lWt[-1],\lct{\mathcal{Y}}{T},\lct{\mathcal{X}}{T}) = f(W_t\mid \lHt[-1])$.
\end{assumption}
\begin{assumption}
\label{assump: overlap} \spacingset{1}
 There exists $\delta_W>0$ such that $e_t(w)>\delta_W f_h(w)$ for all $w\in\mathcal{W}$ for which $f_h(w) > 0$, where $e_t(w)=f(W_t = w \mid \lHt[-1])$.
\end{assumption}

\cref{assump: unconfoundedness} states that, given the observed history $\lHt[-1]$, the treatment assignment at time $t$ does not depend on all past and future potential outcomes and potential time-varying confounders. Since our framework focuses on a single unit observed over time, \cref{assump: unconfoundedness} is more restrictive than the standard sequential ignorability assumption \citep[e.g.,][]{robins2000marginal}, which only involves future potential outcomes.
One of the simplest cases that satisfy \cref{assump: unconfoundedness} is autoregressive spatio-temporal systems where the treatment assignment depends only on observed state variables. However, the assumption would still hold in the presence of latent spatial structure, spillover, or unmeasured dynamics, so long as the realized history contains all information that jointly predicts treatment assignment and unobserved potential outcomes.

\cref{assump: overlap} does not require every location in the study region to have positive probability of receiving treatment. Instead, all feasible treatment patterns under the stochastic intervention must also be feasible under the actual treatment assignment mechanism. If the intervention places positive density on treatment patterns that could not arise under the observed treatment assignment mechanism, the overlap condition would fail and the corresponding causal effect would not be identified. This relaxation is important in our setting because, for spatio-temporal point-pattern treatments, the probability of any fixed treatment pattern can be extremely small, making it impractical to impose positivity conditions on each possible treatment point pattern separately. In our application in \cref{sec:application}, we construct the counterfactual intervention based on historical data, which supports the plausibility of this condition.

\subsection{Two-step estimation procedure}\label{subsec: estimation}

We develop a two-step estimation procedure by building on the existing methodology for the CATE with i.i.d. data \citep[e.g.,][]{kennedy2020towards,fan2022estimation,yang2023forster}.
We first construct a pseudo-outcome based on the estimated propensity score, and then regress their contrasts on the moderators. Importantly, space is not discretized when constructing the pseudo-outcome in Eqn~\cref{eq: y_IPW}. Instead, discretization is introduced in the second stage, allowing us to estimate the CATE without imposing structural assumptions on the interference mechanism.
This procedure maintains a simple CATE analysis regardless of the complexity of the propensity score estimation.
 
We construct a pseudo-outcome by weighting the observed number of events in each pixel at each time period using the estimated propensity score and the density of the stochastic intervention. The (IPW) pseudo-outcome in pixel $S_i$ at time $t$ is given by
\begin{equation}\label{eq: y_IPW}
  \pseudoOutI[][it] = \rho_{t \bm h}(\hat\ggamma) N_{S_i}(Y_t), \quad \text{ where  } \quad \rho_{t\bm h}(\hat\ggamma) = 
  \prod_{j=t-M+1}^{t}\frac{f_h(W_j)}{e_j(W_j;\hat\ggamma)}
\end{equation}
for $i=1,\dots,p$ and $t = M,\dots,T$, where $f_h$ is the density corresponding to the stochastic intervention $F_h$, $e_t(W_t;\ggamma)$ is a propensity score model with a $K$-dimensional vector of parameters $\ggamma\in\mathbb{R}^K$, and $\hat{\ggamma}$ is an estimate of $\ggamma.$ 
The pseudo-outcome construction follows IPW estimators commonly used in the interference literature \citep[e.g.,][]{hudgens2008toward, aronow2013estimating, tchetgen2012causal, papadogeorgou2019causal, leung2022causal}. The key difference is that our analysis treats the data as a single dependent spatio-temporal trajectory and uses martingale arguments to establish identification and valid inference.

To ensure computational tractability, the propensity score model incorporates prior values of variables from a finite temporal lag, chosen based on one's domain-specific knowledge of treatment assignment persistence.

The above IPW-based weighting scheme can produce large weights for some time periods when the estimated propensity score is much smaller than the intervention density.
Therefore, we consider the following H\'ajek-type pseudo-outcome with stabilized weights,
\begin{equation}
  \pseudoOutH[][it] = \dfrac{\rho_{t\bm h}(\hat\ggamma)}{\sum_{t=M}^T \rho_{t\bm h}(\hat\ggamma)/(T-M+1)} N_{S_i}(Y_t). 
\end{equation}
We then compare the outcome under two intervention distributions, $F_{\bm{h''}}$ and $F_{\bm{h'}}$, at each pixel using
$\pseudoEffH[it] = \pseudoOutH[''][it]-\pseudoOutH['][it]$.
Below, for notational simplicity, we denote $\pseudoEffH[it]$ as $\pseudoEffHVec[it,\bm h',\bm h''][](\hat\ggamma)$, and let $\pseudoEffHVec(\hat\ggamma) = \big(\pseudoEffHVec[1t,\bm h',\bm h''][](\hat\ggamma),\dots, \pseudoEffHVec[pt,\bm h',\bm h''][](\hat\ggamma)\big)^\top$ represent the vector of pseudo-outcome contrasts for the $p$ pixels at time period $t$.  

We regress the pseudo-outcome contrast at time $t$, $\pseudoEffHVec(\hat\ggamma)$, on the moderator as shown in Eqn~\cref{eq:model}.
This gives the following H\'ajek estimator of $\bbeta_t$,
\begin{equation}
    \hat{\bbeta}_t^{H}  = \argmin_{\bbeta_t} \ (\pseudoEffHVec(\hat\ggamma)-\mat{Z}_t\bbeta_t)^{\top}(\pseudoEffHVec(\hat\ggamma)-\mat{Z}_t\bbeta_t),
\label{eq:betahat_t}
\end{equation}
where $\mat{Z}_t = \left[\bm z(\effm[1,t-M+1]) \  \dots \  \bm z(\effm[p,t-M+1]) \right]^\top$.
Finally, we obtain the projected CATE estimator by averaging across time periods, 
\begin{equation}\label{eq: estimator_cate}
  \hat\tau_{\bm h',\bm h''}(\seffm) = \tau^{\mathrm{Proj.}}_{\bm h',\bm h''}(\seffm;\hat\bbeta_M^{H},\dots,\hat\bbeta_T^{H})=\bm z(\seffm)^\top\left(\shiftmean \hat\bbeta_t^{H}\right).  
\end{equation}
As evident in Eqn~\cref{eq:betahat_t}, the time-specific estimates $\hat{\bbeta}_t^{H}$ are obtained by fitting a separate regression model for each time period rather than a pooled regression using data across all time periods.
With separate regressions, the resulting estimating equations have a conditional mean of zero for any given time period, forming a martingale difference sequence with respect to the history filtration.
This allows us to derive the asymptotic properties of the estimator, as shown below.

\subsection{Asymptotic properties}\label{subsec: asymptotic properties}

We now show that the proposed H\'ajek-type CATE estimator is consistent and asymptotically normal.
Unfortunately, the asymptotic variance is unidentifiable because we only observe a single time series.
Nevertheless, we derive a variance upper bound and propose its consistent estimator.
All proofs are presented in the appendix.

Our results are new in the spatio-temporal causal inference literature. \cite{papadogeorgou2022causal} derive asymptotic properties for the IPW estimator, but not for stabilized weighting estimators. We also show that when the propensity score is estimated, the asymptotic variance of the H\'ajek estimator is no greater than when the propensity score is known. 
This extends the analogous result of \citet{hirano2003efficient}, which was applicable only to the IPW estimator of the ATE in i.i.d. settings, to CATE estimation with stabilized weights in spatio-temporal settings.

We formally state the asymptotic normality of the H\'ajek estimator based on the estimated propensity score. 
Besides the two assumptions introduced in \cref{subsec:assumption}, we require standard regularity conditions, which are formally stated in the appendix. 
\begin{theorem}
\label{thm: Hajek est} \singlespacing
    Suppose that Assumptions \ref{assump: unconfoundedness} and \ref{assump: overlap} hold, along with the regularity conditions specified in Assumptions \ref{assump:IPW1}, \ref{assump: IPWest}, \ref{assump:Hajek} and \ref{assump: Hajekestadd}. Let $\hat\ggamma$ be the estimate of the propensity score parameters obtained by solving the estimating equation 
    $\sum_{t = M}^T\psi\left(W_t, \lHt[-1] ; \ggamma\right) = 0,$ where $\psi\left(W_t, \lHt[-1] ; \ggamma\right)$ is 
    specified
    in Assumption \ref{assump: IPWest}. Then as $T\to\infty$, we have that $$\frac{1}{\sqrt{T-M+1}}\sum_{t=M}^T(\hat{\bbeta}^{H}_t-\bbeta^\ast_t)\dto N\big(\bm 0,\mat{J}\mat{V}^H\mat{J}^\top\big),$$ where $\mat{V}^H = \widetilde{\mat{V}}^H-\mat{U}^\top \mat{V}_{ps}^{-1}\mat{U}$, matrices $\mat{J}$, $\widetilde{\mat{V}}^H$ that are defined in \cref{thm: Hajek true}, matrix $\mat{V}_{ps}$ defined in \cref{assump: IPWest}, and matrix $\mat{U}$ defined in \cref{assump: Hajekestadd}. 
\end{theorem}
This result relies on correct specification of the propensity score model. To account for possible misspecification, we develop a doubly robust estimator in \cref{a:sec:doublyrobust} that combines the propensity score model with the outcome model. This estimator remains consistent as long as at least one of the two models is correctly specified.  However, in our setting, specifying a reliable outcome model would require modeling how treatment effects propagate across space and time, which is difficult to validate under arbitrary spatio-temporal interference. For this reason, we focus on an IPW-based construction in the main text. Developing efficiency theory for doubly robust estimators under unrestricted spatio-temporal interference is left for future work.

\cref{thm: Hajek true} of \cref{a:sec:Hajek_true} establishes that the estimator based on the true propensity score is asymptotically normal with variance $\mat{J}\widetilde{\mat{V}}^{H}\mat{J}^\top$. Thus, the difference between the asymptotic variance based on the true and the estimated propensity score is $\mat{J}\mat{U}^\top \mat{V}_{ps}^{-1} \mat{U}\mat{J}^\top$.
This quantity is large when the estimating equation corresponding to the causal effect estimator based on the true propensity score is strongly correlated with the score function of the propensity score model.
By showing that this difference is a positive semidefinite matrix, we establish that using the estimated propensity score leads to a more efficient estimator. 
\begin{theorem}
\label{thm: Hajek efficiency} \singlespacing
If the propensity score model is correctly specified, the estimator $\shiftmean\hat{\bbeta}_t^{H}$ based on the estimated propensity score has asymptotic
variance that is no larger than the asymptotic variance of the same estimator using the known
propensity score. That is, for $\mat{J}\widetilde{\mat{V}}^{H}\mat{J}^\top$ in \cref{thm: Hajek true}, and $\mat{J}\mat{V}^{H}\mat{J}^\top$ in \cref{thm: Hajek est},  $\mat{J}(\widetilde{\mat{V}}^{H}-\mat{V}^{H})\mat{J}^\top$is a positive semidefinite matrix.
\end{theorem}

The matrix $\mat{V}^H$ is based on the covariance matrix of the estimating equations conditional on the complete history, i.e., $\mat{V}^H =  \displaystyle\plim_{T\to\infty} \shiftmean \Var[\aat(\ggamma^\ast)\mid\lHt[-M]^\ast]$  for vector $\aat(\ggamma)$ defined in \cref{a:sec:Hajek_regularity}.
As mentioned earlier, $\Var[\aat(\ggamma^\ast)\mid\lHt[-M]^\ast]$ cannot be consistently estimated without further assumptions due to the fact that we only observe a single time series.
However, since $\mat{V}^{H^\ast} =\displaystyle\plim_{T\to\infty} \shiftmean \E[\aat(\ggamma^\ast)\aat(\ggamma^\ast)^{\top}\mid\lHt[-M]^\ast]$ is an upper bound of $\mat{V}^{H}$, the asymptotic variance $\mat{J}\mat{V}^{H}\mat{J}^\top$ is bounded above by $\mat{J}\mat{V}^{H^\ast}\mat{J}^\top$. Moreover, the following proposition shows that we can consistently estimate the upper bound of the asymptotic variance.
\begin{proposition}
\label{cor:2} \singlespacing
Suppose that 
the conditions of \cref{thm: Hajek est}
hold.
   For $\aat(\ggamma)$ defined in \cref{assump:Hajek}, define $\hat{\mat{V}}^{H} = \shiftmean\hat{\mat{V}}^{H}_t$ where $\hat{\mat{V}}_t^{H} = \aat(\hat\ggamma){\aat(\hat\ggamma)}^\top$, and
  \begin{equation*}
       \hat{\mat{J}} = \begin{bmatrix}
    \mat{I} &-\mat{I} &-\shiftmean \projm\tYt(F_{\bm h'};\hat\ggamma) &\shiftmean \projm\tYt(F_{\bm h''};\hat\ggamma)
\end{bmatrix},
  \end{equation*}
  where $\tYt(F_{\bm h};\hat\ggamma) = \left(\pseudoOutH[][1t],\dots,\pseudoOutH[][pt]\right)^\top.$
  Then, for any matrix $\mat{Q}$ that depends on $T$ and converges to the identity matrix $\mat{I}$ in probability as $T\to\infty,$   $\hat{\mat{J}}\mat{Q}\hat{\mat{V}}^{H}\mat{Q}^\top\hat{\mat{J}}^\top$ is a consistent estimator for $\mat{J}\mat{V}^{H^\ast}\mat{J}^\top$.
 \end{proposition}

We use the estimated variance bound to construct confidence intervals whose coverage will be at least as large as the nominal level.

Moreover, the choice of the matrix $\mat{Q}$ affects the performance of our inferential procedure. Even though setting $\mat Q = \mat I$ returns a consistent estimator for the variance upper bound according to \cref{cor:2}, it often performs poorly in finite samples. Building on the same idea as the H\'ajek estimator for the causal effect, $\mat Q$ is chosen to stabilize the variance bound estimator.

First, we consider $F_{h'}$ and $F_{h''}$, the simplifications of $F_{\bm h'}$ and $F_{\bm h''}$ to one time period interventions, $M = 1$. Then, we define the mean IPW weights for $F_{h'}$ and $F_{h''}$ as
$
\xi_{h'} = \frac{1}{T}\sum_{t=1}^T \frac{f_{h'}(W_t)}{e_t(W_t;\hat\ggamma)}$ and  
$\xi_{h''} = \frac{1}{T}\sum_{t=1}^T \frac{f_{h''}(W_t)}{e_t(W_t;\hat\ggamma)}
$, respectively.
We set $\mat{Q}=\diag(\rho_{\bm h'}^{-1}\mat{1}_L,\; \rho_{\bm h''}^{-1}\mat{1}_L,\; \xi_{h'}^{-M},\; \xi_{h''}^{-M}),$ 
where $\diag(\cdot)$ denotes a diagonal matrix with diagonal entries specified as its argument, $\mat{1}_L$ is a $L\times 1$ vector of ones, and $\rho_{\bm h'} = \shiftmean \rho_{t\bm h'}$ and $\rho_{\bm h''} = \shiftmean \rho_{t\bm h''}$ are the mean IPW weights for $F_{\bm h'}$ and $F_{\bm h''},$ respectively.
Since $\rho_{\bm h'}$, $\rho_{\bm h''}$, $\xi_{h'}$, and $\xi_{h''}$ converge to~1 in probability as $T\to\infty$, $\mat{Q}$ converges to~$\mat{I}$ in probability as $T\to\infty.$
Here, $\rho_{\bm h'}^{-1}\mat{1}_L$ and $\rho_{\bm h''}^{-1}\mat{1}_L$ stabilize the pseudo-outcomes, while $\xi_{h'}^{-M}$ and $\xi_{h''}^{-M}$ stabilize the IPW weights, which take a product form across $M$ time periods, using the mean of the one-period weights. This choice of $\mat{Q}$ provides stabilization of the variance upper bound estimator. For stochastic interventions that are dependent across the $M$ time periods, we discuss the choice of $\mat Q$ in \cref{a:sec:extension}.


Our simulation study in \cref{sec:simulation} demonstrates a good empirical coverage rate for the constructed confidence interval, even when $M$ is relatively large. 
Thus, our methodology provides significantly improved inference compared to a heuristic variance bound used in \cite{papadogeorgou2022causal}, where the coverage rate deteriorates quickly as $M$ increases.

\cref{a:sec:Hajek_true} presents additional theoretical results for the H\'ajek estimator, while \cref{a:sec:IPW result} establishes that the IPW estimators are also consistent and asymptotically normal, with consistently estimable variance bounds. In simulations, the IPW estimators show larger variances than the stabilized H\'ajek estimator (see \cref{a: subsec: Hajek_IPW}). Moreover, the mean estimated variance bound across simulations closely matches the Monte Carlo variance when $M$ is large (\cref{a: subsec: different_bound_estimator}).

\subsection{Statistical test of no heterogeneous treatment effect}\label{subsec: test}

We develop a statistical test for treatment effect heterogeneity with the null hypothesis that all coefficients of $\bm z(\seffm)$ in Eqn~\eqref{eq: estimator_cate}, other than the intercept, are equal to zero. For notational simplicity, we describe the proposed test in a setting in which $\bm z(\seffm)$ does not include an intercept.
The test, however, can be easily adopted in models with an intercept. 

Let $\bar\bbeta = \shiftmean\hat\bbeta^H_t$ and $\bar\bbeta^* = \shiftmean\bbeta^\ast_t.$ We are interested in testing the null hypothesis that $H_0: \bar\bbeta^* = \bm 0$, which holds if the treatment effect is not heterogeneous with respect to $\bm z (\seffm)$. The alternative is $H_A: \bar\bbeta^\ast \neq 0$. We propose the following test statistic
$$
T_c = (T-M+1)\bar\bbeta^\top(\hat{\mat{J}}\mat{Q}\hat{\mat{V}}^{H}\mat{Q}^\top\hat{\mat{J}}^\top)^{-1}\bar\bbeta,
$$
and calculate the $p$-value as $\P(\chi^2_{L}>T_c)$, where $\chi^2_L$ denotes the chi-square distribution with $L$ degrees of freedom. The following theorem shows that the limiting rejection probability under the null hypothesis is no greater than $\alpha$ as $T\to\infty.$ 
\begin{theorem}
\label{a:thm:reject_prob} \singlespacing
     Under the assumptions of \cref{thm: Hajek est} and the null hypothesis $\bar\bbeta^\ast = \bm 0$, we have that 
     $ \displaystyle \limsup_{T\to\infty} \P(\text{p-value}<\alpha)\leq \alpha.$
\end{theorem}
By inverting this test, we obtain an $(1-\alpha)100\%$ confidence set for $\bar\bbeta^\ast$ with limiting coverage that is no less than $1-\alpha$ as $T\to\infty$ as:
\begin{equation}\label{eq: CI}
C_{\alpha} = \left\{\bar\bbeta^\prime:(T-M+1)(\bar\bbeta-\bar\bbeta^\prime)^\top(\hat{\mat{J}}\mat{Q}\hat{\mat{V}}^{H}\mat{Q}^\top\hat{\mat{J}}^\top)^{-1}(\bar\bbeta-\bar\bbeta^\prime)< F^{-1}_{\chi^2_L}(1-\alpha)\right\},
\end{equation} 
where $F^{-1}_{\chi^2_L}(q)$ is the $q$-quantile of the chi-square distribution with $L$ degrees of freedom.

\subsection{Adaptive stochastic interventions}\label{sec: adaptive}

The previous sections focus on static interventions, where the same treatment distribution is assigned across all time periods. We extend the framework to adaptive interventions, in which the treatment distribution may depend on prior outcomes and covariates. Such extensions are particularly challenging if interventions span multiple time periods and depend on recent history, including intermediate covariates.  To handle this complexity, we build on literature on dynamic treatment regimes \citep[e.g.][]{moodie2012q, barrett2014doubly} to define the new CATE estimand.

Formally, we consider an adaptive intervention that can be represented by a sequence of one-period intervention distributions $F_{\bm h_t}=(F_{h_{t1}},\dots,F_{h_{tM}})$, where each $F_{h_{tm}}$ is allowed to depend on the complete observed history $\lHt[-m]$, i.e.\ $F_{h_{tm}}$ has density $f_{h_{tm}}(w\mid \lHt[-m])$ for all $w\in\mathcal{W}$ and $m = 1, 2, \dots, M.$  
We first define the expected number of outcome events in pixel $i$ at time $t$ under an adaptive intervention $F_{h_{t1}}$ which may depend on $\lHt[-1]$ as
\begin{equation*}
  N_{it}(F_{h_{t1}};\lHt[-1])
  = \int_{\mathcal{W}}N_{S_i}(Y_t(\lWt[-1],w_t)) f_{h_{t1}}(w_t\mid\lHt[-1])\dd w_t.
\end{equation*}
Next, we define the expected value of $N_{it}(F_{h_{t1}};\lHt[-1])$ where the treatment in the previous period is drawn from $F_{h_{t2}}$ based on $\lHt[-2]$,
\begin{equation}\label{eq:adaptive_def}
    N_{it}(F_{h_{t1}}, F_{h_{t2}};\lHt[-2])
  = \int_{\mathcal{W}} N_{it}(F_{h_{t1}};\lHt[-1])f_{h_{t2}}(w_{t-1}\mid \lHt[-2])\dd w_{t-1}.  
\end{equation}
Making the dependence on the observed history explicit in $N_{it}(F_{h_{t1}};\lHt[-1])$ highlights how the estimand is constructed sequentially and clarifies the integral in Eqn~\cref{eq:adaptive_def}. Repeating this construction sequentially, we define $N_{it}(F_{\bm h_{t}};\lHt[-M]),$ which averages over the treatments at time periods $t, t-1, \cdots, t - M + 1$ according to the adaptive stochastic intervention. For non-adaptive stochastic intervention, $F_{\bm h}$, $N_{it}(F_{\bm h};\lHt[-M])$ coincides with $N_{it}(F_{\bm h})$ given in Eqn~\cref{eq:Nti}. The corresponding CATE estimand can then be defined analogously to \Cref{subsec: estimand}, replacing $N_{it}$ with its adaptive version.
For estimation, we adjust \Cref{assump: overlap} as follows:
\begin{assumption}
\label{assump: overlap-adaptive}
There exists $\delta_W>0$ such that $e_{t-m+1}(w)>\delta_W f_{h_{tm}}(w\mid \lHt[-m])$ for all $w\in\mathcal{W}$, $1\le m\le M$ and $M\le t\le T$ for which $f_{h_{tm}}(w) > 0$, where $e_{t}(w)=f(W_t = w \mid \lHt[-1]).$ 
\end{assumption}

Finally, define the pseudo-outcome $\widetilde{Y}^{I}_{it}(F_{\bm h_t};\hat\ggamma)$ as done in \Cref{subsec: estimation}, using modified IPW weights:
$
\widetilde{Y}_{it}^{I}(F_{\bm h_t};\hat\ggamma) = \rho_{\bm h_t}(\hat\ggamma)\,N_{S_i}(Y_t),
$
where $\displaystyle\rho_{\bm h_t}(\hat\ggamma) 
  = \prod_{m=1}^M\frac{f_{h_{tm}}(W_{t-m+1}\mid \lHt[-m])}{e_{t-m+1}(W_{t-m+1};\hat\ggamma)}.$
The H\'ajek version of the pseudo-outcome is defined analogously. The theoretical results established in the previous sections continue to hold under the adaptive intervention framework. The details and the choice of $\mat Q$ for our variance upper bound estimator are provided in \Cref{a:sec:adaptive}.

\section{Sensitivity analysis to unmeasured confounding}\label{sec:sensitivity}

Our methodology requires unconfoundedness (\cref{assump: unconfoundedness}), implying that all relevant confounders are measured. Here, we develop a sensitivity analysis that evaluates the robustness of their empirical results to unmeasured, potentially time-varying confounding.  Details on the sensitivity analysis and its implementation are provided in \cref{a:sec:sensitivity}. Our approach can also assess sensitivity to misspecification of the functional form of the propensity score model.

\subsection{Unmeasured confounding and the expanded propensity score}

Let $U_t$ denote a time-varying unmeasured confounder. Assume that unconfoundedness holds only after conditioning on this variable and its history, i.e.,  
$f(W_t \mid \overline{\bm{W}}_{t-1}, \overline{\mathcal{Y}}_T, \overline{\mathcal{X}}_T, \overline{\mathcal{U}}_T)
= f(W_t \mid \overline{H}_{t-1}, \overline{U}_t)$,
where $\overline{\mathcal{U}}_T$ is the set of all possible values of $\overline{U}_t$ over all time periods.
In this setting, 
consistent estimation of causal effects requires the expanded propensity score defined as
$e_t^*(W_t) = f(W_t \mid \overline{H}_{t-1}, \overline{U}_t)$.
The corresponding IPW weights (defined in Eqn~\cref{eq: y_IPW}) are equal to,
\begin{equation}\label{eq: expanded_IPW}
\rho_{t\bm{h}}^\ast
=  \prod_{j=t-M+1}^{t} \dfrac{f_{h}(W_j)}{e_j^*(W_j)} = \prod_{j=t-M+1}^{t}\dfrac{e_j(W_j; \hat{\bm{\gamma}})}{e_j^\ast(W_t)}\rho_{t\bm h}(\hat\ggamma) = \alpha_t\rho_{t\bm h}(\hat\ggamma),  
\end{equation}
where $\alpha_t = \prod_{j=t-M+1}^{t}\frac{e_j(W_j; \hat{\bm{\gamma}})}{e_j^\ast(W_t)}$ denotes the cumulative ratio between the estimated propensity score based on measured covariates and the true propensity score based on all confounders.

However, since $U_t$ is not measured, we examine the extent to which estimation based on the estimated propensity score is robust to the presence of $U_t$ by letting the estimated propensity score deviate from the true propensity scores as follows,
\begin{align}
\Gamma^{-1} \le \frac{e_t(W_t; \hat{\bm{\gamma}})}{e_t^*(W_t)} \le \Gamma,
\label{supp_eq:sens_gamma}
\end{align}
where a greater value of the sensitivity parameter $\Gamma > 1$ implies a larger deviation.
Therefore, $\alpha_t$ satisfies that
$
\Gamma^{-M} \le \alpha_t \le \Gamma^M,
$ for  $ t = M, \dots, T.$

\subsection{Sensitivity analysis through zero-attainability}\label{subsec: sensitivity_through_zero_attainability}


The proposed sensitivity analysis assesses whether the CATE could remain constant across covariate values (i.e., homogeneous effect) under violations of unconfoundedness of magnitude $\Gamma$. The CATE coefficients based on the expanded propensity score are given by (see Eqn~\cref{eq: estimator_cate}),
\begin{equation}\label{eq: CATE_coeff}
\shiftmean\hat{\bbeta}_t^{H\ast} = \shiftmean\alpha_t\hat{\bbeta}_t^{H}. 
\end{equation}
The CATE coefficients are \textit{zero-attainable} if for a vector 
$\bm{\alpha} = (\alpha_M,\allowbreak  \dots, \alpha_T) \in [\Gamma^{-M}, \Gamma^M]^{T-M+1}$ 
it holds that $\shiftmean\hat{\bbeta}_t^{H\ast}=0$.  If zero is unattainable, treatment effects would be estimated to be heterogeneous even in the presence of unmeasured confounding of strength $\Gamma$. Thus, we define the robustness value of estimated CATE as the largest value of $\Gamma$ for which zero is unattainable.

When there is a single coefficient, zero-attainability can be evaluated by solving a linear program (see \cref{sub:sec:sensitivity_computation} for details).
In the presence of multiple coefficients (such as the coefficients of the spline basis functions in Eqn~\cref{eq:model}), zero attainability requires all coefficients to equal zero simultaneously. Evaluating the extent to which this is possible is difficult due to the increasing dimensionality. Instead, we evaluate zero attainability for each coefficient separately, using the procedure described in \cref{sub:sec:sensitivity_computation}. We use the smallest value of $\Gamma$ among all elements as the overall robustness value. This component-wise procedure is conservative because even if each component can attain zero separately, all components may not be able to attain zero under a common vector $\bm{\alpha}$. Consequently, the resulting robustness value is a lower bound on the robustness value that would be obtained by jointly evaluating all coefficients. 

Standard sensitivity analyses for the IPW estimator compute worst-case bounds for the treatment effect and subsequently check whether zero is within these bounds \citep[e.g.,][]{zhao2019sensitivity, rosenbaum2002observational}. However, since the H\'ajek estimator is nonlinear in the optimization parameters, computing these bounds often requires conservative approximations even for the ATE \citep{aronow2013interval, papadogeorgou2022causal}. 
Instead, we simplify computation by focusing solely on whether the estimated effect heterogeneity can be zero under a given value of $\Gamma$. As a result, our approach is exact for the sensitivity analysis of the ATE or CATE for a binary effect modifier.

\section{Simulation Studies}\label{sec:simulation}

In this section, we conduct simulation studies. 
To create a realistic data generating process, we base our simulations on the Iraq data described in \cref{subsec:data}.  We focus on a univariate moderator but consider both spatial and spatio-temporal variables. 

\subsection{Study design}\label{subsec:sim design}

For each simulation, we independently generate 500 datasets.
We set $T=500$, which is similar to our Iraq data.
We begin by generating a vector of confounders $\bm X_t = (X_1, X_2, X_{3t}, X_{4t})^\top$, as in \cite{papadogeorgou2022causal}.  Specifically, $X_1$ and $X_2$ are spatial confounders; $X_1$ is based on the distance to the main road network and the border of Iraq, while $X_2$ is based on the distance to Baghdad (see \cref{fig:illustration_s}).
Moreover, $X_{3t}$ and $X_{4t}$ are two spatio-temporal confounders based on the observed patterns of airstrikes and insurgent attacks in our Iraq data. 

We use the locations of all airstrikes in the entire study period to obtain a spatial density estimate of airstrike patterns, $\hat g(\omega)$. 
At each time period $t$, we first generate a point pattern $Z$ from a Poisson point process with the intensity $\exp(\rho_0 + \rho_1 \hat{g}(\omega))$ for $\rho_0\approx -2.7$ and $\rho_1 = 8$. Then, the confounder $X_{3t}$ is computed as a function of the distance to the closest realized point in $Z$, $D(\omega; Z)$, i.e., $X_{3t}(\omega)=\exp(-D(\omega; Z))$. A realization of $X_{3t}$ at a randomly chosen time period is shown in \cref{fig:illustration_st}.
We generate  $X_{4t}$ similarly but estimate the spatial density $\hat g$ based on the observed insurgent violence events with $\rho_0 \approx -3.2$ and $\rho_1 = 7.$ 
We conduct two sets of simulation studies: one with the spatial moderator $X_{2}$ and the other with the spatio-temporal moderator $X_{3t}$. All variables are included as confounders in all simulations.

We generate the treatment point pattern conditional on the covariates 
from a Poisson point process with intensity function
 $\lambda_t^W(\omega) = \exp\left\{\alpha_0 +\bm{\alpha}_{\bm X}^\top \bm X_t(\omega)+{\alpha}_{ W} W^\ast_{t-1}(\omega)+\alpha_{Y}Y_{t-1}^\ast(\omega)\right\}$,   
where $Y^\ast_{t-1}(\omega)$ and $W^\ast_{t-1}(\omega)$ are the smoothed versions of the realized point patterns $Y_{t-1}$ and $W_{t-1}$, respectively. Specifically, $Y^\ast_{t-1}(\omega) = \exp(-2D(\omega; Y_{t-1}))$, where $D(\omega; Y_{t-1})$ denotes the minimum distance of location $\omega\in\Omega$ from the observed outcome events at the previous time period, $Y_{t-1}$.  We define $W^\ast_{t-1}(\omega)=\exp(-2D(\omega; W_{t-1}))$ in a similar manner.

Additionally, we generate the outcome point pattern from a Poisson process with an intensity function $\lambda^Y_t$.  For the simulations with the spatial moderator $X_2$, we use 
 $\lambda_t^Y(\omega) = \exp\big\{\gamma_0 +{\bm\gamma}_{\bm X}^\top\bm X_t(\omega)+{\gamma}_{W} W^\ast_{(t-3):t}(\omega)+\gamma_{Y}Y_{t-1}^\ast(\omega)+\gamma_1X_2 W^\ast_{(t-3):t}(\omega)\big\}$, 
where $W^\ast_{(t-3):t} = \break\exp(-2D(\omega; \bm W_{(t-3):t}))$ and  $D(\omega;\bm W_{(t-3):t})$ denotes the minimum distance of location $\omega$ to treatment locations during any of the time periods $t-3$ to $t$. When the moderator is the spatio-temporal variable $X_{3t}$, we use the  intensity function:
$\lambda_t^Y(\omega) = \exp\{\gamma_0
 +\bm\gamma_{\bm X}^\top\bm X_t(\omega)+ \break \gamma_{W} W^\ast_{(t-3):t}(\omega)+{\bm \gamma}_{Y}Y_{t-1}^\ast(\omega) +\sum_{j=1}^4 \gamma_j X_{3,t-j}(\omega)W^\ast_{t-j+1}(\omega)\}$.
These model specifications yield on average approximately 5 treatment events and 30 outcome events per time period, closely reflecting the observed data, which show an average of 4.9 airstrikes and 31.3 IED attacks per day.

This data-generating process allows for spillover and carryover effects. 
In fact, the potential outcomes at time $t$ depend on the realized treatments and outcomes in all previous time periods either directly or indirectly due to the inclusion of the lagged treatment and outcome variables.

We consider stochastic interventions of the form $F^M_{h} = F_{h}\times\dots\times F_{h}$ where the intensity function is given by $h(\omega) = c\phi(\omega)$. We set $\phi$ equal to a density estimate of the generated treatment events (see \cref{fig:illustration_phi}). Here, the parameter $c$ regulates the expected number of treatment events.  For $M=1,3,7$, we consider the stochastic interventions with $c = 3,5,7$, denoted as $F^M_{h_1}$,$F^M_{h_2}$, and $F^M_{h_3}$ respectively. We estimate the CATE that contrasts between any two of these interventions.

In total, we evaluate four estimators: IPW/H\'ajek estimators based on the true propensity score and IPW/H\'ajek estimators based on the estimated propensity score. 
When computing pseudo-outcomes, we use either the IPW or stabilized H\'ajek weights.
We choose six basis functions $\basef[_1](\seffm[]), \dots, \basef[_L](\seffm[])$ for the working model in Eqn~\cref{eq:model}, i.e., $L = 6$. 
\cref{fig:moderator_hist} presents the histogram of the spatial and the spatio-temporal moderators across all pixels and all time periods. 
Since both moderators $X_2$ and $X_{3t}$ take values in $[0,1]$, we set $\basef[_1](r), \dots, \basef[_6](r)$ to be the basis functions of natural cubic splines on $[0,1]$ with equally spaced knots between 0 and 1.  
At each time period, we regress the pseudo-outcomes on the covariates formed by $\basef[_1](r), \dots, \basef[_L](r)$.

\subsection{Results}\label{subsec: simulation_res}

\begin{figure}[!t]
\centering
\includegraphics[width = 0.9\textwidth,trim = 0 20 0 35, clip]{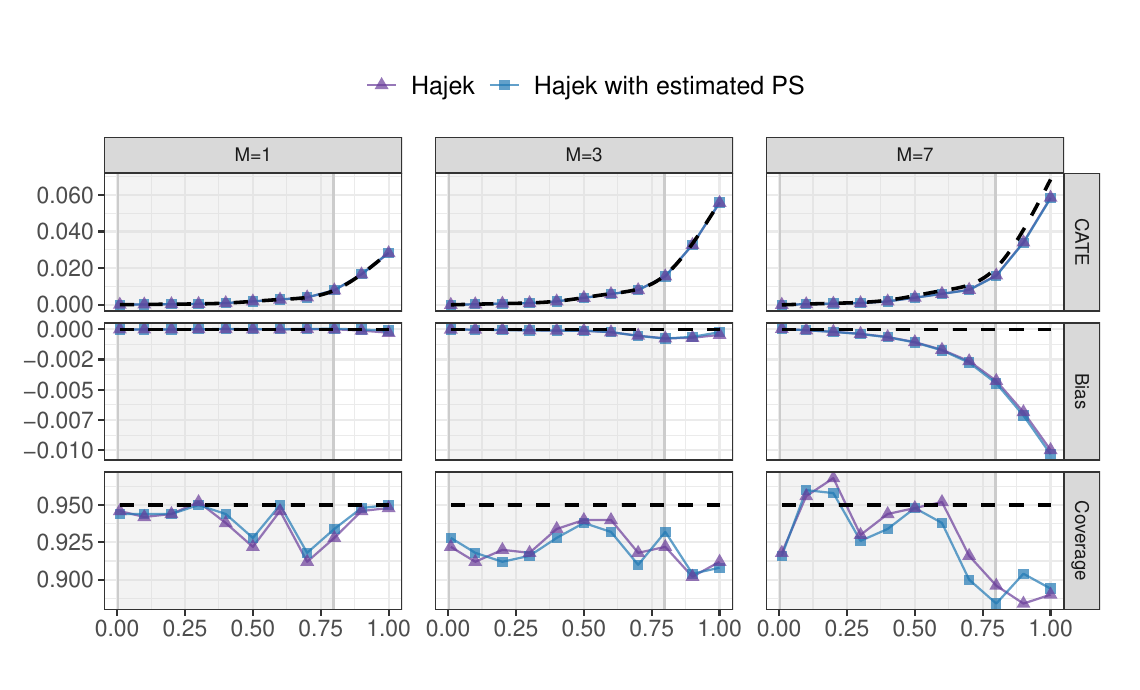}
\caption{The average of estimated CATE (first row), bias (second row), and coverage (third row) based on H\'ajek estimators across 500 simulations. Dashed lines in the first, second, and third rows represent the true CATE, bias of zero, and the nominal coverage of 95\%, respectively. The plot pertains to the stochastic intervention over $M = 1, 3, 7$ time periods. The purple line with triangles denotes the estimator based on the true propensity score, while the blue line with squares represent the estimators based on the estimated propensity scores. The shaded region indicates the 0.025 and 0.975 quantile range of the moderator values. } 
\label{fig:sim_est_13}
\end{figure}

We assess the empirical performance of the H\'ajek estimators under the most challenging scenario, which compares $F^M_{h_1}$ and $F^M_{h_3}$. 
\cref{fig:sim_est_13} presents the true and estimated CATEs using the H\'ajek estimator, the bias and the coverage rates of 95\% intervals, for the spatio-temporal moderator with $M = 1, 3$, and $7$, averaged over 500 Monte Carlo simulations.
The shaded region highlights the 0.025 and 0.975 quantile range of the spatio-temporal moderator values, on which we focus our discussion. Results outside this range are mostly based on extrapolation.
The true value of the CATE is different across data sets.
Since no closed-form expression is available, we use a Monte Carlo approximation for the true CATE values (see
\cref{a:montecarlo} for details).

Overall, the H\'ajek estimators perform well within the shaded region: the estimated CATEs closely track the true CATEs, with minimal bias. Empirical coverage rates remain close to the nominal 95\% level.
Even in the most challenging scenario ($M = 7$), the bias remains below 0.005. 
As the number of intervention time periods increases, empirical coverage decreases slightly for certain moderator values but remains above 0.88. For regions with limited data, particularly near the upper end of the moderator range, the estimated CATEs show small biases, especially when $M = 7$. However, the empirical coverage remains stable across all scenarios, indicating that the variance estimates account well for the limited information in that moderator range.  
The estimator performs equally well regardless of whether true or estimated propensity scores are used.

The results for the spatial moderator exhibit similar patterns, with the estimated CATEs closely following the true CATEs and empirical coverage rates remaining above 0.87 (see \cref{a: additional_sim}). Beyond the spatial moderator, \cref{a: additional_sim} also examines the H\'ajek estimator for alternative CATE contrasts and shows results for IPW estimators. While both perform similarly in low-variance cases with $M=1$, the H\'ajek estimators outperform IPW estimators for $M=3$ and $M=7$, consistently exhibiting smaller variances. Lastly, estimated propensity scores generally improve efficiency over true propensity scores, consistent with our theoretical results.

\section{Empirical Application}\label{sec:application}







We examine how the effects of airstrikes on insurgent violence depend on the prior provision of aid during the Iraq War (2003--2011).
We focus on the period of the so-called ``Surge'' (2007--2008) when US policymakers, alarmed by escalating violence, deployed additional forces, aid, and airstrikes. 
These efforts formed part of the broader {\it hearts and minds} counterinsurgency campaign, in which aid was meant to win over locals while airstrikes sought to deter insurgent groups.

By estimating heterogeneous effects 
we investigate the widespread but largely untested claim that aid and airstrikes work in tandem to reduce insurgent attacks.
Previous studies often aggregate data into coarse temporal windows (e.g., years) or administrative boundaries (e.g., provinces), limiting their ability to 
capture spillover and carryover dynamics \citep[e.g.][]{nunn_14,sexton_16}.  
Our analysis provides a more robust test in the presence of spatial and temporal interference.

\subsection{Data}\label{subsec:data}

We analyze newly declassified high-frequency data recording the date and precise location of US airstrikes and insurgent attacks against Coalition forces. 
Our data cover the time period from January 2007 to July 2008.
The information about insurgent attacks is based on the Significant Act 
data collected by Coalition forces in Iraq. 
During this period, some 68,573 insurgent attacks against US forces were recorded across 57 different types of events, including Small Arms Fire (SAF) and Improvised Explosive Device (IED). Each incident report includes the time, date, and nature of an attack, along with precise coordinates that permit spatio-temporal analysis at a fine-grained level. \cref{fig:violence_spatial} plots the spatial distribution of all insurgent attacks.  

We use declassified data on airstrikes obtained from the US Air Force's Combined Air Operations Center 
as our treatment variable. A total of 2,446 airstrikes were recorded during our study period. 
Each event records the date, coordinates, weapon type and number, and aircraft type(s) involved in the airstrike. \cref{fig:airstrikes_spatial} plots the spatial distribution of these airstrikes. 

We draw on a dataset of aid spending (in US dollars) as recorded by the US military and USAID. Some 23,204 aid projects totaling \$4.97 billion were delivered during this period. Two programs, the Commander's Emergency Response Program (CERP) and its USAID equivalent, Economic Support Funds (ESF), dominate these aid expenditures, representing about 50\% (n=11,714) and nearly 45\% (n=10,370) of all completed projects, respectively. 

\begin{figure}[!t]
\centering
\includegraphics[width=0.73\textwidth,trim = 0 0 0 0, clip]{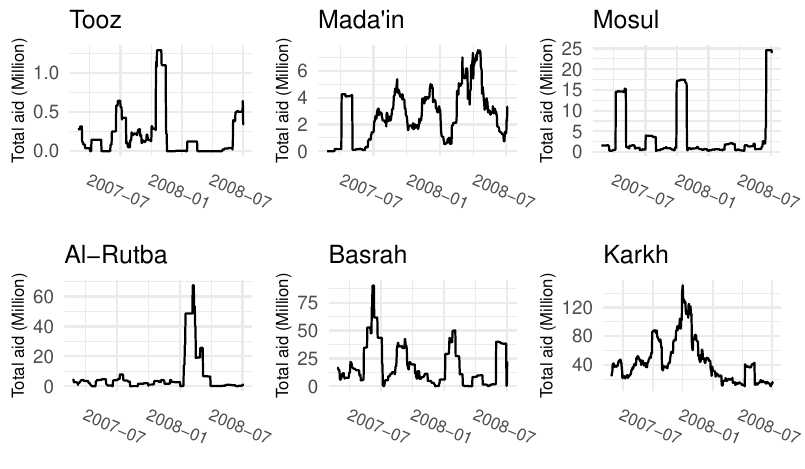}
\caption{Total aid amount (in million US dollars) over the previous month for six selected districts across Iraq, spanning from February 23, 2007, to July 5, 2008. Aid spending shows significant variation across districts and over time throughout the study period.}
\label{fig:aid_total}
\end{figure}

As \cref{fig:aid_total} illustrates, aid spending during the previous month varied widely across districts and over time during the study period due to shifting priorities, budgetary cycles, and patterns of insurgent violence.
The data are publicly available as part of the \texttt{geocausal} R package \citep{geocausal} and on the Harvard Dataverse \citep{DVN/IU8RQK_2023}.

\subsection{Design and Implementation}\label{subsec:design}

We analyze how the effects of US airstrikes on insurgent violence change with the amount of recent aid spending. We first consider as a moderator a binary indicator for whether a district received any US aid spending in the previous month. We also consider the aid spending per capita in the previous month as a continuous moderator. Our analysis focuses on the two most frequent types of insurgent attacks, IED and SAF. Since IED and SAF have different operational dynamics, airstrikes may affect them in different ways. 

We consider two types of interventions. We first construct static interventions following our simulation studies. The baseline density of airstrikes, denoted by $\phi_0(\omega)$, is estimated using airstrikes prior to the surge period.  We then estimate the causal contrast between the two stochastic interventions, $F_{\bm h_1} = F^M_{h_1}$ versus $F_{\bm h_2} = F^M_{h_2}$, where $h_1 = \phi_0(\omega)$ and $h_2 = 6\phi_0(\omega)$ for $M=1,\dots,10$. That is, we estimate the effects of increasing airstrike intensity sixfold for a period ranging from one to ten days. The scaling factors (1 and 6) are calibrated to the $25^{\mathrm{th}}$ and $75^{\mathrm{th}}$ percentiles of the observed airstrike counts. These counterfactual interventions allow us to estimate the effect of increasing the number of airstrikes while holding their spatial distribution fixed.

We also construct adaptive interventions that allow the counterfactual treatment assignment to depend on prior observed outcomes, among other variables. The details of constructing adaptive interventions are provided in \cref{a:sec:add_adaptive}.


Our propensity score model is a non-homogeneous Poisson point process with intensity $\lambda(\omega) = \exp(\bbeta^\top \bm X_t(\omega))$, 
where $\bm X_t(\omega)$ includes temporal splines, an indicator for
a surge period of airstrikes, and 32 spatial or spatio-temporal surfaces. The spatio-temporal surfaces are observed airstrikes, insurgent attacks, and shows of force (i.e., simulated bombing raids designed to
deter insurgents) patterns
from the last day, week, and month (9 surfaces), 
and the amount of aid spent (in US dollars) in each Iraqi district in the past month (1 surface). Time-invariant spatial covariates are distance from major cities, road networks and rivers (3 surfaces), population size (logged, measured in 2003; 
1 surface), and distances from local
settlements in each of the Iraqi districts (18 surfaces). 
These covariates account for changes in treatment assignment over time and differences across locations, including the possibility that recent violence affects where and when airstrikes occur. If the variables included in the propensity score model capture the factors that determine airstrike assignment, our procedure can identify the causal effects of airstrikes on subsequent violence. In \cref{a:sec:ps_prediction}, we evaluate the propensity score model using out-of-sample prediction. We find that the fitted model captures the general trend of airstrikes in different governorates.

Using estimated propensity scores, we construct pseudo-outcomes by truncating IPW weights at the $95^{th}$ percentile and using stabilized weights. Truncation is a common practice in applied research to avoid extreme weights. We divide Iraq's enclosing rectangle into a $128 \times 128$ grid of equally-sized pixels (each covering approximately 9 km $\times$ 8 km), yielding 8,451 pixels within Iraq. Pseudo-outcomes are computed for each pixel, with moderator values assigned by the Iraqi district containing the pixel's centroid. While this grid choice reflects a reasonable balance between computational cost and spatial resolution, other geographic units, such as administrative districts, could also be used. We therefore assess the sensitivity of our findings to alternative pixel resolutions and to a district-based discretization. Details are provided in \cref{a:subsec:pixel}, and the results show that our main findings are largely unchanged across these choices.

\subsection{Findings}\label{subsec: findings}

In this section, we present the results for static interventions. The corresponding results for adaptive interventions are shown in \cref{a:sec:add_adaptive}. For the analysis with binary aid, we consider the model $
  \tau_{t,\bm{h}_1,\bm{h}_2}(r;\bbeta_t) =\beta_{t,0}+\beta_{t,1} r
$
for each time period, where $r$ is the indicator variable for receiving aid in the previous month.  Given that the moderator is a binary variable, we need not consider the projection estimand in this case. 
Under this model, the CATE, $\shiftmean \beta_{t,1}$, represents the average difference, between pixels that received aid in the previous month and those that did not, in the effect of changing the stochastic intervention from $F^M_{h_1}$ to $F^M_{h_2}$. We multiply this value by the average number of pixels in a district, so that estimated quantities are interpreted as the estimated average moderation effect at the district level.

\begin{figure}[!t]
\centering
\includegraphics[width = 0.75\textwidth,trim = 0 10 0 0, clip]{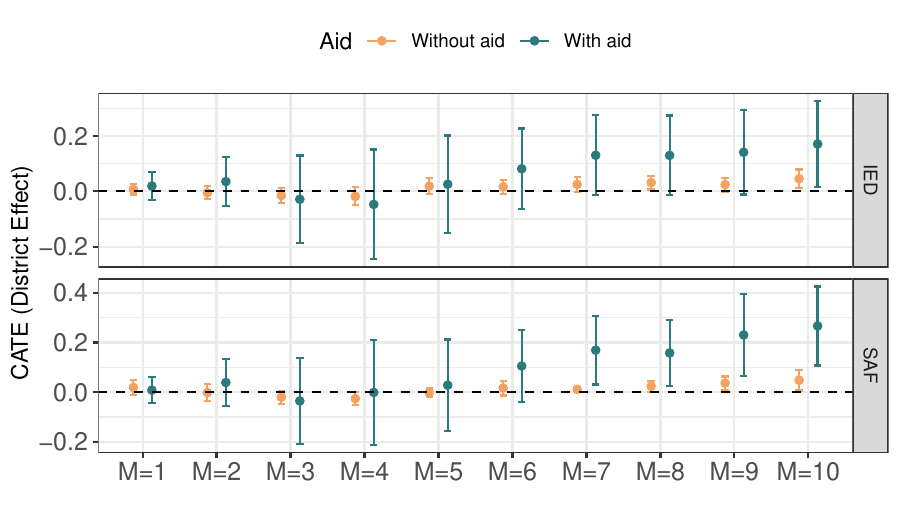}
\caption{Estimated CATEs of increasing airstrike intensity on the number of insurgency attacks in districts with and without aid in the previous month. The corresponding 95\% confidence intervals are also shown.}
\label{fig:app_binary_cate_aid}
\end{figure}



\cref{fig:app_binary_cate_aid} presents the estimated CATEs and 95\% confidence intervals for $r = 0$ and $r = 1$, for interventions that span from $1$ to $10$ days. The results of IED and SAF attacks show similar patterns. An increase in airstrikes over multiple consecutive days leads to an increase in subsequent insurgent violence. 
Districts with aid would experience a significant increase in SAF attacks for an increase in airstrikes spanning at least 7 days, while districts with or without aid would respond with a statistically significant increase in IED attacks for interventions spanning 10 days. 

Importantly, the estimated CATE for $r = 1$ is generally higher than that for $r = 0$, indicating locations that received prior aid respond more strongly to intensified airstrikes.
\cref{fig:app_binary_beta_aid} shows the difference in the estimated CATE for $r=0$ and $1$ with 95\% confidence intervals. The confidence intervals are calculated based on Eqn~\cref{eq: CI}. For IED, the estimation uncertainty is too large to draw a definitive conclusion about whether regions with and without aid exhibit different reactions to the increase of airstrikes. 
For SAF, however, the estimated difference is positive and statistically significant for the intervention period of at least seven days. This finding suggests that when the intervention period is sufficiently long, increasing the intensity of airstrikes sixfold leads to a greater number of insurgent attacks in districts with prior aid compared to those without.

Since the baseline distribution of airstrikes is not uniform across space, increasing airstrike intensity relative to the baseline leads to some regions experiencing a greater increase in airstrikes than others. \cref{fig:airstrike_diff} illustrates the estimated difference in the average number of airstrikes per district between the two interventions. One concern is that the significant differences observed in \cref{fig:app_binary_beta_aid} might be driven by this uneven increase in airstrike intensity. To address this, we incorporate the expected difference in airstrikes into the regression model and perform the analysis while adjusting for this expected difference. Most results remain consistent. The confidence interval for the CATE for $r = 0$ becomes wider, but the difference between the CATE for $r = 0$ and $r = 1$ becomes significant for IED when airstrike intensity is increased sixfold for 10 days after controlling for the expected difference in airstrikes. Detailed results are provided in \cref{a: application_inten}.

\begin{figure}[!t]
\centering
\includegraphics[width = 0.8\textwidth,trim = 0 0 0 0, clip]{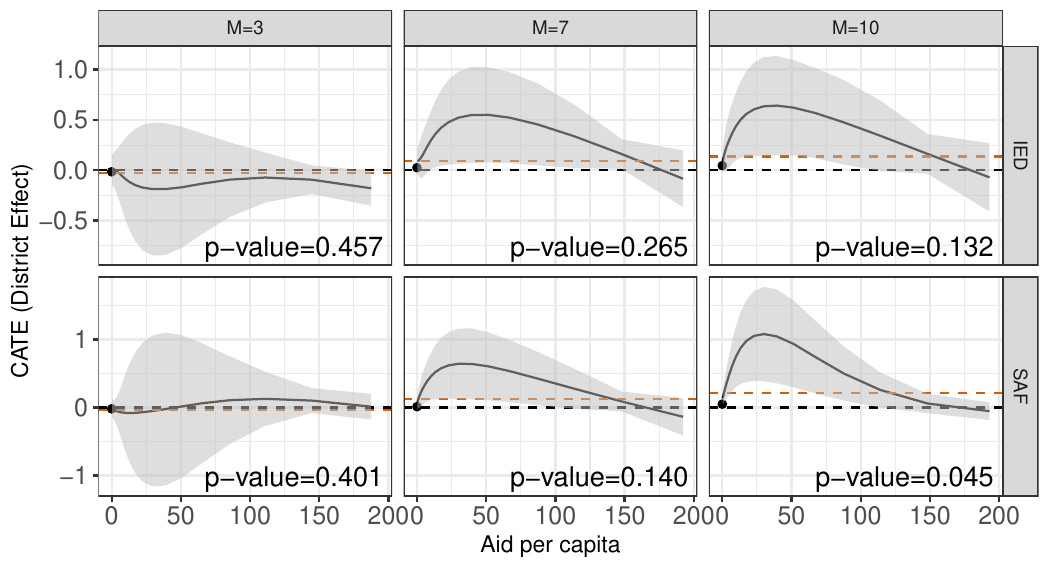}
\caption{Estimated CATEs of increasing airstrike intensity on the number of insurgency attacks in for different values of aid per capita received in the previous month with the shaded region indicating the 95\% confidence intervals. The red lines represent the estimated average treatment effects. The p-values correspond to tests for the overall heterogeneity effect.} 
\label{fig:app_continuous_aid}
\end{figure}

We next analyze the aid per capita. The histogram of continuous aid (\cref{fig:hist_aid}) reveals that many pixels have no aid during some time periods. To account for this, we include an indicator for zero aid in our working model. Consistent with simulation studies, we use four natural cubic spline basis functions, $z_l(r)$, based on strictly positive aid values, using equally spaced knots. Additionally, we set $\basef(0) = 0$ for $l = 1, 2, 3, 4.$ In sum, we consider the following working model:
  $\tau^{\mathrm{Proj.}}_{t,\bm{h}_1,\bm{h}_2}(r;\bbeta_t) =\beta_{t,0}+\sum_{l=1}^4\beta_{t,l}\basef(r)+\beta_{t,5} I\{r=0\}$, 
where $r$ represent the aid per capita received during the previous month.   We consider the projection CATE under this model and multiply all the estimates by the average number of pixels in a district to obtain a district-level effect.

\cref{fig:app_continuous_aid} displays the estimated CATEs for varying amounts of prior aid $r$, along with the average treatment effect when the expected number of airstrikes over the previous $M$ days increases from one to six per day. The p-values on the plot correspond to tests for the overall heterogeneous effect, with their calculation detailed in \cref{subsec: test}.
The results show similar patterns for both IED and SAF attacks. The average treatment effects are positive for $M = 7$ and $M = 10$, increasing as the intervention duration extends.
For interventions lasting three days, the estimated CATE is relatively constant with the amount of economic aid per capita. However, for interventions lasting one week and 10 days, the estimated CATE increases as the amount of aid per capita rises from \$0 to \$25, followed by a gradual decline as the aid increases from \$25 to \$200. 
There is statistical evidence for the heterogeneous effect of increasing airstrikes on SAF attacks (p-value $<$ 0.05), highlighting that areas with different levels of prior aid respond differently to increased airstrikes.

In sum, intensifying airstrikes for one week or more leads to a greater number of insurgent attacks. Among districts with relatively low levels of aid, those with a greater amount of aid in the previous month tend to exhibit a stronger response to intensified airstrikes.  However, when the amount of aid is sufficiently large, increasing the intensity of airstrikes appears to have a smaller effect on the number of insurgent attacks. We assess sensitivity to unmeasured confounding using the method introduced in \cref{sec:sensitivity}, and the largest robustness value across all such cases is $\Gamma = 1.16$. The relatively small $\Gamma$ value suggests that the estimated causal effect may be sensitive to violations of the unconfoundedness assumption. However, part of this sensitivity may reflect the high variability in estimating causal effects from spatio-temporal point pattern data. We further evaluate the robustness of our findings to alternative specifications in \cref{a:subsec:aid_lag,a:subsec:trunc_level}.


\section{Discussion}
\label{sec:discussion}

In this paper, we propose a method for analyzing treatment effect heterogeneity in spatio-temporal settings.  The proposed method estimates the CATE without imposing structural assumptions about spillover and carryover effects.
We establish asymptotic results for the proposed two-stage estimator based on stabilized weights, extending beyond existing work that considered only average treatment effects. We develop a new consistent estimator of the variance bound, which enables valid confidence intervals and shows improved finite-sample performance. In addition, we propose a statistical test of no heterogeneous treatment effects and establish a sensitivity analysis framework for the untestable unconfoundedness assumption. The proposed estimator performs well in simulation studies and exhibits lower variance than the standard IPW estimator.

Our application shows that airstrikes increased insurgent attacks in the short term, even when aid had been disbursed in the prior month at substantial levels. 
Aid moderated these negative effects, but only after high and likely unsustainable levels of aid had been delivered. In addition, these large aid outlays were only sufficient to offset the increased attacks created by the airstrikes themselves. Taken together, our findings question whether aid and airstrikes, the two central planks in counterinsurgency, actually work to reduce short-term insurgent attacks.  
 
There are several directions for future research. First, our asymptotic properties rely on the correct specification of the propensity score model. Future work could explore estimators robust to model misspecification. Second, while we employed a parametric propensity score model, future research could consider semi-parametric or nonparametric models.

\spacingset{1.88}
\bibliographystyle{chicago}
\bibliography{references,spatial_cate}


\newpage

\doparttoc 
\faketableofcontents 
\part{} 

\setcounter{page}{1}

\vspace{-20pt}
\begin{center}
{\sc \LARGE Supplementary Appendix for ``Estimating Heterogeneous Treatment Effects for Spatio-Temporal Causal Inference''}
\end{center}

\allowdisplaybreaks
\appendix
\setcounter{equation}{0}
\renewcommand{\theequation}{A.\arabic{equation}}

\setcounter{table}{0}
\renewcommand{\thetable}{A.\arabic{table}}

\setcounter{figure}{0}
\renewcommand{\thefigure}{A.\arabic{figure}}

\setcounter{theorem}{0}
\renewcommand{\thetheorem}{A.\arabic{theorem}}

\setcounter{assumption}{0}
\renewcommand{\theassumption}{A.\arabic{assumption}}

\setcounter{corollary}{0}
\renewcommand{\thecorollary}{A.\arabic{corollary}}

\setcounter{proposition}{0}
\renewcommand{\theproposition}{A.\arabic{proposition}}

\setcounter{lemma}{0}
\renewcommand{\thelemma}{A.\arabic{lemma}}

\setcounter{remark}{0}
\renewcommand{\theremark}{A.\arabic{remark}}

\makeatletter
\renewcommand{\theHequation}{A.\arabic{equation}}
\renewcommand{\theHtable}{A.\arabic{table}}
\renewcommand{\theHfigure}{A.\arabic{figure}}
\renewcommand{\theHtheorem}{A.\arabic{theorem}}
\renewcommand{\theHassumption}{A.\arabic{assumption}}
\renewcommand{\theHcorollary}{A.\arabic{corollary}}
\renewcommand{\theHproposition}{A.\arabic{proposition}}
\renewcommand{\theHlemma}{A.\arabic{lemma}}
\renewcommand{\theHremark}{A.\arabic{remark}}
\makeatother

\vspace{-40pt}
\setstretch{1.2}
\addcontentsline{toc}{section}{Supplement} 
\part{ } 
\parttoc
\clearpage

\setstretch{1.3}
\section{Table of notation}

Definitions for notations used in the manuscript are given in \cref{supp_tab:notation}.

\begin{table}[H]
    \centering
    \caption{Table of Notation}
    \label{supp_tab:notation}
    \begin{tabular}{|c|c|p{10cm}|}
    \hline
     & Symbol & Description \\
    \hline
    \multirow{4}{*}{Paths} & $\lWt$ & Treatmemt over the time periods $1,\dots,t$ \\
    & $\lwt$ & Realized treatment assignments for time periods $1\dots,t$ \\
    & $\lct{\mathcal{Y}}{t}$ & Collection of all potential outcomes for time periods $1\dots,t$ \\
    & $\lYt$ & Observed outcomes for time periods $1\dots,t$ \\
    \hline
    \multirow{3}{*}{Covariates} & $\lct{\mathcal{X}}{t}$ & Collection of all potential values of confounders over the time periods $1,\dots,t$ \\
    & $\lct{\bm{X}}{t}$ & Observed confounders for time periods $1\dots,t$ \\
    & $\effm$ & Observed values of moderators for time periods $1\dots,t$\\
    \hline
    \multirow{2}{*}{History} & $\lct{H}{t}$ & Observed history preceding the treatment at time $t + 1$\\
    & $\lct{H}{t}^\ast$ &  Complete history including all counterfactual values\\
    \hline
    \multirow{2}{*}{Intervention} & $M$ & The number of time periods over which we intervene \\
    &$h$ & Poisson point process intensity defining the stochastic intervention\\
    \hline
    \multirow{5}{*}{Estimands} & $\Nti$ & The expected number of outcome events in pixel $S_i$ at time $t$ for an intervention with intensity $\bm h$\\
    & $\tauti$ &  Expected difference in the number of outcome events for interventions $F_{\bm h'}$  over interventions $F_{\bm h''}$ at time $t$ in pixel $S_i$\\
    &$\catet$ &The conditional average treatment effect of intervention with intensity $\bm{h}'$ v.s. intervention with intensity $\bm{h}''$ for time period $t$\\
    &$\catetproj$ & The projected conditional average treatment effect for time period $t$\\
    & $\tau^\mathrm{Proj.}_{\bm h',\bm h''}(\seffm;\bbeta_1,\dots,\bbeta_T)$ & Conditional average treatment effect of intervention $F_{\bm h'}$ over $F_{\bm h''}$\\
    \hline
    \multirow{2}{*}{Estimators} & $\hat\bbeta^{I}$ & The IPW estimator for $\bbeta_t$\\
    & $\hat\bbeta^{H}$ &The Hajek estimator for $\bbeta_t$\\
    \hline
    \end{tabular}
\end{table}

\section{Asymptotic properties of the IPW estimator}\label{a:sec:IPW result}

In this section, we present the asymptotic properties of the IPW estimator. We first formally present the regularity conditions (\cref{subsec: regularityIPW}) and then show the asymptotic normality, the variance bound, and its consistent estimation (\cref{a:IPW_normal} and \cref{a:IPW_bound}). Those results are similar to the results for the H\'ajek estimator shown in \cref{subsec: asymptotic properties}.

\subsection{Regularity conditions}\label{subsec: regularityIPW}

We first state the regularity assumptions for establishing the asymptotic normality of the IPW estimator when the true propensity score is known.
For distribution $F_{\bm{h''}}$ and $F_{\bm{h'}}$, we define the IPW pseudo-outcome contrasts as $\pseudoEffI[it] = \pseudoOutI[''][it]-\pseudoOutI['][it]$. 
For notational simplicity, we denote $\pseudoEffI[it]$ as $\pseudoEffIVec[it,\bm h',\bm h''][](\hat\ggamma)$, and let $\pseudoEffIVec(\hat\ggamma) = \big(\pseudoEffIVec[1t,\bm h',\bm h''][](\hat\ggamma),\dots, \pseudoEffIVec[pt,\bm h',\bm h''][](\hat\ggamma)\big)^\top$ represent the vector of pseudo-outcome contrasts for the $p$ pixels at time period $t$. These contrasts are the IPW analogue of the H\'ajek pseudo-outcomes defined in \Cref{subsec: estimation}.

\begin{assumption}[Regularity conditions for the IPW estimator with the true propensity score] \singlespacing The following two conditions hold.
\begin{enumerate}[label=(\alph*)]
    \item\label{assump: IPW1a}(Bounded outcome and covariates) There exist positive constants $\delta_{Y},\delta_{z}<\infty$ such that $N_\Omega(Y_t)<\delta_{Y}$ and $|\basef[_j](\effm)|<\delta_{z}$ for all $j\in\{1,\dots,L\}$, $Y_t\in\lct{\mathcal{Y}}{T}$ and $\effm\in\lct{\mathcal{X}}{T}$.
    \item \label{assump: IPW1b} (Non-singularity and asymptotic variance) For all $t$, $\mat{Z}_t^\top \mat{Z}_t$ is non-singular and there exists a positive definite matrix $\widetilde{\mat{V}}^{I}$ such that 
    $$\frac{1}{T-M+1}\sum_{t=M}^T \Var[\aat\mid \lHt[-M]^\ast]\pto \widetilde{\mat{V}}^{I}$$ as $T\to \infty,$ where $\aat = \projm\pseudoEffIVec(\ggamma^\ast)$
\end{enumerate}
\label{assump:IPW1}
\end{assumption}
\cref{assump:IPW1}.\ref{assump: IPW1a} states that there exist uniform upper bounds on the number of outcome events and the absolute value of the moderator over all time period. In our empirical application, it is reasonable to assume that the number
of insurgent attacks occurring in Iraq and the amount of aid does not diverge to infinity as $t$ increases. \cref{assump:IPW1}.\ref{assump: IPW1b} requires that the moderators for each time period are linearly independent and that the sequence of covariance matrices converges to a positive definite matrix. The assumption about the convergence of the covariance matrix is relatively mild, given that the random vector $\aat$ is bounded above and bounded away from $\bm 0$. 

The following two sets of assumptions are needed for the asymptotic normality of the IPW estimator with the estimated propensity score. Those assumptions are also made in \cite{papadogeorgou2022causal}.  The first set of assumptions is about the behavior of the propensity score model.
\begin{assumption}[Regularity conditions for the propensity score model]\label{assump: IPWest} \singlespacing
Assume that the parametric form of the propensity score indexed by $\ggamma, f\left(W_t=w_t \mid \lHt[-1] ; \ggamma\right)$, is correctly specified and differentiable with respect to $\ggamma\in\mathbb{R}^K$, and let $$\psi\left(W_t, \lHt[-1] ; \ggamma\right)=\frac{\partial}{\partial \ggamma} \log f\left(W_t=w_t \mid \lHt[-1]=\lct{h}{t-1} ; \ggamma\right)$$ be twice continuously differentiable score functions. Let $\ggamma^\ast$ denote the true values of the parameters, where $\ggamma^\ast$ is in an open subset of the Euclidean space. Denote $\mathcal{F}_t=\lHt[-M+1]^\ast$. We assume that the following conditions hold:
\begin{enumerate}
    \item \label{assump: IPWest1}
    \begin{enumerate}[label = (\alph*)]
        \item $\E_{\ggamma^\ast}\left[\left\|\psi\left(W_t, \lHt[-1] ; \ggamma^\ast\right)\right\|^2\right]<\infty$,
        \item \label{assump: IPWest1b}There exists a positive definite matrix $\mat{V}_{p s}$ such that
        $$
        \frac{1}{T} \sum_{t=1}^T \E_{\ggamma^\ast}\left(\psi\left(W_t, \lHt[-1] ; \ggamma^\ast\right) \psi\left(W_t, \lHt[-1] ; \ggamma^\ast\right)^{\top} \mid \mathcal{F}_{t-1}\right) \pto \mat{V}_{ps}
        $$
        \item $\frac{1}{T} \sum_{t=1}^T \E_{\ggamma^\ast}\left[\left\|\psi\left(W_t, \lHt[-1] ; \ggamma^\ast\right)\right\|^2 I\left(\left\|\psi\left(W_t, \lHt[-1] ; \ggamma^\ast\right)\right\|>\epsilon \sqrt{T}\right) \mid \mathcal{F}_{t-1}\right] \pto 0$, for all $\epsilon>0$, as $T\to\infty.$
    \end{enumerate}
\item\label{assump: IPWest2} For all $k, j$, if we denote the $k^{th}$ element of the $\psi\left(w_t, \lHt[-1] ; \ggamma\right)$ vector by $\psi_k\left(w_t, \lHt[-1] ; \ggamma\right)$ and $P_{kj t}=\frac{\partial}{\partial \ggamma_j} \psi_k\left(W_t, \lHt[-1] ; \ggamma\right)|_{\ggamma^\ast}$, then $\E_{\ggamma^\ast}\left[\left|P_{k j t}\right|\right]<\infty$ and there exists $0<r_{k j} \leq 2$ such that $\sum_{t=1}^T \frac{1}{t^{r_{k j}}} \E_{\ggamma^\ast}\left(\left|P_{k j t}-\E_{\ggamma^\ast}\left(P_{k j t} \mid \mathcal{F}_{t-1}\right)\right|^{r_{k j}} \mid \mathcal{F}_{t-1}\right) \pto 0$
\item\label{assump: IPWest3} There exists an integrable function $\ddot{\psi}\left(w_t, \lct{h}{t-1}\right)$ such that $\ddot{\psi}\left(w_t, \lct{h}{t-1}\right)$ dominates the second partial derivatives of $\psi\left(w_{t}, \lct{h}{t-1} ; \ggamma\right)$ in a neighborhood of $\ggamma^\ast$ for all $\left(w_{t}, \lct{h}{t-1}\right)$
\end{enumerate}

\end{assumption}
\Cref{assump: IPWest} states that the model is correctly specified and the score function is bounded in the $L^2$-norm. That is, the model has finite variance, which is a mild requirement under standard positivity and common parametric models. Moreover, the expectation of the product of the score function averaged over all time periods stabilizes to a specific positive definite matrix. This condition guarantees that the asymptotic variance is well-defined and nonsingular. The Lindeberg-type tail condition in part 1(c) rules out rare, extremely large score contributions. This is a usual device for martingale central limit theorems in time-series settings. The assumption also limits the extent to which the derivative of the score function varies around its conditional expectation.  Lastly, the assumption ensures that the second partial derivatives of the score function are bounded in magnitude by an integrable function, which controls their growth and guarantees stability in a neighborhood of $\ggamma^\ast$. Overall, these assumptions represent standard regularity conditions of large-sample M-estimation with dependent data (smooth model, bounded moments, and stable information), and they are typically satisfied by common propensity score specifications.

\begin{assumption}[Regularity conditions on the score function of propensity score model for IPW estimators] \label{assump: IPWestadd} \singlespacing Let $\pseudoEffIVec(\ggamma)$ denote the pseudo-effect depend on the parameter $\ggamma$.
For $$s(\lHt[-1],W_t,Y_t;\ggamma) = \projm \pseudoEffIVec(\ggamma)-\bbeta_t^\ast,$$ and a propensity score function $\scorefunction$ satisfying \cref{assump: IPWest}, the following conditions hold.
\begin{enumerate}[label=(\alph*)]
    \item\label{assump: IPWestadda} There exists $\mat{U}\in\mathbb{R}^{K\times L}$ such that $$\shiftmean \E_{\ggamma^\ast}[\psi(W_t,\lHt[-1];\ggamma^\ast)s(\lHt[-1],W_t,Y_t;\ggamma^\ast)^\top\mid\lHt[-M]^\ast]\pto \mat{U},\ \mathrm{and}\ $$
    $\mat{V}^\ast = \begin{bmatrix}
        \widetilde{\mat{V}}^I & \mat{U}^\top\\
        \mat{U} & \mat{V}_{ps}
        \end{bmatrix}$ is positive definite.
    \item\label{assump: IPWestaddb} If $P_{jt} = \frac{\partial}{\partial \ggamma_j} s(\lHt[-1],W_t,Y_t;\ggamma)\Big|_{\ggamma^\ast}$, where $\ggamma_j$ is the $j^{th}$ entry of $\ggamma$, then there exists $r_j\in(0,2]$ such that
 $$\shiftmean \frac{1}{t^{r_j}}(|P_{jt}-\E_{\ggamma^\ast}[P_{jt}\mid\lHt[-M]^\ast]|)\pto 0.$$
\end{enumerate} 
\end{assumption}
\Cref{assump: IPWestadd} regulates the behavior of the propensity score$\,\psi(W_t,\overline H_{t-1};\gamma)\,$ and the function $s(\overline H_{t-1},W_t,Y_t;\gamma).$ Specifically, it states that the average expectation of the outer product of $s(\lHt[-1],W_t,Y_t;\ggamma)$ and $s(\lHt[-1],W_t,Y_t;\ggamma)$ converges to a matrix as $T$ grows and $\mat{V}^\ast$ is positive definite. This is the common requirement from large-sample M-estimation, which ensures that the limiting covariance matrix is nonsingular. Furthermore,  part (b) imposes a mild condition that the derivative of  $s(\lHt[-1],W_t,Y_t;\ggamma)$
 does not fluctuate wildly on average.

 The regularity conditions in Assumptions \ref{assump:IPW1}, \ref{assump: IPWest}, and  \ref{assump: IPWestadd} are formulated to accommodate complex spatio-temporal dependencies without requiring explicit spatial or temporal correlation decay rates. By treating events as spatial point patterns and constructing a Martingale Difference Sequence (MDS), we allow for the possibility that current treatments or outcomes depend on the entire observed history. This MDS framework is more flexible than traditional mixing-based approaches, as it only requires that the conditional moments of the score functions are well-behaved. These conditions are satisfied under the standard assumption that the joint process is strictly stationary and ergodic with finite $2+\delta$ moments. In non-stationary settings, these assumptions remain valid if the process is $\alpha$-mixing with a sufficiently fast decay rate.

\subsection{Asymptotic normality}\label{a:IPW_normal}

We now establish the asymptotic normality of the IPW estimator using the true and estimated propensity score.
\begin{theorem}[Asymptotic normality of the IPW estimator using the true propensity score]\label{thm:IPW true}
Suppose that Assumptions~\ref{assump: unconfoundedness},~\ref{assump: overlap},~and~\ref{assump:IPW1}.  Then
$$\frac{1}{\sqrt{T-M+1}}\sum_{t=M}^T\big(\hat{\bbeta}_t^{I}-\bbeta^\ast_t\big)\overset{d}{\to}N(\bm 0,\widetilde{\mat{V}}^{I}),$$ where $\widetilde{\mat{V}}^{I}$ represents the probability limit of $\frac{1}{T-M+1}\sum_{t=M}^T \mat{V}_t^{I}$ as $T\to \infty$ with
$\mat{V}_t^{I} = \Var[\hat\bbeta_t^{I}\mid \lHt[-M]^\ast]$ for $t\ge M$.
\end{theorem}

\begin{theorem}[Asymptotic normality of the IPW estimator using the estimated propensity score]\label{thm: IPW est} \singlespacing
Suppose that Assumptions~\ref{assump: unconfoundedness},~\ref{assump: overlap},~\ref{assump:IPW1},~\ref{assump: IPWest},~and~\ref{assump: IPWestadd} hold. Let $\hat\ggamma$ be the estimate obtained by solving the estimating equation $\sum_{t = M}^T\psi\left(W_t, \lHt[-1] ; \ggamma\right) = 0,$ where $\psi\left(W_t, \lHt[-1] ; \ggamma\right)$ is defined in Assumption~\ref{assump: IPWest}.  Then, as $T\to\infty$, we have $$\frac{1}{\sqrt{T-M+1}}\sum_{t=M}^T(\hat{\bbeta}^{I}_t-\bbeta^\ast_t)\dto N(\bm 0,\mat{V}^{I}).$$ 
\end{theorem}

\begin{theorem}[Asymptotic efficiency of the IPW estimator using the estimated propensity score]\label{thm: IPW efficiency}
If the propensity score model is correctly specified, the estimator $\shiftmean\hat{\bbeta}_t^{H}$ based on the estimated propensity score has an asymptotic variance that is no greater than the asymptotic variance of the same estimator using the known propensity score. That is, for $\widetilde{\mat{V}}^{I}$ defined in \cref{thm:IPW true} and
$\mat{V}^{I}$ defined in \cref{thm: IPW est}, $(\widetilde{\mat{V}}^{I}-\mat{V}^{I})$ is a positive semidefinite matrix.
\end{theorem}

\subsection{Consistent estimation of variance bound for the IPW estimator} \label{a:IPW_bound}

Similarly to the case of the H\'ajek estimator, it is not possible to consistently estimate the asymptotic variance of the IPW estimator. Thus, we consider a variance bound and propose a consistent estimator. 

\begin{proposition}[Consistent estimator of the variance upper bound]\label{prop: variance estimator} Suppose that Assumptions~\ref{assump: unconfoundedness},~\ref{assump: overlap},~and~\ref{assump:IPW1} hold.
  Let  $$\hat{\mat{V}}^{I^\ast} = \frac{1}{T-M+1}\sum_{t=M}^T\hat\bbeta_t^{I}\hat\bbeta_t^{I^\top}.$$ Then $\hat {\mat{V}}^{I}$ is a consistent estimator of ${\mat{V}}^{I^\ast}.$
\end{proposition}

\begin{corollary}[Consistent estimator of the variance upper bound using the estimated propensity score]\label{cor:1} Suppose that Assumptions~\ref{assump: unconfoundedness},~\ref{assump: overlap},~\ref{assump:IPW1},~\ref{assump: IPWest},~and~\ref{assump: IPWestadd} hold.
Moreover, let
$$\hat{\mat{V}}^{I} = \shiftmean \hat\bbeta_t^{I}\hat\bbeta_t^{I^\top}$$ Then, $\hat{\mat{V}}^{I}$ is a consistent estimator of ${\mat{V}}^{I^\ast}.$
\end{corollary}

\subsection{Proof of \cref{thm:IPW true}}\label{a: subsec: IPW proof true}

We will apply Theorem~A.2 of \cite{papadogeorgou2022causal}, which is a multivariate version of Theorem~4.16 of \cite{van2010time}. Define $\aat^\ast = \aat-\bbeta_t^\ast$.  We need to verify the existence of a filtration $\mathcal{F}_t$ such that the following conditions hold:
\begin{enumerate}
    \item $\E_{\bbeta^\ast_t}[\aat^\ast\mid \mathcal{F}_{t-1}]=0$ and $\E_{\bbeta^\ast_t}[||\aat^\ast||]<\infty.$
    \item  There exists a positive definite $\widetilde{\mat{V}}^{I}$ such that $$\frac{1}{T-M+1}\sum_{t=M}^T \E_{\bbeta^\ast_t}[\aat^\ast{\aat^\ast}^{\top}\mid \mathcal{F}_{t-1}]\pto\widetilde{\mat{V}}^{I}$$
    \item  $$\frac{1}{T-M+1}\sum_{t=M}^T \E_{\bbeta^\ast_t}\left[||\aat^\ast||^2I\left(||\aat^\ast||>\epsilon\sqrt{T}\right)\ \bigg | \ \mathcal{F}_{t-1}\right]\pto0$$ as $T\to \infty$ for all $\epsilon>0.$
\end{enumerate}
    
To check the first condition, take $\mathcal{F}_t = \lct{H}{t-M+1}^\ast$. Then for a stochastic intervention $F_{\bm h}$ over $M$ time periods, 
\begin{align*}
    & \E_{\bbeta^\ast_t}\left[\pseudoOutI[][it]\mid \mathcal{F}_{t-1}\right]\\
    = \ & \E_{\bbeta^\ast_t}\left[\prod_{j=t-M+1}^t\frac{f_{\bm {h}}(W_j)}{e_j(W_j)}N_{S_i}(Y_t)\ \bigg | \ \mathcal{F}_{t-1}\right]\\
    = \ &\int_{\mathcal{W}^M} \prod_{j=t-M+1}^t\frac{f_{\bm {h}}(w_j)}{e_j(w_j)}N_{S_i}(Y_t(\lWt[-M],w_{t-M+1},\dots,w_t))\times\\
    &\hspace{25pt}f(w_{t-M+1}\mid \lHt[-M]^\ast)f(w_{t-M+2}\mid \lHt[-M+1]^\ast)\dots f(w_{t}\mid \lHt[-1]^\ast)\dd \bm{w}_{(t-M+1):t}\\
    = \ &\int_{\mathcal{W}^M} \prod_{j=t-M+1}^t\frac{f_{\bm {h}}(w_j)}{e_j(w_j)}N_{S_i}(Y_t(\lWt[-M],w_{t-M+1},\dots,w_t))\prod_{j=t-M+1}^t e_j(w_j)\dd \bm{w}_{(t-M+1):t} \\
    = \  & \int_{\mathcal{W}^M} N_{S_i}(Y_t)\prod_{j=t-M+1}^t f_{\bm {h}}(w_j)\dd \bm{w}_{(t-M+1):t}\\
    = \ & \Nti,
\end{align*}
where the third equality follows from \cref{assump: unconfoundedness}. Let $\bm N_{t}(F_{\bm h}) = (\Nti[][\bm][1t],\dots,\Nti[][\bm][pt]),$ and $\tYt[I](F_{\bm h};\ggamma) = \left(\pseudoOutI[][1t][\bm][],\dots,\pseudoOutI[][pt][\bm][]\right)^\top.$ Then
\begin{align*}
    \projm \E_{\bbeta^\ast_t}[\pseudoEffIVec(\ggamma^\ast)\mid  \mathcal{F}_{t-1}]
    \ = \  &\projm \E_{\bbeta^\ast_t}\left[\tYt[I](F_{\bm h''};\ggamma)-\tYt[I](F_{\bm h'};\ggamma)\mid \mathcal{F}_{t-1}\right]\\
    = \ &\projm\left( \bm N_{t}(F_{\bm h''})-\bm N_{t}(F_{\bm h'})\right)\\
    = \ & \bbetat.
\end{align*}
Thus, 
\begin{align*}
     \E_{\bbeta^\ast_t}[\aat^\ast\mid \mathcal{F}_{t-1}]
    \ = \ & \projm \E_{\bbeta^\ast_t} [\pseudoEffIVec(\ggamma^\ast)-\bbeta^\ast_t\mid \mathcal{F}_{t-1}]\\
    \ = \ &\bbetat-\bbetat\\
    \ =\ & \bm 0
\end{align*}
Moreover, by Assumptions~\ref{assump: overlap}~and~\ref{assump:IPW1}\ref{assump: IPW1a}, each term in $\aat^\ast$ is bounded, and so is $\E_{\bbeta^\ast_t}[||\aat^\ast||]$. 

Next, Condition~2 is satisfied by \cref{assump:IPW1}\ref{assump: IPW1b}, while Condition~3 follows from the fact that $\E_{\bbeta^\ast_t}[||\aat^\ast||^2\mid \mathcal{F}_{t-1}]$ is bounded.
Hence, as $T\to\infty$, we have,
$$\frac{1}{\sqrt{T-M+1}}\sum_{t=M}^T\aat^\ast\overset{d}{\to}N(\bm 0,\mat{V})$$
and 
$$\frac{1}{T-M+1}\sum_{t=M}^T \E_{\bbeta^\ast_t}[\aat^\ast{\aat^\ast}^{\top}\mid\mathcal{F}_{t-1}] \ = \ \frac{1}{T-M+1}\sum_{t=M}^T \Var[\aat\mid\mathcal{F}_{t-1}]\pto\widetilde{\mat{V}}^{I}.
$$ 
Note that 
\begin{align*}
    &\frac{1}{\sqrt{T-M+1}}\sum_{t=M}^T\aat^\ast=
    \frac{1}{\sqrt{T-M+1}}\sum_{t=M}^T(\hat\bbeta_t^{I}-\bbetat)
\end{align*}
Thus, as $T\to\infty$, we have:
$$
\shiftfrac(\hat{\bbeta}_t^{I}-\bbetat)\overset{d}{\to}N(\bm 0,\widetilde{\mat{V}}^{I}).$$
\qed

\subsection{Proof of \cref{prop: variance estimator}}
\label{proof: prop1}

Let $\hat{\mat{V}}^{I}_t = \hat\bbeta_t^{I}\hat\bbeta_t^{I^\top}$ and $\mat{V}^{I}_t = \E_{\bbeta^\ast_t}[\hat\bbeta_t^{I}\hat\bbeta_t^{I^\top}\mid\lct{H}{t-M}^\ast]$ . Consider the sequence  $\hat{\mat{V}}_t^{I}-\mat{V}^{I}_t.$ Note that $\E[\hat{\mat{V}}_t^{I}-\mat{V}_t^{I}\mid\lHt[-M]^\ast] = 0.$
Since $\hat{\mat{V}}_t^{I}-\mat{V}_t^{I^\ast}$ is bounded, $\E|\hat{\mat{V}}_t^{I}-\mat{V}_t^{I^\ast}|<\infty.$ Also, $\sum_{t =1}^\infty t^{-2}\E[(\hat{\mat{V}}_t^{I}-\mat{V}_t^{I^\ast})^2]<\infty.$ Then, Theorem~1 of \cite{csorgHo1968strong} implies:
$$\frac{1}{T-M+1}\sum_{t=M}^T \big(\hat{\mat{V}}_t^{I}-\mat{V}_t^{I^\ast} \big)= \frac{1}{T-M+1}\sum_{t=M}^T \hat{\mat{V}}^{I}_t-\frac{1}{T-M+1}\sum_{t=M}^T \mat{V}_t^{I^\ast}\pto0.$$ Since ${\mat{V}}^{I^\ast}$ is the probability limit of $\frac{1}{T-M+1}\sum_{t=M}^T \mat{V}_t^{I^\ast}$ as $T\to\infty$, we have $\frac{1}{T-M+1}\sum_{t=M}^T \hat{\mat{V}}^{I}_t\pto {\mat{V}}^{I^\ast}$.
\qed

\subsection{Proof of \cref{thm: IPW est}}\label{a: subsec: IPW proof estimated}

We begin by introducing the following lemma. The proof of this lemma is omitted because it is similar to that of Lemma~A.2 given in \cite{papadogeorgou2022causal}.
\begin{lemma}\label{lemma1}
Suppose that \cref{assump: unconfoundedness} holds. Let $\sscorefunction$ be the score functions of a propensity score model that satisfy \cref{assump: IPWest}. Recall the definition of $\esteq[][][\ggamma]$ given in \cref{assump: IPWestadd}. Let $\mathcal{F}_t = \lHt[-M+1]^\ast$. Then, we have 
\begin{enumerate}
    \item $\E_{\ggamma^\ast}[\esteq[^\ast][][\ggamma]\scorefunction[^\ast]^\top\mid \ft] = -\E_{\ggamma^\ast}\left[\partialggamma \esteq[^\ast][][\ggamma]\Big|_{\ggamma^\ast}\ \bigg | \ \ft\right]$.
    \item $\partialgamma[_l]\sesteq[][][\ggamma] = -\projm \pseudoEffIVec\sum_{j = t-M+1}^t\sscorefunction[][_l]$, where $\sscorefunction[][_l]$ is the $l^{th}$ element of $\sscorefunction$.
    \item \begin{align*}
     & \partialgamma[_m]\partialgamma[_l]\sesteq[][][\ggamma] \\
    = &-\projm\pseudoEffIVec\Bigg\{\left[\sum_{j=t-M+1}^t\partialgamma[_m]\scorefunction[][_l]\right] \\
    & \hspace{1.5in} -\left[\sum_{j=t-M+1}^t\scorefunction[][_m]\right]\left[\sum_{j=t-M+1}^t\scorefunction[][_l]\right]\Bigg\}.
    \end{align*}
\end{enumerate}
\end{lemma}

\begin{proof}[Proof of \cref{thm: IPW est}]
Let $\ggamma\in\mathbb{R}^K$ be the parameters in the propensity score model whose score function is denoted by $\psi(w_t,\lct{h}{t-1};\ggamma).$  Consider the following estimating equations:
$$\esteq =  \begin{pmatrix}
    \aat^\ast-\mmu\\
    \scorefunction
\end{pmatrix}$$ 
where $\aat^\ast = \projm \pseudoEffIVec(\ggamma)-\bbeta_t^\ast$ and $\bm \theta^\top = (\mmu^\top,\bm \ggamma^\top).$
We apply Theorem~A.3 of \cite{papadogeorgou2022causal} to establish the asymptotic normality of the estimating equations. This requires verifying the following conditions. 
\begin{enumerate}
\item \label{condition: 1}
    \begin{enumerate}
    \item $\E_{\ttheta^\ast}[\esteq[^\ast]\mid\mathcal{F}_{t-1}]=0$ and $\E_{\ttheta^\ast}[||\esteq[^\ast]||]<\infty.$
    \item  There exists a positive definite $\mat{V}_a$ such that, as $T\to \infty$,  $$\frac{1}{T-M+1}\sum_{t=M}^T \E_{\ttheta^\ast}[\esteq[^\ast]\esteq[^\ast]^{\top}\mid\mathcal{F}_{t-1}]\pto \mat{V}_a$$
    
    \item As $T\to \infty$ for all $\epsilon>0$, we have: $$\frac{1}{T-M+1}\sum_{t=M}^T \E_{\ttheta^\ast}\left[||\esteq[^\ast]||^2I(||\esteq[^\ast]||>\epsilon\sqrt{T})\ \bigg | \ \mathcal{F}_{t-1}\right]\pto0$$ 
    \end{enumerate}
    \item\label{condition: 2} As $T\to \infty$, we have:  $$\shiftmean\E_{\ttheta^\ast}\left[\frac{\partial}{\partial \ttheta^\top}\esteq\Big|_{\ttheta^\ast}\ \bigg | \ \mathcal{F}_{t-1}\right]\pto \mat{V}_d$$ where $\mat{V}_d$ is invertible.
       
    \item\label{condition: 3} For all $k,j$, if we denote $P_{kjt} = \frac{\partial}{\partial \theta_j}\esteq[][_k]\Big|_{\ttheta^\ast}$, we have $\E_{\ttheta^\ast}[|P_{kjt}|]<\infty$, and there exists $0<r_{kj}\leq 2$ such that $\sum_{t=1}^T\frac{1}{t^{r_{kj}}}\E_{\ttheta^\ast}[|P_{kjt}-\E_{\ttheta^\ast}\left[P_{kjt}\mid\mathcal{F}_{t-1}]|^{r_{kj}}\mid \mathcal{F}_{t-1}\right]\pto0$ as $T\to\infty.$

    \item \label{condition: 4}There exists an integrable function $\ddot{\psi}(x)$ such that $\ddot{\psi}(x)$ dominates second partial derivatives of $\esteq{}$ in a neighborhood of $\ttheta^\ast$ for all $\{\lct{h}{t-1},w_t,y_t\}$    
\end{enumerate}
We check each of these four conditions in turn.
\paragraph*{Condition~\ref{condition: 1}.} Recall the definition of $\aat$ given in the proof of \cref{thm:IPW true} under the true value of $\ggamma$.  As shown in Appendix~\ref{a: subsec: IPW proof true}, we have $\E_{\ttheta^\ast}[\aat^\ast\mid\ft] = 0.$ Therefore, under ${\ttheta^\ast}^\top = ({\mu^\ast}^\top,{\ggamma^\ast}^\top) = (\bm 0^\top,{\ggamma^\ast}^\top)$, we have 
$$
    \E_{\ttheta^\ast}[\esteq[^\ast]\mid\ft]=0.
$$ 
According to Lemma~A.1 of \cite{papadogeorgou2022causal}, we have $\E_{\ggamma^\ast}[\scorefunction[^\ast]\mid\ft] = 0.$ Therefore, $\E_{\ttheta^\ast}[\esteq[^\ast]{}\mid\ft] = 0.$ Moreover, by Jensen's inequality, we have
\begin{align*}
    \E_{\ttheta^\ast}[||\esteq[^\ast]{}||]^2&\leq \E_{\ttheta^\ast}[||\esteq[^\ast]{}||^2]\\&\leq \E_{\ttheta^\ast}[||\aat^\ast||^2] + \E_{\ttheta^\ast}[||\scorefunction[^\ast]||^2].
\end{align*}
The first term is bounded due to \cref{assump:IPW1}.\ref{assump: IPW1a} and the seconded term is bounded due to \cref{assump: IPWest}.\ref{assump: IPWest1}. Thus, $\E_{\ttheta^\ast}[||\esteq[^\ast]||] <\infty.$ Note that $$
\begin{aligned}
    & \E_{\ttheta^\ast}[\esteq[^\ast]\esteq[^\ast]^\top\mid \ft]\\
    =& \begin{bmatrix}
    \E_{\ttheta^\ast}[\aat^\ast{\aat^\ast}^\top\mid \ft] & \E_{\ttheta^\ast}[\aat^\ast\scorefunction[^\ast]^\top\mid \ft]\\
    \E_{\ttheta^\ast}[\scorefunction[^\ast]{\aat^\ast}^\top\mid \ft] & \E_{\ttheta^\ast}[\scorefunction[^\ast]\scorefunction[^\ast]^\top\mid\ft]
    \end{bmatrix}.
\end{aligned}
$$ 
By Assumptions~\ref{assump:IPW1}.\ref{assump: IPW1b},~\ref{assump: IPWest}.\ref{assump: IPWest1}~and~\ref{assump: IPWestadd}, we have 
\begin{align*}
    \frac{1}{T-M+1}\sum_{t=M}^T \E_{\ttheta^\ast}[\esteq[^\ast]\esteq[^\ast]^{\top}\mid \mathcal{F}_{t-1}]
    \pto\begin{bmatrix}
            \mat{V} & \mat{U}^\top\\
            \mat{U} & \mat{V}_{ps}
    \end{bmatrix},
\end{align*} which is positive definite. For Condition~(c), we observe that for $\epsilon>0$,
\begin{align}
    &\shiftmean \E_{\ttheta^\ast}[||\esteq[^\ast]||^2I(||\esteq[^\ast]||>\epsilon\sqrt{T})\mid \ft]\nonumber\\
    = &\shiftmean \E_{\ttheta^\ast}[||\aat^\ast||^2 I(||\aat^\ast||^2+||\scorefunction[^\ast]||^2>\epsilon^2T)\mid \ft]\label{a:eq 1}\\
    &+\shiftmean \E_{\ttheta^\ast}[||\scorefunction[^\ast]||^2I(||\scorefunction[^\ast]||^2>\epsilon^2T-||\aat^\ast||^2)\mid \ft].\label{a:eq2}
\end{align}
By \cref{assump: IPWest}.\ref{assump: IPWest1} and the fact that $||\aat^\ast||^2$ is bounded, the term in Equation~\eqref{a:eq 1} converges in probability to 0. Since $$I(||\aat^\ast||^2+||\scorefunction[^\ast]||^2>\epsilon^2T)\leq I(||\aat^\ast||^2>\epsilon^2 T/2)+I(||\scorefunction[^\ast]||^2>\epsilon^2 T/2),$$ we have 
\begin{align*}
    & \E_{\ttheta^\ast}[||\aat^\ast||^2 I(||\aat^\ast||^2+||\scorefunction[^\ast]||^2>\epsilon^2T)\mid\ft]\\
    \leq \ &
    \E_{\ttheta^\ast}[||\aat^\ast||^2 I(||\aat^\ast||^2>\epsilon^2 T/2)\mid\ft]\\
    &+\E_{\ttheta^\ast}[||\aat^\ast||^2 I(||\scorefunction[^\ast]||^2>\epsilon^2 T/2)\mid \ft].
\end{align*}
By the arguments in the Proof of \cref{thm:IPW true}, we have $\shiftmean \E_{\ttheta^\ast}[||\aat^\ast||^2 I(||\aat^\ast||^2>\epsilon^2 T/2)\mid \ft]\pto 0.$ Let $B$ be the event that $||\aat^\ast||^2\leq||\scorefunction[^\ast]||^2.$ Then 
\begin{align*}
    & \E_{\ttheta^\ast}[||\aat^\ast||^2 I(||\scorefunction[^\ast]||^2>\epsilon^2 T/2)\mid \ft]\\
    = \ &  \E_{\ttheta^\ast}[||\aat^\ast||^2 I(||\scorefunction[^\ast]||^2>\epsilon^2 T/2)\mid B,\ft]\P(B\mid \ft)\\
    & \quad + \E_{\ttheta^\ast}[||\aat^\ast||^2 I(||\scorefunction[^\ast]||^2>\epsilon^2 T/2)\mid B^c,\ft]\P(B^c\mid \ft)\\
    \leq \ & \E_{\ttheta^\ast}[||\scorefunction[^\ast]||^2 I(||\scorefunction[^\ast]||^2>\epsilon^2 T/2)\mid B,\ft]\P(B\mid \ft)\\
    & + \E_{\ttheta^\ast}[||\aat^\ast||^2 I(||\aat^\ast||^2>\epsilon^2 T/2)|B^c,\ft]\P(B^c\mid \ft)\\
    \leq \ & \E_{\ttheta^\ast}[||\scorefunction[^\ast]||^2 I(||\scorefunction[^\ast]||^2>\epsilon^2 T/2)\mid \ft]\\
    & \quad + \E_{\ttheta^\ast}[||\aat^\ast||^2 I(||\aat^\ast||^2>\epsilon^2 T/2)\mid \ft]    
\end{align*}
where the last inequality follows from the law of total expectation. Since the mean of the last two terms converge in probability to 0 under \cref{assump: IPWest}.\ref{assump: IPWest1}, Condition~1.(c) holds.

\paragraph*{Condition~\ref{condition: 2}.} Observe that 
\begin{align}
    \frac{\partial}{\partial \ttheta^\top}\esteq &= 
    \begin{bmatrix}
        -\mat{I}_L &\frac{\partial}{\partial\ggamma^\top}\esteq[][_{1:L}]\\
        \frac{\partial}{\partial\mmu}\scorefunction & \frac{\partial}{\partial\ggamma^\top}\scorefunction
    \end{bmatrix}\nonumber\\&=
    \begin{bmatrix}
        -\mat{I}_L &\frac{\partial}{\partial\ggamma^\top}\esteq[][_{1:L}]\\
        \bm 0 & \frac{\partial}{\partial\ggamma^\top}\scorefunction
    \end{bmatrix}\label{A:eq3}
\end{align}
By \cref{lemma1}, we have 
\begin{align*}
   & - \shiftmean \E_{\ttheta^\ast}\left[\tfrac{\partial}{\partial\ggamma^\top}\esteq[^\ast][_{1:L}]\mid\ft\right]\\ = \ &\shiftmean \E_{\ttheta^\ast}\left[\esteq[^\ast][_{1:L}]\scorefunction[^\ast]^\top\mid \ft\right] \pto \mat{U}^\top.
\end{align*}
Moreover, by Lemma A.1 of \cite{papadogeorgou2022causal}, we have
$$ - \shiftmean \E_{\ggamma^\ast}\left[\tfrac{\partial}{\partial \ggamma^\top}\scorefunction\big|_{\ggamma^\ast}\ \bigg | \ \ft\right]\pto \mat{V}_{ps}.$$ Since $\mat{V}_{ps}$is positive definite, it is also invertible. Hence 
$$- \shiftmean \E_{\ttheta^\ast}\left[ \frac{\partial}{\partial \ttheta^\top}\esteq\Big|_{\ttheta^\ast} \right]\pto\begin{bmatrix}
    \mat{I}_L & \mat{U}^\top\\
    \bm 0 & \mat{V}_{ps}
\end{bmatrix},$$ and this limit matrix is invertible.

\paragraph*{Condition~\ref{condition: 3}.} Let $\esteq[][_k]$ be the $k^{th}$ element of $\esteq$. We need to show that for $j,k = 1,\dots,L+K$, if we denote $$P_{kjt}=\frac{\partial}{\partial\theta_j}\esteq[][_k]\Big|_{\ttheta^\ast},$$ then $\E_{\ttheta^\ast}[|P_{kjt}|]<\infty$, and there exists $0<r_{kj}\leq 2$ such that $$\sum_{t=1}^T\frac{1}{t^{r_{kj}}}\E_{\ttheta^\ast}\left[|P_{kjt}-\E_{\ttheta^\ast}[P_{kjt\mid \mathcal{F}_{t-1}}]|^{r_{kj}}\mid \mathcal{F}_{t-1}\right]\pto0$$ as $T\to\infty.$ We will consider four cases.
\begin{enumerate}
\item[]\textbf{Case1:} For $L<k,j\leq K+L $, the desired condition holds by \cref{assump: IPWest}.\ref{assump: IPWest1}. 
\item[]\textbf{Case2:} Let $1\leq k,j\leq L$. By Equation~\cref{A:eq3}, we have $P_{jkt} = -1$ or $0$. Thus, $P_{kjt} = \E_{\ttheta^\ast}[P_kjt\mid \ft]$ and $\E_{\ttheta^\ast}[|P_{jkt}|]<\infty.$
\item[]\textbf{Case3:} Let $1\leq j\leq L$ and $L<k\leq K+L$. Then $$\frac{\partial}{\partial \theta_{j}}\esteq[][_k] = 0.$$ Therefore, the condition is satisfied.
\item[] \textbf{Case4:} For $1\leq k\leq L$ and $L<j\leq K+L$, there exists $0<r_{kj}\leq 2$ such that $$\sum_{t=1}^T\frac{1}{t^{r_{kj}}}\E_{\ttheta^\ast}[|P_{kjt}-\E_{\ttheta^\ast}[P_{kjt|\mathcal{F}_{t-1}}]|^{r_{kj}}\mid \mathcal{F}_{t-1}]\pto0$$ by \cref{assump: IPWestadd}.\ref{assump: IPWestaddb}. Moreover, \cref{lemma1} implies that
\begin{align*}
    \E_{\ttheta^\ast}[|P_{kjt}|] &= \E_{\ttheta^\ast}\left[\left|\partialgamma[_{j-L}]\esteq[][_k]\Big|_{\ttheta^\ast}\right|\right]\\
    &= \E_{\ttheta^\ast}\left[\left|-\projm\pseudoEffIVec(\ggamma^\ast)\sum_{t' = t-M+1}^t\psi_{j-L}(W_{t'},\lct{H}{t'-1};\ggamma^\ast)\right|\right]
\end{align*}
By \cref{assump:IPW1}, $\mat{Z}_t$ and $\pseudoEffIVec(\ggamma^\ast)$ are bounded. Since \begin{align*}
     \E_{\ggamma^\ast}\left[\left|\psi_{j-L}(W_{t'},\lct{H}{t'-1};\ggamma^\ast)\right|\right]^2 &\leq \E_{\ggamma^\ast}\left[\psi_{j-L}(W_{t'},\lct{H}{t'-1};\ggamma^\ast)^2\right]\\
     &\leq \E_{\ggamma^\ast}\left[||\psi(W_{t'},\lct{H}{t'-1};\ggamma^\ast)||^2\right]<\infty,
\end{align*}
where the first inequality follows from Jensen's inequality and the second inequality follows from Assumption~\ref{assump: IPWest}.  Therefore, it follows that $\E_{\ttheta}[| P_{kjt}|]<\infty$.
\end{enumerate}
 
\paragraph*{Condition~\ref{condition: 4}.} We need to show that there exists an integrable function $\ddot{\psi}(x)$ such that $\ddot{\psi}(x)$ dominates the second partial derivatives of $\esteq{}$ in a neighborhood of $\ttheta^\ast$ for all $\{\lct{h}{t-1},w_t,y_t\}$. Let $k,l,m = 1,\dots,K+L$. We consider the following three cases.
\begin{enumerate}
\item[] \textbf{Case1:} If $l\leq L$ or $m\leq L$, then $\partialtheta[_m]\partialtheta[_j]\scorefunction[_k] = 0$, implying that the condition is satisfied.

\item[] \textbf{Case2:} If $k,l,m>L$, the condition holds by \cref{assump: IPWest}.\ref{assump: IPWest3}.

\item[] \textbf{Case3:} Suppose $k\leq L$ and $m,l>L.$  \cref{lemma1} implies:
\begin{align*}
     \partialgamma[_{m}]\partialgamma[_{l}]\sesteq[][_k][\ttheta] = &-\projmk\pseudoEffIVec(\ggamma)\left\{\sum_{j=t-M+1}^t\partialgamma[_{m-L}]\sscorefunction[][_{l-L}][j]\right.\\
     &\left. \quad -\sum_{j=t-M+1}^t\sscorefunction[][_{m-L}][j]\sum_{j=t-M+1}^t\sscorefunction[][_{l-L}][j]\right\}.
 \end{align*}
Note that $|\projmk\pseudoEffIVec(\ggamma)|\leq \delta$ for some $\delta\in\mathbb{R}$, so 
\begin{align*}
    & \left|\partialgamma[_{m}]\partialgamma[_{l}]\sesteq[][_k][\ttheta] \right|\\
    \leq \ & \delta\left|\sum_{j=t-M+1}^t\partialgamma[_{m-L}]\sscorefunction[][_{l-L}][j]\right|\\
    & \quad +\delta\left|\sum_{j=t-M+1}^t\sum_{j'=t-M+1}^t\sscorefunction[][_{m-L}][j]\sscorefunction[][_{l-L}][j']\right|\\
    \leq \ & \sum_{j=t-M+1}^t\delta\Bigg|\partialgamma[_{m-L}]\sscorefunction[][_{l-L}][j]\Bigg|\\
    & \quad +\sum_{j=t-M+1}^t\sum_{j'=t-M+1}^t \delta\left|\sscorefunction[][_{m-L}][j]\sscorefunction[][_{l-L}][j']\right|
\end{align*}
Consider the first term. \cref{assump: IPWest} implies that the second derivatives of $\sscorefunction$ is dominated by $\ddot{\psi}(w_t,\lct{h}{t-1})$ in a neighborhood of $\ggamma^\ast$. Let $B_\epsilon(\ggamma^\ast)$ be an open ball contained in the neighborhood. By Taylor's expansion, we have $$\left|\partialgamma[_{m-L}]\sscorefunction[][_{l-L}]\right| \ \leq \ \left|\partialgamma[_{m-L}]\sscorefunction[^\ast][_{l-L}]\right|+\epsilon K\ddot{\psi}(w_t,\lct{h}{t-1}),$$ on the ball $B_\epsilon(\ggamma^\ast).$ By \cref{assump: IPWest}.\ref{assump: IPWest2}, the quantity on the right hand side is an integrable function fixed in $\ggamma$, where we denote the maximum of these functions over $m,l$ by $\ddot{\psi_1}(w_t,\lct{h}{t-1}).$

Next, consider the second term. On the open ball $B_{\epsilon}(\ggamma^\ast)$, we have
 $$\left|\sscorefunction[][_{m-1}]\right|\leq \left|\sscorefunction[^\ast][_{m-1}]\right|+\epsilon K \ddot{\psi_1}(w_t,\lct{h}{t-1}),$$ and $\E_{\ggamma^\ast}[|\sscorefunction[^\ast][_{m-1}]|]<\infty$ by \cref{assump: IPWest}. Then, the right hand side is an integrable function fixed in $\ggamma$ where we denote the maximum of these functions over $m$ by $\ddot{\psi_2}(w_t,\lct{h}{t-1}).$ Putting these together, we have
 $$\left|\partialgamma[_{m}]\partialgamma[_{l}]\sesteq[][_k][\ttheta] \right|\leq M \delta\ddot{\psi_1}(w_t,\lct{h}{t-1})+M^2\delta[\ddot{\psi_2}(w_t,\lct{h}{t-1})]^2,$$ where the right hand side is an integrable function that satisfy the condition.
\end{enumerate}
 
Finally, we need to show that the solution of the following equation is consistent for $\ttheta^\ast$, $$\shiftmean \esteq = 0.$$ Let $\esteq[][_{1:L}]$ denote the first $L$ entries of $\esteq.$ \cref{thm:IPW true} shows that the IPW estimator based on the true propensity score is consistent, while the parameter estimates of the propensity score model $\hat\ggamma$ are also consistent for $\ggamma$.  Furthermore, $\esteq[][_{1:L}]$ is a continuous function of the propensity score, which is itself continuous in $\ggamma$.  Thus, Slutsky’s theorem implies that solving the following equation with the estimated propensity score parameters is also consistent, $$\shiftmean \esteq[][_{1:L}]=0.$$

Hence, all the conditions of Theorem A.3 in \cite{papadogeorgou2022causal} are satisfied, and as $T\to\infty$, we have
$$\sqrt{T}(\hat\ttheta_T-\ttheta^\ast)\dto N(0,\mat{V}_{\ttheta}),$$ where $\hat\ttheta_T$ is the solution to $$\shiftmean \esteq=0$$ and $\mat{V}_{\ttheta} = \mat{A}^{-1}\mat{B}(\mat{A}^{-\top})$ for
\begin{equation}\label{a:eq4}
   \mat{A} = \begin{bmatrix}
 \mat{I}_L &\mat{U}^\top\\
 \bm 0 &\mat{V}_{ps}
 \end{bmatrix}\hspace{4mm}\mathrm{and}\hspace{4mm}\mat{B} = 
 \begin{bmatrix}
     \widetilde{\mat{V}}^I &\mat{U}^\top\\
     \mat{U} & \mat{V}_{ps}
 \end{bmatrix}.  
\end{equation}
Note that $\shiftmean\hat\bbeta_t^{I}-\bbeta_t^\ast$ is the first $L$ entries of $\hat\ttheta_T$. As a result, we have $$\shiftfrac(\hat\bbeta_t^{I}-\bbeta_t^\ast)\dto N(\bm 0,\mat{V}^{I}).$$
\end{proof}

\subsection{Proof of \cref*{thm: IPW efficiency}}

The asymptotic variance $\mat{V}^{I}$ corresponds to the $L\times L$ submatrix located in the upper-left corner of matrix $\mat{A}^{-1}\mat{B}\mat{A}^{-\top}$ where $\mat{A},\mat{B}$ are defined in Equation~\cref{a:eq4}. We have
\begin{align*}
       \mat{A}^{-1}\mat{B}\mat{A}^{-\top} &= \begin{bmatrix}
           \mat{I}_L & \mat{U}^\top\\
           \bm 0 & \mat{V}_{ps}
       \end{bmatrix}^{-1} 
       \begin{bmatrix}
     \widetilde{\mat{V}}^{I} &\mat{U}^\top\\
     \mat{U} & \mat{V}_{ps}
 \end{bmatrix}
 \begin{bmatrix}
       \mat{I}_L & \mat{U}^\top\\
       \bm 0 & \mat{V}_{ps}
   \end{bmatrix}^{-\top} \\
   &= \begin{bmatrix}
           \mat{I}_L^{-1} & -\mat{I}_L^{-1}\mat{U}^\top \mat{V}_{ps}^{-1}\\
           \bm 0 & \mat{V}_{ps}^{-1}
       \end{bmatrix} 
       \begin{bmatrix}
     \widetilde{\mat{V}}^{I} &\mat{U}^\top\\
     \mat{U} & \mat{V}_{ps}
 \end{bmatrix}
 \begin{bmatrix}
       \mat{I}_L^{-\top} & \bm 0\\
       -\mat{V}_{ps}^{-1\top}\mat{U}^\top \mat{I}_L^{-1\top} & \mat{V}_{ps}^{-\top}
   \end{bmatrix}\\
   & =  \begin{bmatrix}
       \widetilde{\mat{V}}^{I} -\mat{U}^\top \mat{V}^{-1}_{ps} \mat{U}^{\top} & \dots\\
       \dots & \dots
   \end{bmatrix}.
\end{align*} Note that $\mat{V}^{I}$ is the asymptotic variance of the estimator based on the true propensity score. Let $\bm x$ be a nonzero vector in $\mathbb{R}^L.$ Then $\bm x^{*} = \mat{U}^{\top}\bm x\in\mathbb{R}^K$. Since $\mat{V}_{ps}$ is positive definite, so is $\mat{V}_{ps}^{-1},$ implying $\bm x^\top \mat{U}^\top \mat{V}^{-1}_{ps} \mat{U}^{\top}\bm x = {\bm{x}^{*}}^\top \mat{V}^{-1}_{ps} \bm x^{*} \geq 0.$ Thus, $\mat{U}^\top \mat{V}^{-1}_{ps} \mat{U}^{\top}$ is a positive semidefinite matrix, completing the proof.
\qed

\subsection{Proof of \cref*{cor:1}}
 
The proof follows directly from the proof of \cref{prop: variance estimator} and the fact that $\hat\ggamma$ is consistent for $\ggamma.$
\qed


\section{Asymptotic properties of the H\'ajek estimator}\label{a:sec:Hajek_true}

In this section, we present the asymptotic properties of H\'ajek estimator based on either the true or estimated propensity score.  We begin by stating the required regularity conditions and then establish that the H\'ajek estimator is asymptotically normal.  Although the asymptotic variance is not identifiable, we derive its upper bound and propose a consistent estimator of the bound.

\subsection{Regularity conditions}\label{a:sec:Hajek_regularity}

We represent an additional assumption that is required to achieve the asymptotic normality of the H\'ajek estimator. For $M\leq t\leq T$, define
$$ 
\bbeta_t^{'} =\solvem \bm N_t(F_{\bm h'}) \ \mathrm{and}\ 
  \bbeta_t^{''} =\solvem \bm N_t(F_{\bm h''}).
$$
\begin{assumption} \singlespacing
Define $$\aat=\begin{pmatrix}
    \projm\tYt[I](F_{\bm h'};\ggamma^\ast)\\ 
    \projm\tYt[I](F_{\bm h''};\ggamma^\ast)\\
    \rho_{t\bm h'}(\ggamma^\ast) \\
    \rho_{t\bm h''}({\ggamma^\ast})
\end{pmatrix}.$$ 
Then, there exists a positive definite matrix $\mat{V}^{H}$ such that as $T\to \infty$, 
$$\frac{1}{T-M+1}\sum_{t=M}^T \Var[\aat\mid \lHt[-M]^\ast]\pto \widetilde{\mat{V}}^{H}.$$ 
\label{assump:Hajek}
\end{assumption}
\cref{assump:Hajek} is slightly stronger than \cref{assump:IPW1}.\ref{assump: IPW1b}, since $\pseudoEffHVec(\ggamma) = \tYt(F_{\bm h''};\ggamma)-\tYt(F_{\bm h'};\ggamma).$

The next assumption is similar to \cref{assump: IPWestadd} with a different form of $s(\lHt[-1],W_t,Y_t;\ggamma).$
\begin{assumption}[Regularity conditions on score function of propensity score model for H\'ajek estimators]\label{assump: Hajekestadd} \singlespacing
For a stochastic intervention $F_{\bm h}$, let $\rho_{t\bm h}(\ggamma) = \prod_{j=t-M+1}^{t}\frac{f_{\bm {h}}(W_t)}{e_t(W_t;\ggamma)}$ and $\tYt(F_{\bm h};\ggamma) = \rho_{t\bm h}(\ggamma)\left(N_{S_1}(Y_t),\dots,N_{S_p}(Y_t)\right)^\top,$ and $\pseudoEffHVec(\ggamma) = \tYt(F_{\bm h''};\ggamma)-\tYt(F_{\bm h'};\ggamma)$.  Define
$$s(\lHt[-1],W_t,Y_t;\ggamma)=\begin{pmatrix}
    \projm\tYt[I](F_{\bm h'};\ggamma)-\bbeta_t^{'}\\ 
    \projm\tYt[I](F_{\bm h''};\ggamma)-\bbeta_t^{''} \\
    \rho_{t\bm h'}(\ggamma)-1 \\
    \rho_{t\bm h''}(\ggamma)-1
\end{pmatrix}, $$ 
and a propensity score function $\scorefunction$ satisfying \cref{assump: IPWest}, the following conditions hold.
\begin{enumerate}[label=(\alph*)]
    \item\label{assump: Hajekestadda} There exists $\mat{U}\in\mathbb{R}^{K\times 2L}$ such that $$\shiftmean \E_{\ggamma^\ast}[\psi(W_t,\lHt[-1];\ggamma^\ast)s(\lHt[-1],W_t,Y_t;\ggamma^\ast)^\top\mid \lHt[-M]^\ast]\pto \mat{U},\ \mathrm{and}\ $$
    $\mat{V}^\ast = \begin{bmatrix}
            \widetilde{\mat{V}}^H & \mat{U}^\top\\
            \mat{U} & \mat{V}_{ps}
    \end{bmatrix}$ is positive definite.
    \item\label{assump: Hajekestaddb} If $P_{jt} = \frac{\partial}{\partial \ggamma_j} s(\lHt[-1],W_t,Y_t;\ggamma)\Big|_{\ggamma^\ast}$, where $\ggamma_j$ is the $j^{th}$ entry of $\ggamma$, then there exists $r_j\in(0,2]$ such that 
        $$\shiftmean \frac{1}{t^{r_j}}\left(|P_{jt}-\E_{\theta^\ast}[P_{jt}\mid \lHt[-M]^\ast]|\right)\pto 0.$$
    \end{enumerate} 
 \end{assumption}

\subsection{Asymptotic normality}

The asymptotic normality of the H\'ajek estimator based on the estimated propensity score is presented as \cref{thm: Hajek est}.  Here, we state the analogous result for the H\'ajek estimator with the true propensity score.

\begin{theorem}[Asymptotic normality of the H\'ajek estimator using the true propensity score]\label{thm: Hajek true} \singlespacing
Suppose that Assumptions~\ref{assump: unconfoundedness},~\ref{assump: overlap},~\ref{assump:IPW1},~and~\ref{assump:Hajek} hold.  Then,
$$
\frac{1}{\sqrt{T-M+1}}\sum_{t=M}^T\big(\hat{\bbeta}_t^{H}-\bbeta^\ast_t\big)\overset{d}{\to}N(\bm 0,\mat{J}\widetilde{\mat{V}}^{H}\mat{J}^\top),
$$ 
where $\widetilde{\mat{V}}^{H}$ represents the probability limit of $\frac{1}{T-M+1}\sum_{t=M}^T \widetilde{\mat{V}}_t^{H}$ as $T\to \infty$ with $\widetilde{\mat{V}}_t^{H} = \Var[\aat(\ggamma^\ast)\mid\lHt[-M]^\ast]\  \mathrm{for}\  t\ge M,$ where $\aat$ and $\mat{J}$ are defined as, 
$$  \aat(\ggamma)=\begin{pmatrix}
    \projm\tYt[I](F_{\bm h'};\ggamma^\ast)\\ 
    \projm\tYt[I](F_{\bm h''};\ggamma^\ast)\\
    \rho_{t \bm h'}(\ggamma) \\
    \rho_{t \bm h''}({\ggamma})
\end{pmatrix},$$
and $\mat{J} = \begin{bmatrix}
    \mat{I} &-\mat{I} &-\shiftmean \bbeta_t^{'} &\shiftmean \bbeta_t^{''}
\end{bmatrix}.$

\end{theorem}

\subsection{Consistent estimation of variance bound for the H\'ajek-estimator}

For the case of the estimated propensity score, the result is stated as \cref{cor:2}.  We state the result for the H\'ajek estimator with the true propensity score.
\begin{proposition}[Consistent estimation of variance bound for the H\'ajek estimator with the true propensity score]\label{prop: variance Hajek} \singlespacing
Suppose that Assumptions~\ref{assump: unconfoundedness},~\ref{assump: overlap},~\ref{assump:IPW1},~and~\ref{assump:Hajek} hold.  Let
  \begin{align*}
       &\hat{\mat{V}}^{H} = \shiftmean\hat{\mat{V}}^{H}_t\text{ with }\hat{\mat{V}}_t^{H} = \aat(\ggamma^\ast){\aat(\ggamma^\ast)}^\top\ \mathrm{and}\ \\
       &\hat{\mat{J}} = \begin{bmatrix}
    \mat{I} &-\mat{I} &-\shiftmean \projm\tYt(F_{\bm h'};\ggamma^\ast) &\shiftmean \projm\tYt(F_{\bm h''};\ggamma^\ast)
\end{bmatrix} 
\end{align*}
  Then $\hat{\mat{J}}\hat{\mat{V}}^{H} \hat{\mat{J}}^\top$ is a consistent estimator for $\mat{J}\mat{V}^{H^\ast}\mat{J}^\top.$
\end{proposition}

\subsection{Proof of \cref{thm: Hajek true}}

The proof is similar to that of \cref{thm:IPW true}. The key difference is the definition of $\bm A_t^\ast$.
We apply Theorem~$A$.2 of \cite{papadogeorgou2022causal}. Set $\mathcal{F}_t = \lct{H}{t-M+1}^\ast$, and define $\aat^\ast = \aat(\ggamma^\ast)-({\bbeta_t^{'}}^\top,{\bbeta_t^{''}}^\top,\rho_{\bm h'},\rho_{\bm h''})^\top$ where $\rho_{\bm h'} = \rho_{\bm h''} = 1.$ 

We first show that $\aat^\ast$ is a martingale difference sequence, i.e., $\E[\aat^\ast] = 0$ and $\E[||\aat^\ast||]<\infty.$ By similar arguments used in the proof of \cref{thm:IPW true} (see \cref{a: subsec: IPW proof true}),  for any $F_{\bm h}$, we have
$\E[\tYt[I](F_{\bm h})\mid \mathcal{F}_{t-1}] = N_t(F_{\bm h})$, and thus,
\begin{align*}
  \E[\projm\tYt[I](F_{\bm h'})\mid \mathcal{F}_{t-1}] &= \projm \bm N_t(F_{\bm h})=\bbeta_t^{'}.
\end{align*} 
We have 
\begin{align*}
 \E[\rho_{t\bm h'}\mid \mathcal{F}_{t-1}] 
 \ = \ &\int_{\mathcal{W}^M} \prod_{j=t-M+1}^t  \frac{f_{\bm h'}(w_j)}{e_j(w_j)} f(w_{j}\mid \lct{H}{j-1}^\ast)\dd \bm{w}_{(t-M+1):t}\\
 \ = \ &\int_{\mathcal{W}^M}  \prod_{j=t-M+1}^t \frac{f_{\bm h'}(w_j)}{e_j(w_j)} e_j(w_j) \dd \bm{w}_{(t-M+1):t} \\
 \ = \ & \int_{\mathcal{W}^M} \prod_{j=t-M+1}^t f_{\bm h'}(w_j)\dd \bm{w}_{(t-M+1):t} \\
 \ = \ & 1.
\end{align*}
where the second equality follows from \cref{assump: unconfoundedness}.

Similarly, $\E[\projm\tYt[I](F_{\bm h''})\mid \mathcal{F}_{t-1}]=\bbeta_t^{''}$ and $\E[\rho_{t\bm h''}\mid \mathcal{F}_{t-1}] = 1.$
Thus, $\E[\aat^\ast\mid \mathcal{F}_{t-1}] = 0.$
Furthermore, since every element in $\aat^\ast$ is bounded by Assumptions~\ref{assump: overlap}~and~\ref{assump:IPW1}.\ref{assump: IPW1a}, $||\aat^\ast||$ is bounded and hence $\E[||\aat^\ast||]<\infty.$ For an arbitrary $\epsilon>0$, $I(||\aat^\ast||>\epsilon\sqrt{T})=0$ for large $t$, since $||\aat^\ast||$ is bounded. Therefore, as $T\to \infty$ for all $\epsilon>0$, we have
$$\frac{1}{T-M+1}\sum_{t=M}^T E\left[||\aat^\ast||^2I(||\aat^\ast||>\epsilon\sqrt{T})\ \bigg | \ \mathcal{F}_{t-1}\right]\pto0.$$  
The remaining condition of the multivariate Martingale central limit theorem is satisfied by \cref{assump:Hajek}. Hence, 
$$\frac{1}{\sqrt{T-M+1}}\sum_{t=M}^T\aat^\ast\overset{d}{\to}N(\bm 0,\widetilde{\mat{V}}^{H})$$ where, as $T\to\infty$,
$$\frac{1}{T-M+1}\sum_{t=M}^T \E[\aat^\ast{\aat^\ast}^{\top}\mid \mathcal{F}_{t-1}]\pto\widetilde{\mat{V}}^{H}.$$  
Then, as $T\to\infty$, we have 
$$\sqrt{T-M+1}(\hat\ttheta-\ttheta^\ast)\dto N(\bm 0,\widetilde{\mat{V}}^{H}),$$ where 
\begin{align*}
    &\ttheta = \frac{1}{T-M+1} \left(\sum_{t = M}^T \bbeta_t^{'}, \sum_{t = M}^T \bbeta_t^{''},\rho_{\bm h'},\rho_{\bm h''}\right)^\top,\\
    &\hat\ttheta = \frac{1}{T-M+1} \sum_{t = M}^T\left( \projm\tYt[I](F_{\bm h'}), \projm\tYt[I](F_{\bm h''}),  \rho_{t\bm h'},  \rho_{t\bm h''}\right)^\top.
\end{align*}

Finally, we apply the Delta method.
Define $$h(\ttheta) = \frac{\shiftmean\bbeta_t^{'}}{\rho_{\bm h'}}-\frac{\shiftmean\bbeta_t^{''}}{\rho_{\bm h''}}.$$ The Jacobian matrix is 
$$
\mat{J}(\ttheta) = 
\begin{bmatrix}
    \frac{1}{\rho_{\bm h'}} I &-\frac{1}{ \rho_{\bm h''}} I & -\frac{\shiftmean \bbeta_t^{'}}{\rho_{\bm h'}^2} & \frac{\shiftmean \bbeta_t^{'}}{\rho_{\bm h''}^2}
\end{bmatrix}
$$ where $\mat{I}$ is the identity matrix. Denote $$\mat{J} = \mat{J}(\ttheta^\ast) = \begin{bmatrix}
        \mat{I} &-\mat{I} &-\shiftmean \bbeta_t^{'} &\shiftmean \bbeta_t^{''}
\end{bmatrix}$$ Then, we obtain, 
$$\sqrt{T-M+1}(h(\hat\ttheta)-h(\ttheta^\ast))\dto N(\bm 0,\mat{J}\widetilde{\mat{V}}^{H}\mat{J}^\top)$$ as $T\to\infty.$ Thus, as $T\to\infty$, we have 
$$\frac{1}{\sqrt{T-M+1}}\sum_{t=M}^T\big(\hat{\bbeta}_t^{H}-\bbeta^\ast_t\big)\overset{d}{\to}N(\bm 0,\mat{J}\widetilde{\mat{V}}^{H}\mat{J}^\top).$$
\qed

\subsection{Proof of \cref{prop: variance Hajek}}
 
Assuming the same notations in the previous proof. Note that 
\begin{align*}
 \hat\ttheta = \Big[\shiftmean \big(&\projm\tYt(F_{\bm h'};\ggamma^\ast)\big)^\top ,\\ & \shiftmean \big(\projm\tYt(F_{\bm h''};\ggamma^\ast\big)^\top,1,1 \Big]^\top   
\end{align*}
is a consistent estimator of $\ttheta.$ So $\hat{\mat{J}} = \mat{J}(\hat\ttheta)$ is a consistent estimator of $\mat{J}(\ttheta).$ By similar arguments used in \cref{proof: prop1}, we have $\hat{\mat{V}}^{H}\pto \mat{V}^{H^\ast}.$ Hence, $\hat{\mat{J}}\hat{\mat{V}}^{H} \hat{\mat{J}}^\top\pto \mat{J} \mat{V}^{H^\ast} \mat{J}^\top. $
\qed

\begin{remark}\label{a: remark_bound}
 By Slutsky's theorem, if $\mat{Q}$ is a matrix such that $\mat{Q}\pto \mat{I}$ as $T\to\infty$, then $\hat{\mat{J}}\mat{Q}\hat{\mat{V}}^{H} \hat{\mat{Q}}^\top\hat{\mat{J}}^\top$ is also a consistent estimator for $\mat{J}\mat{V}^{H^\ast}\mat{J}^\top.$
\end{remark}

\begin{remark}\label{a: remark_bound2}
There are several possible choices for $\mat{Q}$. In addition to the one presented in \cref{subsec: asymptotic properties}, another natural choice is
\[
\mat{Q} = \diag(\rho_{\bm h'}^{-1}\mat{1}_L,\; \rho_{\bm h''}^{-1}\mat{1}_L,\; \rho_{\bm h'}^{-1},\; \rho_{\bm h''}^{-1}).
\]
It is straightforward to show that $\mat{Q}$ converges to $\mat{I}$ in probability as $T \to \infty$. Although this choice also helps stabilize the elements of $\hat{\mat{V}}^{H}$, we find that the estimated variance bound based on this $\mat{Q}$ generally performs worse than the proposed variance bound using the choice of $\mat{Q}$ in \cref{subsec: asymptotic properties}. The bias introduced by this stabilization (bias which occurs when using stabilized weights in general) can be non-negligible when $M$ is relatively large in finite samples, resulting in under-coverage of the confidence intervals.

\end{remark}

\subsection{Proofs of \cref{thm: Hajek est}, \cref{thm: Hajek efficiency} and \cref{cor:2}}\label{a:Hajek_est_proof}

The proofs of \cref{thm: Hajek est} and \cref{thm: Hajek efficiency} follow directly from those of \cref{thm: IPW est} and \cref{thm: IPW efficiency} with the following alternative definition, 
    $$\esteq=\begin{pmatrix}
    \projm\tYt[I](F_{\bm h'};\ggamma)-\bbeta_t^{'}-\mmu_1\\ 
    \projm\tYt[I](F_{\bm h''};\ggamma)-\bbeta_t^{''}-\mmu_2 \\
    \rho_{t\bm h'}(\ggamma)-1-\mu_3 \\
    \rho_{t\bm h''}(\ggamma)-1-\mu_4\\
    \scorefunction
\end{pmatrix}.$$
Thus, we omit the details.  In addition, the proof of \cref*{cor:2} follows directly from \cref{prop: variance Hajek} and the fact that $\hat\ggamma$ is a consistent estimator of $\ggamma^\ast.$

\section{Empirical performance of the proposed estimator}\label{a: subsec: different_bound_estimator}

In this section, we examine the empirical performance of the proposed variance bound estimator through a simulation study. We generate data of different lengths $T$ (200, 500, 1000), consider interventions over different time periods $M$ (1, 3, 7), and analyze two types of moderators: spatial and spatio-temporal. For each scenario, we conducted 500 simulations, where the data-generating process follows the same design as described in \cref{subsec:sim design}. 

\begin{figure}[p]
\centering
\subfloat[Spatial moderator]{
\includegraphics[width = \textwidth,trim = 0 0 0 60, clip]{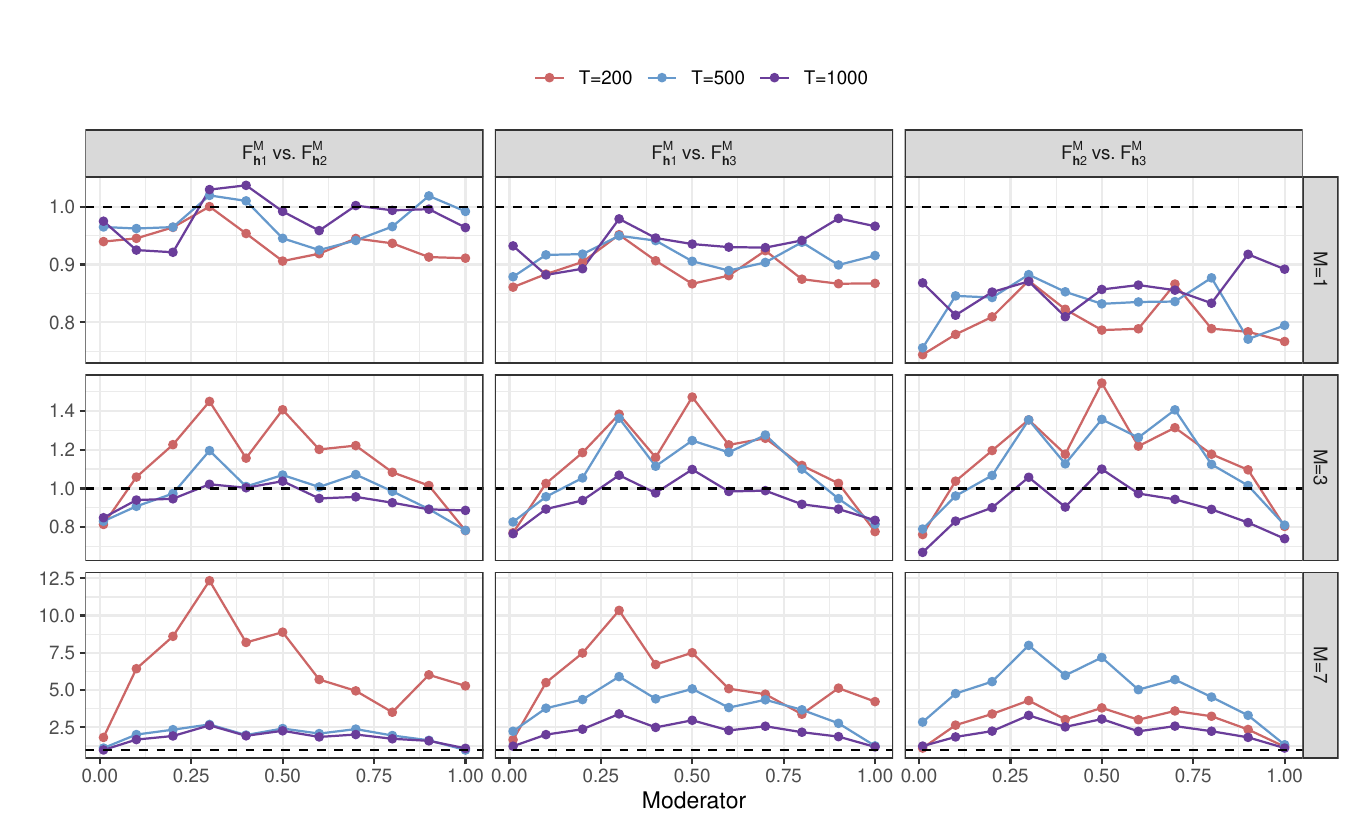}
\label{fig:variance_bound_s}
}  \\
\subfloat[Spatio-temporal moderator]{
\includegraphics[width = \textwidth,trim = 0 0 0 0, clip]{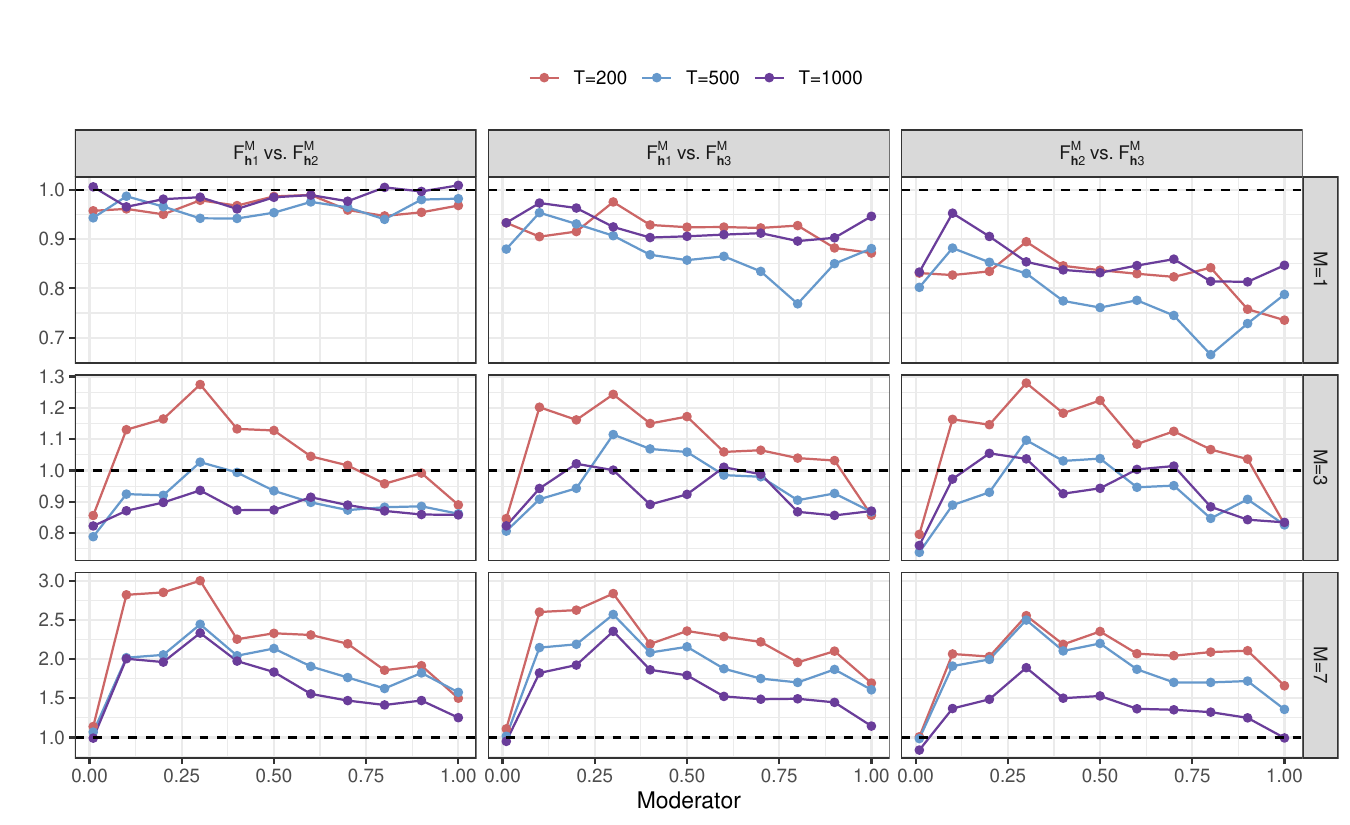}
\label{fig:variance_bound_st}
}
\caption{Ratio of the square root of the estimated variance bound, averaged across 500 simulations, to the mean standard deviation of the H\'ajek estimator based on the estimated propensity score with a spatio-temporal moderator. The standard deviation is computed by the Monte Carlo approximation. Interventions are considered over different time periods ($M=1, 3, 7$) and for time series of varying lengths ($T=200, 500, 1000$).}
\label{fig:variance_bound}
\end{figure}

For each simulation, we compute the square root of the estimated variance bound for the H\'ajek estimator, averaged over all simulations. We also calculate the Monte Carlo standard deviation of the H\'ajek estimator. We report the ratio of these two values in \cref{fig:variance_bound}. Theoretically, this ratio should be greater than 1.

\cref{fig:variance_bound_s} examines the empirical performance of the variance bound estimator when the moderator is spatial. Each row corresponds to a different intervention time period $M$ (1, 3, 7), and each column compares the results under different interventions. The results show that the ratio of the square root of the variance bound to the Monte Carlo standard deviation generally exceeds 1, aligning with theoretical expectations. However, for $M = 1$, the ratio occasionally falls below 1, particularly when $T$ is small. As $T$ increases from 200 to 1000, the ratio becomes more stable and approaches 1 across all moderator values, reflecting the improved accuracy of the estimator with longer time periods. Furthermore, for larger $M$ (3 and 7), the trends are smoother, suggesting that longer intervention periods reduce variability in the performance of the variance bound estimator.

\cref{fig:variance_bound_st} presents the results for the spatio-temporal moderator. Similarly to the results for the spatial moderator, the rows correspond to different intervention time periods $M$, while the columns compare the results under different interventions. The ratio of square root of the variance bound to the Monte Carlo standard deviation remains close to or above 1, particularly for larger values of $M$ and $T$. As $T$ increases, the ratio stabilizes and approaches 1, especially when $M = 3$ or $M = 7$. Compared to the spatial moderator case, the spatio-temporal results exhibit less fluctuation, even for small $T$, indicating that the variance bound estimator performs more reliably under the spatio-temporal setting.

\section{Statistical test of no heterogeneity}\label{a:sec:test}

In this section, we prove Theorem 3. To do this, we first prove the following lemma, which shows the limiting reference distribution of the test statistic under the null hypothesis.
\begin{lemma}\label{a:lemma_test}
    Suppose that the assumptions of Theorem 1 hold and $\bar\bbeta^\ast = \bm 0.$ Then, as $T\to\infty$, we have: 
    \begin{equation*}
(T-M+1)(\bar\bbeta-\bar\bbeta^\ast)^\top(\mat{J}\mat{V}^H\mat{J}^\top)^{-1}(\bar\bbeta-\bar\bbeta^\ast)\dto \chi^2_L.
\end{equation*} 
\end{lemma}

\subsection{Proof of \Cref{a:lemma_test}} \label{a:subsec:reject_prob}

According to Theorem 1, we have, as $T\to\infty$, 
\begin{equation*}
\sqrt{T-M+1}(\bar\bbeta-\bar\bbeta^\ast)\dto N(\bm 0,\mat{J}\mat{V}^H\mat{J}^\top), 
\end{equation*}
Since $\mat{J}\mat{V}^H\mat{J}^\top$ is positive definite, by Cholesky decomposition, we have $(\mat{J}\mat{V}^H\mat{J}^\top)^{-1} = \mat{L}\mat{L}^\top$ for some lower triangular matrix $\mat{L}.$ Then, as $T\to \infty$, 
\begin{equation*}
\mat{L}^\top\sqrt{T-M+1}(\bar\bbeta-\bar\bbeta^\ast)\dto N(\bm 0,\bm I), 
\end{equation*} 
By continuous mapping theorem, we have, as $T\to\infty$,
\begin{align*}
(T-M+1)(\bar\bbeta-\bar\bbeta^\ast)^\top(\mat{L}\mat{L}^\top)^{-1}(\bar\bbeta-\bar\bbeta^\ast)
\ = \ &(T-M+1)(\bar\bbeta-\bar\bbeta^\ast)^\top(\mat{J}\mat{V}^H\mat{J}^\top)^{-1}(\bar\bbeta-\bar\bbeta^\ast)\\
\dto \ & \chi^2_L.   
\end{align*}
\qed

\subsection{Proof of Theorem \ref{a:thm:reject_prob}}

By \Cref{a:lemma_test}, we have
\begin{align*}
    \lim_{T\to\infty} \P\left((T-M+1)(\bar\bbeta-\bm 0)^\top(\mat{J}\mat{V}^H\mat{J}^\top)^{-1}(\bar\bbeta-\bm 0)>\chi^2_{1-\alpha,L}\right) = \alpha 
\end{align*}
Since $\mat{V}^{H^\ast}-\mat{V}^H$ is positive semidefinite matrix, 
$$(T-M+1)\bar\bbeta^\top(\mat{J}\mat{V}^{H^\ast}\mat{J}^\top)^{-1}\bar\bbeta\le (T-M+1)\bar\bbeta^\top(\mat{J}\mat{V}^{H}\mat{J}^\top)^{-1}\bar\bbeta.$$
Therefore,
$$\P\left((T-M+1)\bar\bbeta^\top(\mat{J}\mat{V}^{H^\ast}\mat{J}^\top)^{-1}\bar\bbeta>\chi^2_{1-\alpha,L}\right)\le \P\left((T-M+1)\bar\bbeta^\top(\mat{J}\mat{V}^{H}\mat{J}^\top)^{-1}\bar\bbeta> \chi^2_{1-\alpha,L}\right).$$ 
This implies,
$$\limsup_{T\to\infty} \P\left((T-M+1)\bar\bbeta^\top(\mat{J}\mat{V}^{H^\ast}\mat{J}^\top)^{-1}\bar\bbeta>\chi^2_{1-\alpha,L}\right) \le \alpha.$$ 

Let $n\in\mathbb{N}$ be given. Define $\mat{D} = (\hat{\mat{J}}\mat{Q}\hat{\mat{V}}^{H}\mat{Q}^\top\hat{\mat{J}}^\top)^{-1}-(\mat{J}\mat{V}^{H^\ast}\mat{J}^\top)^{-1}.$ Since $\hat{\mat{J}}\mat{Q}\hat{\mat{V}}^{H}\mat{Q}^\top\hat{\mat{J}}^\top$ is a consistent estimator of $\mat{J}\mat{V}^{H^\ast}\mat{J}^\top$, $(\hat{\mat{J}}\mat{Q}\hat{\mat{V}}^{H}\mat{Q}^\top\hat{\mat{J}}^\top)^{-1}$ is a consistent estimator of $(\mat{J}\mat{V}^{H^\ast}\mat{J}^\top)^{-1}.$ Therefore, we have $\lim_{T\to\infty}\P(||\mat{D}||>\frac{1}{n})=0.$ Since $\bar\bbeta$ is bounded, we have 
$$\lim_{T\to\infty}\P\left(\mid\bar\bbeta^\top \mat{D} \bar\bbeta\mid>\frac{1}{n}\right)\leq \lim_{T\to\infty}\P\left(||\bar\bbeta||^2||\mat{D}||>\frac{1}{n}\right) = 0.$$ 
Note that
\begin{align*}
    & \P\left((T-M+1)\bar\bbeta^\top(\hat{\mat{J}}\hat{\mat{V}}^{H}\hat{\mat{J}}^\top)^{-1}\bar\bbeta>\chi^2_{1-\alpha,L}+\frac{1}{n}\right)\\
    =\  &\P\left((T-M+1)\bar\bbeta^\top(\mat{J}\mat{V}^{H^\ast} \mat{J}^\top)^{-1}\bar\bbeta+\bar\bbeta^\top \mat{D}\bar\bbeta>\chi^2_{1-\alpha,L}+\frac{1}{n}\right)\\
    \leq\  &\P\left((T-M+1)\bar\bbeta^\top(\mat{J}\mat{V}^{H^\ast} \mat{J}^\top)^{-1}\bar\bbeta>\chi^2_{1-\alpha,L}\right)+\P\left(\mid\bar\bbeta^\top \mat{D}\bar\bbeta\mid>\frac{1}{n}\right)\\
    \le \  &\alpha+\P\left(\mid\bar\bbeta^\top \mat{D}\bar\bbeta\mid>\frac{1}{n}\right).
\end{align*}
Taking the lim sup of both sides, we obtain
$$\limsup_{T\to\infty} \P\left((T-M+1)\bar\bbeta^\top(\hat{\mat{J}}\hat{\mat{V}}^{H}\hat{\mat{J}}^\top)^{-1}\bar\bbeta>\chi^2_{1-\alpha,L}+\frac{1}{n}\right)\leq \alpha+0=\alpha.$$ 
Let $E_n$ denote the event that 
$$(T-M+1)\bar\bbeta^\top(\hat{\mat{J}}\hat{\mat{V}}^{H}\hat{\mat{J}}^\top)^{-1}\bar\bbeta>\chi^2_{1-\alpha,L}+\frac{1}{n}.$$
Since $\{E_n\}_{n\ge 1}$ is an increasing sequence of events, taking the limit of $n\to\infty$, we have  $$\limsup_{T\to\infty} \P\left((T-M+1)\bar\bbeta^\top(\hat{\mat{J}}\hat{\mat{V}}^{H}\hat{\mat{J}}^\top)^{-1}\bar\bbeta>\chi^2_{1-\alpha,L}\right)\leq\alpha.$$ 
Note that $$\P(\text{p-value}<\alpha) = \P(T_c>\chi^2_{1-\alpha,L}) = \P\left((T-M+1)\bar\bbeta^\top(\hat{\mat{J}}\hat{\mat{V}}^{H}\hat{\mat{J}}^\top)^{-1}\bar\bbeta>\chi^2_{1-\alpha,L}\right).$$ We obtain the desired result, 
    $$\limsup_{T\to\infty} \P(\text{p-value}<\alpha)\leq \alpha.$$
\qed

\section{Doubly robust estimator}\label{a:sec:doublyrobust}
In this section, we introduce a doubly robust estimator and its asymptotic properties. We present the results assuming the true propensity score is unknown. Similar results can be established for the easier case where the true propensity score is known with a similar proof. 

\subsection{Regularity conditions}

We first introduce the form of the doubly robust estimator and state the regularity conditions for establishing its consistency. Define $w\in\mathcal{W}$ and $\bm w = (w_{t-M+1},\dots,w_t)\in\mathcal{W}^M$. 

Under the fixed potential outcome framework, let $m_{ti}(\bm w) = N_{S_i}(Y_t(W_{t-M},w_{t-M+1},\dots,w_t))$ denote the true potential outcome for the $i$-th component. We define the true outcome vector as $\bm m_t(\bm w) = (m_{t1}(\bm w),\dots,m_{tp}(\bm w))^\top$ and the observed outcome vector as $\bm N(Y_t) = (N_{S_1}(Y_t),\dots, N_{S_p}(Y_t))^\top$.

Let $e_t(w, \ggamma)$ and $\bm m_t(\bm w, \eeta)$ be the posited propensity score and outcome models parameterized by $\ggamma$ and $\eeta$. Let $\ggamma^\ast$ and $\eeta^\ast$ denote the true parameters when these models are correctly specified. In this case, $e_t(w, \ggamma^\ast) = P(W_t = w\mid\lHt[-1])$ exactly captures the treatment assignment mechanism, and $\bm m_t(\bm w, \eeta^\ast) = \bm m_t(\bm w)$ perfectly recovers the true potential outcomes.

Let $\hat\ggamma$ and $\hat\eeta$ be parameter estimators based on the observed trajectory up to time $T$. Let $\bm W_{(t-M+1):t} = (W_{t-M+1}, \dots, W_t)$ denote the observed treatment sequence. For interventions $F_{\bm h'}$ and $F_{\bm h''}$, define $\hat\tau^{dr}_{\bm h',\bm h''}(\seffm)= \bm z(\seffm)^\top\left(\shiftmean \hat\bbeta_t^{dr}\right)$, where 
\begin{align*}
  \hat\bbeta_t^{dr} &= \projm\Bigg(\dfrac{f_{\bm h''}(\bm W_{(t-M+1):t})-f_{\bm h'}(\bm W_{(t-M+1):t})}{\prod_{j=t-M+1}^t e_j(W_j, \hat\ggamma)}\left(\bm N(Y_t)-\bm m_t(\bm W_{(t-M+1):t}, \hat\eeta)\right)\\&\hspace{3cm}+\int_{\bm w\in\mathcal{W}^M}(f_{\bm h''}(\bm w)-f_{\bm h'}(\bm w))\bm m_t(\bm w, \hat\eeta)\dd \bm w\Bigg)
\end{align*}

\begin{assumption}\label{assump: dr} \singlespacing
We assume the following conditions hold.
\begin{enumerate}
    \item \label{assump: dr1} There exist limiting values $\ggamma^\dagger $ and $\eeta^\dagger$ such that $\hat\ggamma \pto \ggamma^\dagger$ and $\hat\eeta \pto \eeta^\dagger$ as $T\to \infty$. The functions $e_t(w, \ggamma)$ and $\bm m_t(\bm w, \eeta)$ are continuous in $\ggamma$ and $\eeta$ respectively.
        
    \item\label{assump: dr2} At least one of the models is correctly specified. This means either $\ggamma^\dagger = \ggamma^\ast$ or $\eeta^\dagger = \eeta^\ast$.

    \item\label{assump: dr3} There exists $\delta_e>0$ such that $e_t(w, \ggamma^\dagger)>\delta_{e} f_h(w)$ for all $w\in\mathcal{W}$ for which $f_h(w) > 0$.

    \item\label{assump: dr4} There exists $\delta_m>0$ such that $\E\left[||\bm m_t(\bm W_{(t-M+1):t}, \eeta^\dagger)||^2\right]<\delta_m$ for all $M\le t\le T$, where the expectation is taken with respect to the random treatment assignment.
\end{enumerate}
\end{assumption}

\subsection{Consistency of the doubly robust estimator}
The following theorem states the consistency of the proposed estimator.

\begin{theorem}[Consistency of the doubly robust estimator]
\label{thm: dr_consistency}
Suppose \Cref{assump: overlap,assump: unconfoundedness,assump:IPW1,assump: dr} hold. Then 
$$ \shiftmean(\hat\bbeta_t^{dr}-\bbeta_t^\ast) \pto \bm 0, \qquad T \to \infty. $$
\end{theorem}

\begin{proof}[Proof of \cref{thm: dr_consistency}]
    \paragraph*{Case 1.} Assume the propensity score model is correctly specified, meaning $\ggamma^\dagger = \ggamma^\ast$. We first show the proposed estimator evaluated at the limits $\ggamma^\ast$ and $\eeta^\dagger$ is consistent. According to the proof of \cref{thm:IPW true}, 
    $$ \E\left[\projm \frac{f_{\bm h''}(\bm W_{(t-M+1):t})-f_{\bm h'}(\bm W_{(t-M+1):t})}{\prod_{j=t-M+1}^t e_j(W_j, \ggamma^\ast)}\bm N(Y_t)\mid \lHt[-M]^\ast\right] = \bbeta_t^\ast, $$ for all $M\le t\le T.$
    Moreover, weighting any arbitrary function evaluated at $\eeta^\dagger$ by the true inverse probability weights properly integrates over the target density. Thus,
    $$ \E\left[\frac{f_{\bm h}(\bm W_{(t-M+1):t})}{\prod_{j=t-M+1}^t e_j(W_j, \ggamma^\ast)}\bm m_{t}(\bm W_{(t-M+1):t}, \eeta^\dagger)\mid \lHt[-M]^\ast\right] = \int_{\bm w\in\mathcal{W}^M} f_{\bm h}(\bm w) \bm m_{t}(\bm w, \eeta^\dagger)\dd \bm w, $$ for $h = h', h''.$ We can then construct two mean-zero martingale difference sequences,
    \begin{align*}
        &\aat(\ggamma^\ast) = \projm \left( \frac{f_{\bm h''}(\bm W_{(t-M+1):t})-f_{\bm h'}(\bm W_{(t-M+1):t})}{\prod_{j=t-M+1}^t e_j(W_j, \ggamma^\ast)}\bm N(Y_t) \right) -\bbeta_t^\ast\\
        &\bm B_t(\ggamma^\ast, \eeta^\dagger) = \projm\Bigg(\frac{f_{\bm h''}(\bm W_{(t-M+1):t})-f_{\bm h'}(\bm W_{(t-M+1):t})}{\prod_{j=t-M+1}^t e_j(W_j, \ggamma^\ast)}\bm m_t(\bm W_{(t-M+1):t}, \eeta^\dagger)\\
        &\hspace{3cm}- \int_{\bm w\in\mathcal{W}^M} (f_{\bm h''}(\bm w)- f_{\bm h'}(\bm w))\bm m_t(\bm w, \eeta^\dagger)\dd \bm w\Bigg).
    \end{align*}
    By \cref{assump:IPW1}, \cref{assump: dr}.\ref{assump: dr3} and \cref{assump: dr}.\ref{assump: dr4}, there exists a constant $c$ such that $\E[||\aat(\ggamma^\ast)||^2] <c$ and $\E[||\bm B_t(\ggamma^\ast, \eeta^\dagger)||^2] < c$ for $t\in\{M,M+1,\dots,T\}.$ Therefore $\shiftmean\aat(\ggamma^\ast)\pto \bm 0$ and $\shiftmean\bm B_t(\ggamma^\ast, \eeta^\dagger)\pto \bm 0$ as $T\to\infty.$  
    
    By \cref{assump: dr}, the parameter estimators are consistent for their limits. Since $\aat(\ggamma)$ and $\bm B_t(\ggamma, \eeta)$ are continuous functions of the parameters, we can apply the continuous mapping theorem. Substituting $\hat\ggamma$ and $\hat\eeta$ into the algebraic decomposition $\hat\bbeta^{dr}_t - \bbeta_t^\ast = \aat(\hat\ggamma) - \bm B_t(\hat\ggamma, \hat\eeta)$ yields
    $$ \shiftmean\left(\hat\bbeta^{dr}_t-\bbeta_t^\ast\right) \pto \shiftmean\left(\aat(\ggamma^\ast) - \bm B_t(\ggamma^\ast, \eeta^\dagger)\right) \pto \bm 0, $$ as $T\to\infty.$

    \paragraph*{Case 2.} Assume the outcome model is correctly specified, meaning $\eeta^\dagger = \eeta^\ast$. Under the fixed potential outcome framework, this implies the posited outcome model perfectly recovers the true potential outcomes with $\bm m_t(\bm w, \eeta^\ast) = \bm m_t(\bm w)$. Consequently, when evaluated at the observed treatment sequence $\bm W_{(t-M+1):t}$, the model yields exactly the observed outcome $\bm m_t(\bm W_{(t-M+1):t}, \eeta^\ast) = \bm N(Y_t)$.

    Let $\bm C_t(\eeta)$ be the integral term of the estimator,
    $$ \bm C_t(\eeta) = \projm \int_{\bm w\in\mathcal{W}^M} (f_{\bm h''}(\bm w)- f_{\bm h'}(\bm w))\bm m_t(\bm w, \eeta)\dd \bm w. $$
    By definition of the target parameter $\bbeta_t^\ast$ under fixed potential outcomes, we have
    \begin{align*}
        \bbeta_t^\ast &= \projm\left( \bm N_t(F_{\bm h''})-\bm N_t(F_{\bm h'})\right)\\
        &=\projm \int_{\bm w\in\mathcal{W}^M} (f_{\bm h''}(\bm w)- f_{\bm h'}(\bm w))\bm m_t(\bm w, \eeta^\ast)\dd \bm w = \bm C_t(\eeta^\ast).
    \end{align*}
    Thus $\shiftmean (\bm C_t(\eeta^\ast)-\bbeta_t^\ast) = \bm 0.$ 
    
    Now let $\bm D_t(\ggamma, \eeta)$ represent the IPW residual term of the doubly robust estimator,
    $$ \bm D_t(\ggamma, \eeta) = \projm \left( \frac{f_{\bm h''}(\bm W_{(t-M+1):t})-f_{\bm h'}(\bm W_{(t-M+1):t})}{\prod_{j=t-M+1}^t e_j(W_j, \ggamma)}\left(\bm N(Y_t)-\bm m_t(\bm W_{(t-M+1):t}, \eeta)\right) \right) $$
    Since $\bm N(Y_t) - \bm m_t(\bm W_{(t-M+1):t}, \eeta^\ast) = \bm 0$ exactly, the residual term evaluates to zero at the parameter limit with $\bm D_t(\ggamma^\dagger, \eeta^\ast) = \bm 0$ irrespective of the limit of the propensity score model $\ggamma^\dagger$.

    The doubly robust estimator evaluated with the estimated parameters can be written as $\hat\bbeta^{dr}_t = \bm D_t(\hat\ggamma, \hat\eeta) + \bm C_t(\hat\eeta)$. By the continuous mapping theorem, given $\hat\ggamma \pto \ggamma^\dagger$ and $\hat\eeta \pto \eeta^\ast$, $\bm D_t(\hat\ggamma, \hat\eeta) \pto \bm D_t(\ggamma^\dagger, \eeta^\ast) = \bm 0$ and $\bm C_t(\hat\eeta) \pto \bm C_t(\eeta^\ast)$. Thus,
    \begin{align*}
       &\shiftmean(\hat\bbeta^{dr}_t-\bbeta_t^\ast)\\ =& \shiftmean(\bm D_t(\hat\ggamma, \hat\eeta) + \bm C_t(\hat\eeta)-\bbeta_t^\ast)\\ \pto& \shiftmean(\bm 0 + \bm C_t(\eeta^\ast)-\bbeta_t^\ast) = \bm 0, 
    \end{align*}
    $$  $$ as $T\to \infty.$

\end{proof}

\section{Extensions to dependent interventions}\label{a:sec:extension}

In this section, we present detailed extensions of the framework to allow intervention distributions that are dependent across time periods. 

Consider a dependent (joint) intervention distribution over $M$ time periods, denoted by $F_{\bm h}$, with density $f_{\bm h}(w_1,w_2,\dots,w_M)$ for $(w_1,w_2,\dots,w_M)\in \mathcal{W}^M.$ Under this more general intervention scheme, the estimand of the expected number of outcome events in pixel $S_i$ at time $t$ remains identical to the one introduced in \Cref{subsec: estimand}:
\[
  \Nti
  = \int_{\mathcal{W}^M} N_{S_i}(Y_t(\lWt[-M],w_{t-M+1},\dots,w_t)) \dd F_{\bm h}(w_{t-M+1},\dots, w_t).
\]
The CATE estimand is defined in the same manner, replacing the independent intervention distribution by $F_{\bm h}$.For estimation, we modify \Cref{assump: overlap} as follows.
\begin{assumption}
\label{assump: overlap-dependent}
There exists $\delta_W>0$ such that
\[
  e_{t-M+1}(w_1)\,e_{t-M+2}(w_2)\,\cdots\,e_{t}(w_M) > \delta_W\,f_{\bm h}(w_1,\dots,w_M)
\]
for all $(w_1,\dots,w_M)\in\mathcal{W}^M$ for which $f_{\bm h}(w_1,\dots,w_M) > 0$, where $e_t(w)=f(W_t = w \mid \lHt[-1]).$
\end{assumption}
As expected, when $F_{\bm h}=F_{h}\times\cdots\times F_{h}$ (identical and independent distribution across periods), Assumption~\ref{assump: overlap-dependent} reduces to \Cref{assump: overlap}. The pseudo-outcomes $\pseudoOutI[][it]$ and $\pseudoOutH[][it]$ can be defined in the same way as done in \Cref{subsec: estimation}, using modified inverse-probability weights:
\[
  \rho_{t\bm h}(\hat\ggamma) = \frac{f_{\bm h}(W_{t-M+1},\dots, W_{t})}{\prod_{j=t-M+1}^{t} e_j(W_j;\hat\ggamma)}.
\]

None of the proofs in the main sections depend on the intervention distribution being independent over time. Therefore, all theoretical results remain valid under the dependent intervention distribution, provided the modified overlap assumption holds.

The dependent intervention presented here extends the framework in the main text by allowing treatments to be correlated across time periods. This setup can be regarded as a special case of the adaptive intervention, where the current treatment depends only on past treatments. Accordingly, the choice of matrix $\mat{Q}$ for constructing the estimated variance bound is the same as that described for adaptive interventions in \Cref{a:sec:adaptive}.

\section{Extending proofs to adaptive interventions}\label{a:sec:adaptive}

In this section, we discuss how to extending proofs  the previous sections in the adaptive case. Consider an adaptive intervention over $M$ time periods, $F_{\bm h_t}=(F_{h_{t1}},\dots,F_{h_{tM}})$, where each $F_{h_{tm}}$ is allowed to depend on the observed history, i.e.\ $F_{h_{tm}}$ has density $f_{h_{tm}}(w\mid \lHt[-m])$ for all $w\in\mathcal{W}$ and $1\le m\le M.$  We first show that
$\widetilde{Y}_{it}^{I}(F_{\bm h_t};\ggamma)$ defined in \cref{sec: adaptive} is unbiased to $N_{it}(F_{\bm h_t};\lHt[-M])$ conditional on $\lHt[-M]^\ast$, under the true propensity score.

\begin{lemma}\label{lemma3}
    Under \cref{assump: unconfoundedness}, 
    $$
    \E\left[\widetilde{Y}_{it}^{I}(F_{\bm h_t};\ggamma^\ast)\mid\lHt[-M]^\ast\right] = N_{it}(F_{\bm h_t};\lHt[-M])
    $$
\end{lemma}

\begin{proof}
    We will prove the statement for $M=2.$ The proofs for $M=1$ and $M>2$ are similar. When $M=2$, we have
    \begin{align*}
    &\E\left[\widetilde{Y}_{it}^{I}(F_{h_{t1}},F_{h_{t2}};\ggamma^\ast)\mid\lHt[-2]^\ast\right]\\
    &\E\left[\frac{f_{h_{t1}}(W_t\mid \lHt[-1])\,f_{h_{t2}}(W_{t-1}\mid \lHt[-2])}{e_t(W_t;\ggamma^\ast)\;e_{t-1}(W_{t-1};\ggamma^\ast)}N_{S_i}(Y_t)\;\Big|\;\lHt[-2]^\ast\right]\\
=&\int_{\mathcal{W}}\int_{\mathcal{W}}\frac{f_{h_{t1}}(W_t\mid \lHt[-1])\,f_{h_{t2}}(W_{t-1}\mid \lHt[-2])}{e_t(W_t;\ggamma^\ast)\;e_{t-1}(W_{t-1};\ggamma^\ast)}N_{S_i}(Y_t(\lWt[-2],w_{t-1},w_t))\\
&\quad \times f(w_{t-1}\mid\lHt[-2]^\ast)\,f(w_t\mid \lHt[-1]^\ast)\,\dd w_{t-1}\dd w_{t}\\
=&\int_{\mathcal{W}} \left[\int_{\mathcal{W}}f_{h_{t1}}(W_t\mid \lHt[-1])\,N_{S_i}(Y_t(\lWt[-2],w_{t-1},w_t))\,\dd w_{t}\right]\,f_{h_{t2}}(W_{t-1}\mid \lHt[-2])\, \dd w_{t-1} \  \text{(Assumption 1)}\\
=&\int_{\mathcal{W}} N_{it}(F_{h_{t1}};\lHt[-1])\,f_{h_{t2}}(W_{t-1}\mid \lHt[-2])\, \dd w_{t-1}\\
=& N_{it}(F_{h_{t1}},F_{h_{t2}};\lHt[-2]),
\end{align*}
as desired.
\end{proof}

According to \cref{lemma3}, we can construct martingale difference sequence based on $\widetilde{Y}_{it}^{I}(F_{\bm h_{t}};\ggamma^\ast) - N_{it}(F_{\bm h_{t}};\lHt[-M]).$  Then the proofs from the previous sections carry through in this adaptive case. 

For the estimated variance bound, we can use a similar choice of $\mat{Q}$ as the one introduced in \cref{subsec: asymptotic properties}. Specifically, when comparing two adaptive interventions $F_{\bm h'_t} = (F_{h'_{t1}}, \dots, F_{h'_{tM}})$ and $F_{\bm h''_t} = (F_{h''_{t1}}, \dots, F_{h''_{tM}})$, we define
\[
\xi_{h'_{tm}} = \frac{1}{T} \sum_{t=1}^T \frac{f_{h'_{tm}}(W_t \mid \lHt[-1])}{e_t(W_t; \hat\ggamma)}
\quad \text{and} \quad
\xi_{h''_{tm}} = \frac{1}{T} \sum_{t=1}^T \frac{f_{h''_{tm}}(W_t \mid \lHt[-1])}{e_t(W_t; \hat\ggamma)},
\quad m = 1, \dots, M.
\]
We then choose
\[
\mat{Q} = \diag\big(\rho_{\bm h'_t}^{-1}\mat{1}_L,\; \rho_{\bm h''_t}^{-1}\mat{1}_L,\; \prod_{m=1}^M \xi_{h'_{tm}}^{-1},\; \prod_{m=1}^M \xi_{h''_{tm}}^{-1}\big).
\]
Since non-adaptive and dependent interventions are special cases of the adaptive framework, this choice of $\mat{Q}$ also applies to these settings. When the interventions are independent and identical across time periods, $\mat{Q}$ reduces to the one introduced in \Cref{subsec: asymptotic properties}.

\section{Sensitivity analysis for propensity score misspecification}\label{a:sec:sensitivity}


In this section, we describe in detail the implementation of the sensitivity analysis introduced in \cref{sec:sensitivity}. In \cref{sub:sec:sensitivity_computation}, we show how zero attainability can be checked using linear programming. In \cref{sub:sec:sensitivity_application}, we apply this procedure to the empirical study and report the robustness values for the estimated effects of airstrikes across different effect modifiers and intervention lengths.

\subsection{Computation of zero attainability}\label{sub:sec:sensitivity_computation}

To determine whether there exists a vector $\bm{\alpha}\in[\Gamma^{-M}, \Gamma^M]^{T-M+1}$ such that the estimated CATE coefficient vector in \cref{eq: CATE_coeff} is equal to $\bm 0$, we reformulate the zero-attainability condition as an optimization problem.  Recall that the estimated CATE coefficient vector based on the estimated propensity score, $\shiftmean\hat{\bbeta}_t^{H}$, can be written as
\begin{equation}
\shiftmean\hat{\bbeta}_t^{H}
=\dfrac{\sum_{t=M}^T\hat\bbeta_{t\bm h''}}{\sum_{t=M}^T \rho_{t\bm h''}(\hat\ggamma)}-\dfrac{\sum_{t=M}^T\hat\bbeta_{t\bm h'}}{\sum_{t=M}^T \rho_{t\bm h'}(\hat\ggamma)},
\end{equation}
where $\hat\bbeta_{t\bm h}
= (\mat Z_t^\top \mat Z_t)^{-1}
\mat Z_t^\top\rho_{t\bm h}(\hat\ggamma)\bm N_t$,
and $\bm N_t$ denotes the vector of event counts across pixels at time $t$.
Therefore, zero attainability of the estimated CATE coefficient vector under the expanded propensity score is equivalent to
\begin{equation}\label{eq: beta_equal}
\dfrac{\sum_{t=M}^T\alpha_t\hat\bbeta_{t\bm h''}}{\sum_{t=M}^T\alpha_t \rho_{t\bm h''}(\hat\ggamma)}-\dfrac{\sum_{t=M}^T\alpha_t\hat\bbeta_{t\bm h'}}{\sum_{t=M}^T \alpha_t\rho_{t\bm h'}(\hat\ggamma)}=\bm 0,
\end{equation}
for all non-intercept elements and for some $\bm{\alpha} \in [\Gamma^{-M}, \Gamma^M]^{T-M+1}.$ 

Let $\hat\beta^{(l)}_{t\bm h}$ denote the $l$-th element of $\hat\bbeta_{t\bm h}$ where $$\hat{\beta}_{\bm{h}}^{(l)}(\bm\alpha) = \dfrac{\sum_{t=M}^T\alpha_t\hat\beta^{(l)}_{t\bm h}}{\sum_{t=M}^T \alpha_t\rho_{t\bm h}(\hat\ggamma)}.$$ We note that \cref{eq: beta_equal} holds for the $l$-th element if and only if there exists a scalar $\lambda^{(l)}$ 
and a vector $\bm{\alpha}\in [\Gamma^{-M}, \Gamma^M]^{T-M+1}$ such that
\begin{equation}\label{eq:lambda}
 \hat{\beta}_{\bm{h}''}^{(l)}(\bm{\alpha}) = \lambda^{(l)},
\quad
\hat{\beta}_{\bm{h}'}^{(l)}(\bm{\alpha}) = \lambda^{(l)}.   
\end{equation}

The vector $\bm{\alpha}$ in \cref{eq: expanded_IPW} reflects deviations between the estimated and true propensity scores and does not depend on the intervention $F_{\bm{h}}$. Therefore, zero attainability is equivalent to the existence of a value $\lambda^{(l)}$ that can be achieved by both weighted averages under the same vector $\bm{\alpha}$.
For a fixed $\lambda^{(l)}$, each of these equalities in \cref{eq: beta_equal} can be rewritten as
\begin{equation}\label{eq: beta_equal2}
\sum_{t=M}^T \alpha_t \left( \hat\beta_{t\bm{h}}^{(l)} - \lambda^{(l)} \rho_{t\bm{h}}(\hat\ggamma) \right) = 0.    
\end{equation}
Thus, for a given $\lambda^{(l)}$, zero attainability reduces to checking whether there exists $\bm{\alpha}$ satisfying two linear conditions.

We check \cref{eq: beta_equal2} elementwise for each non-intercept coefficient index $l$. This elementwise procedure is conservative because there may exist vectors $\bm\alpha^{(l)}=(\alpha_M^{(l)},\ldots,\alpha_T^{(l)})$ that make the equation hold for each component $l$ separately, even when no common vector $\bm\alpha$ makes all components equal to zero simultaneously.

According to the definition of $\lambda^{(l)}$ in \cref{eq:lambda}, to determine whether \cref{eq: beta_equal2} holds for some $\lambda^{(l)}$, we only need to consider values of $\lambda^{(l)}$ that can arise as common values of
$\hat{\beta}_{\bm{h}''}^{(l)}(\bm{\alpha})$ and
$\hat{\beta}_{\bm{h}'}^{(l)}(\bm{\alpha})$
for $\bm{\alpha} \in [\Gamma^{-M}, \Gamma^M]^{T-M+1}$.
For each intervention $F_{\bm h}$, we therefore compute the minimum and maximum values of this ratio over all admissible $\bm{\alpha}$:
\[
\lambda^{(l)}_{lower,\bm{h}} = \min_{\bm{\alpha}} \hat{\beta}_{\bm{h}}^{(l)}(\bm{\alpha}), 
\quad
\lambda^{(l)}_{upper,\bm{h}} = \max_{\bm{\alpha}} \hat{\beta}_{\bm{h}}^{(l)}(\bm{\alpha}).
\]
These optimization problems involve ratios of linear functions of $\bm{\alpha}$. To solve them efficiently, we apply the Charnes-Cooper transformation \citep{charnes1962programming}, which converts each problem into a standard linear program.

For each candidate $\lambda^{(l)}$, we then check whether the corresponding linear constraints can be satisfied. We do this by solving a linear program that minimizes a nonnegative variable $s$ measuring the magnitude of constraint violations:
\[
\left| \sum_t \alpha_t \bigl(\hat\beta_{t\bm{h}}^{(l)} - \lambda^{(l)} \rho_{t\bm{h}}(\hat\ggamma)\bigr) \right| \le s.
\]
If the constraints can be satisfied exactly, the optimal value is $s=0$. In practice, we declare zero attainable if the optimal value of $s$ is below a small tolerance level $\varepsilon$.

The algorithm for checking zero attainability is summarized below:
\begin{algorithm}[H]
\caption{Sensitivity analysis for zero attainability}
\begin{algorithmic}[1]

\STATE \textbf{Input:} Sensitivity level $\Gamma$, coefficient index $l$, tolerance $\varepsilon$
\STATE Set bounds $L = \Gamma^{-M}$ and $U = \Gamma^M$

\medskip

\FOR{$\bm{h} \in \{\bm{h}', \bm{h}''\}$}

    \STATE Compute
    \[
    \lambda^{(l)}_{lower,\bm{h}} = \min_{\bm{\alpha}} \frac{\sum_t \alpha_t \hat\beta_{t\bm{h}}^{(l)}}{\sum_t \alpha_t \rho_{t\bm{h}}(\hat\ggamma)},
    \quad
    \lambda^{(l)}_{upper,\bm{h}} = \max_{\bm{\alpha}} \frac{\sum_t \alpha_t \hat\beta_{t\bm{h}}^{(l)}}{\sum_t \alpha_t \rho_{t\bm{h}}(\hat\ggamma)}
    \]
    subject to $L \le \alpha_t \le U$.

\ENDFOR

\STATE Define the common feasible interval
\[
\Lambda^{(l)} =
[\lambda^{(l)}_{lower,\bm{h}'}, \lambda^{(l)}_{upper,\bm{h}'}]
\cap
[\lambda^{(l)}_{lower,\bm{h}''}, \lambda^{(l)}_{upper,\bm{h}''}]
\]

\IF{$\Lambda^{(l)} = \emptyset$}
    \STATE \textbf{return} zero is not attainable
\ENDIF

\medskip

\FOR{each $\lambda^{(l)} \in \Lambda^{(l)}$ (grid search)}

    \STATE Solve the linear program
    \[
    \min_{\bm{\alpha},\, s} \; s
    \]
    subject to
    \[
    \left| \sum_t \alpha_t \bigl(\hat\beta_{t\bm{h}''}^{(l)} - \lambda^{(l)} \rho_{t\bm{h}''}(\hat\ggamma)\bigr) \right| \le s,
    \]
    \[
    \left| \sum_t \alpha_t \bigl(\hat\beta_{t\bm{h}'}^{(l)} - \lambda^{(l)} \rho_{t\bm{h}}(\hat\ggamma)\bigr) \right| \le s,
    \]
    \[
    L \le \alpha_t \le U.
    \]

    \IF{$s \le \varepsilon$}
        \STATE \textbf{return} zero is attainable
    \ENDIF

\ENDFOR

\STATE \textbf{return} zero is not attainable

\end{algorithmic}
\end{algorithm}

\subsection{Sensitivity analysis in the empirical study}\label{sub:sec:sensitivity_application}

We conduct the proposed sensitivity analysis on our empirical study, which examines how the effects of US airstrikes on insurgent violence attacks (SAF and IED) are moderated by aid spending. Specifically, we evaluate the robustness of our findings about whether the causal effect is moderated by a binary indicator about a district's receipt of any US aid in the previous month, and a continuous variable of aid spending per capita.

For both moderators and outcomes, we determine the robustness value, which is defined as the largest value of $\Gamma$ such that the zero vector of CATE coefficients remains unattainable. We report the results across different lengths of stochastic intervention, denoted by $M$.

\cref{tab:sens_binary} presents the robustness values when the moderator is the binary presence of prior aid, for intervention durations $M = 1, \dots, 10$ days. The largest robustness value is $\Gamma = 1.16$ (for IED attacks at $M = 2$). \cref{tab:sens_cont} shows the corresponding values for the continuous aid per capita moderator at intervention lengths $M \in \{3, 7, 10\}$. The relatively small magnitude of these values suggests that even a modest amount of unobserved confounding can explain away the estimated heterogeneous treatment effects. 

\begin{table}[ht]
\centering
\caption{Robustness values ($\Gamma$) for the binary aid moderator across intervention lengths ($M$).}
\label{tab:sens_binary}
\begin{tabular}{l cccccccccc}
\hline
$M$ & 1 & 2 & 3 & 4 & 5 & 6 & 7 & 8 & 9 & 10 \\
\hline
SAF & 1.07 & 1.13 & 1.03 & 1.04 & 1.03 & 1.07 & 1.13 & 1.10 & 1.12 & 1.11 \\
IED & 1.08 & 1.16 & 1.03 & 1.04 & 1.01 & 1.07 & 1.09 & 1.07 & 1.07 & 1.06 \\
\hline
\end{tabular}
\end{table}

\begin{table}[ht]
\centering
\caption{Robustness values ($\Gamma$) for the continuous aid per capita moderator.}
\label{tab:sens_cont}
\begin{tabular}{l ccc}
\hline
$M$ & 3 & 7 & 10 \\
\hline
SAF & 1.09 & 1.05 & 1.09 \\
IED & 1.09 & 1.06 & 1.05 \\
\hline
\end{tabular}
\end{table}

\section{Moderator variables and baseline density in simulations}

In this section, we present additional visualizations of the spatial and spatio-temporal moderator variables used in the simulation study, as well as the baseline density for the intervention distribution.

\begin{figure}[!t]
\centering
\subfloat[Spatial moderator $X_2(\omega)$]{
\includegraphics[width = 0.27\textwidth,trim = 60 75 10 75, clip]{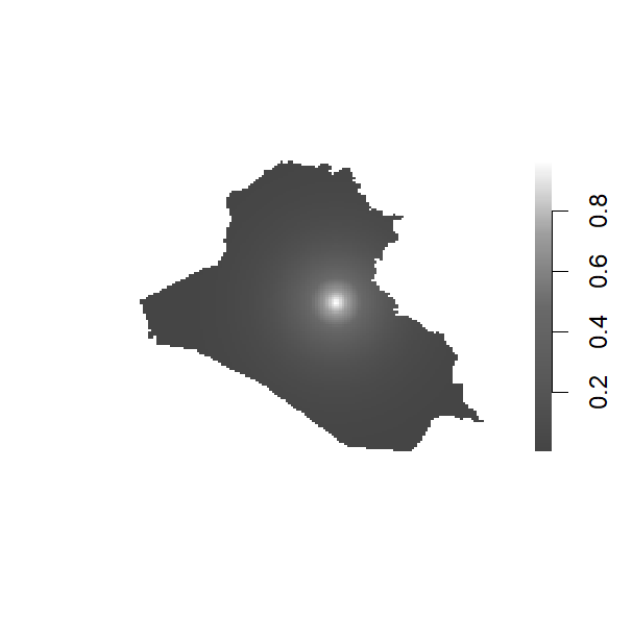}
\label{fig:illustration_s}
}\hfill
\subfloat[A realization of the spatio-temporal moderator $X_{3t}(\omega)$]{
\includegraphics[width = 0.27\textwidth,trim = 60 75 10 75, clip]{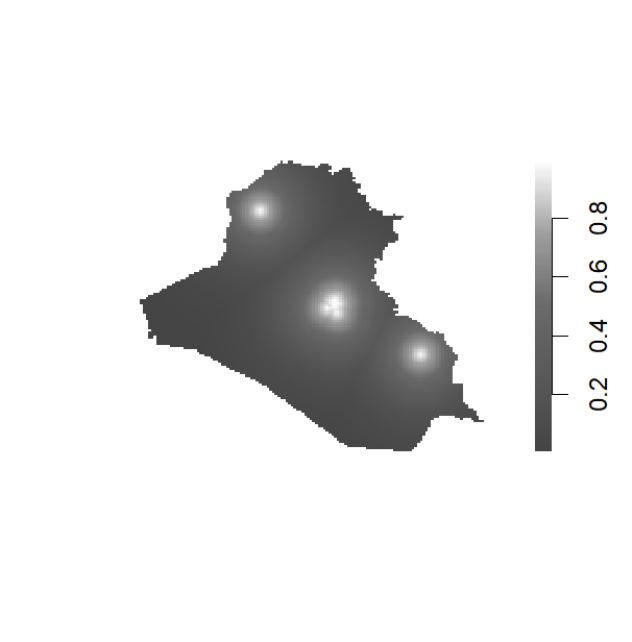}
\label{fig:illustration_st}
}\hfill
\subfloat[Distribution of treatment point patterns]{
\includegraphics[width = 0.27\textwidth,trim = 60 75 10 75, clip]{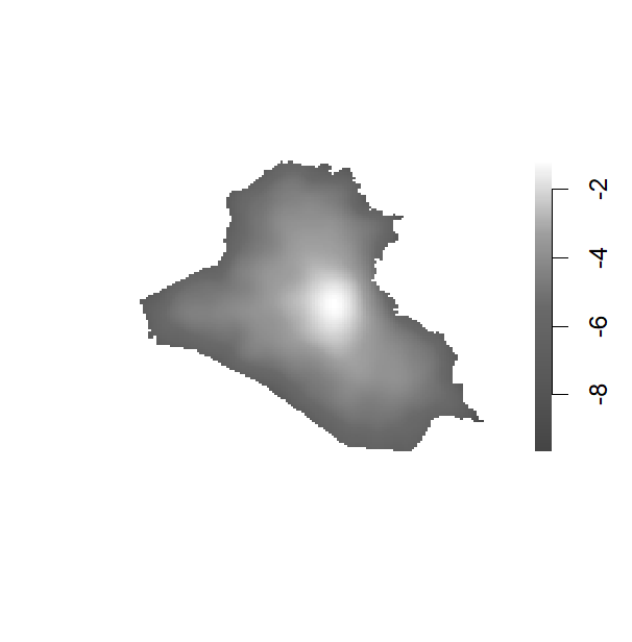}
\label{fig:illustration_phi}
}

\caption{Panels~(a)~and~(b) present the spatial and spatio-temporal moderator variables. Panel~(c) shows the logarithm of the baseline density for the intervention distribution.}
\label{fig:data_generation}
\end{figure}

\begin{figure}[t]
\centering
\includegraphics[width = \textwidth,trim = 0 20 0 0, clip]{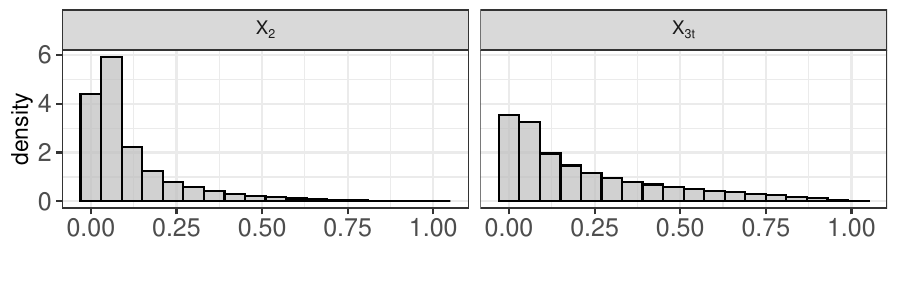}
\caption{Histograms of spatial moderator $X_2$ and spatio-temporal temporal moderator $X_{3t}$ across all pixels and all time periods.}
\label{fig:moderator_hist}
\end{figure}

Figure~\ref{fig:data_generation} provides an illustration of the moderators used in the simulation study. Panel~(a) depicts the spatial moderator \( X_2 \), which is derived based on the distance to Baghdad. Panel~(b) shows a realization of the spatio-temporal moderator \( X_{3t} \), which is generated using a Poisson point process based on estimated airstrike density. Panel~(c) presents the logarithm of the baseline density for the intervention distribution, which is used to model the spatial allocation of treatment. In addition, Figure~\ref{fig:moderator_hist} presents the distribution of the spatial moderator \( X_2 \) and the spatio-temporal moderator \( X_{3t} \) in the simulation.

\section{Application distributions}

In this section, we visualize the spatial distributions of the main events, the expected airstrike difference under the interventions considered in the empirical study, and the moderator variable used in the empirical analysis.

\begin{figure}[!t]
\centering
\subfloat[Insurgent violence]{
\includegraphics[width = 0.43\textwidth,trim = 80 75 25 65, clip]{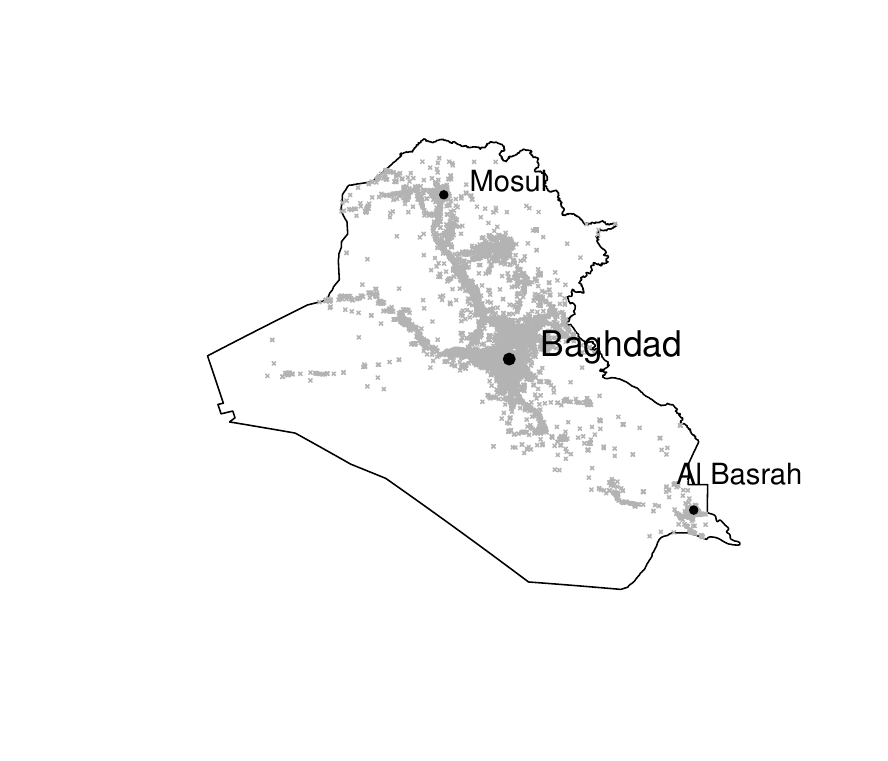}
\label{fig:violence_spatial}
}
\subfloat[Airstrikes]{
\includegraphics[width = 0.43\textwidth,trim = 80 75 25 65, clip]{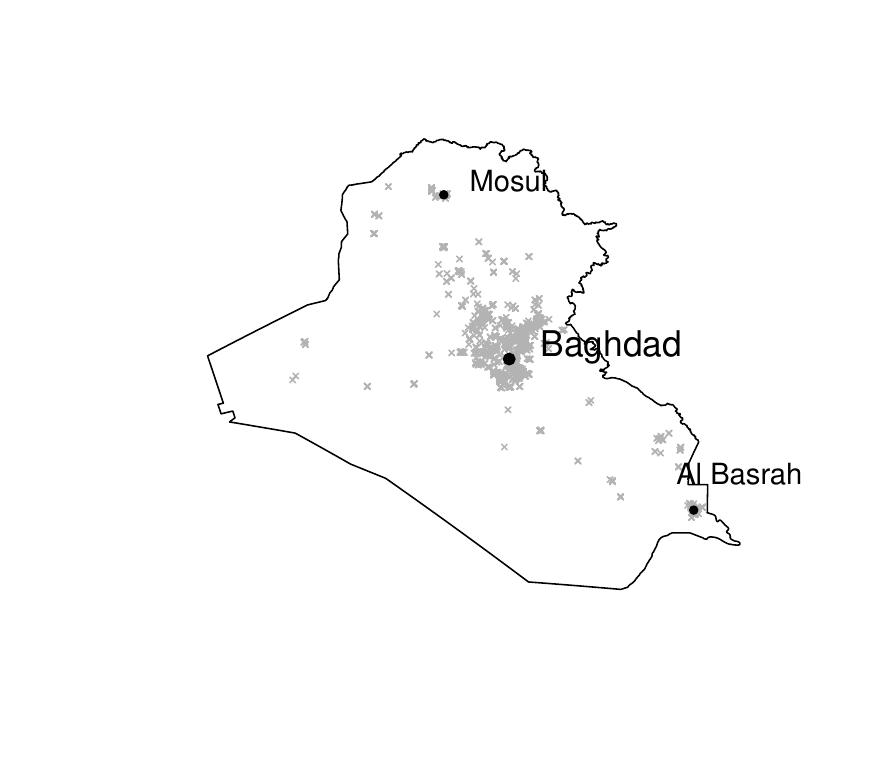}
\label{fig:airstrikes_spatial}
}
\caption{Distribution of US airstrikes and insurgent attacks. Panels (a) and (b) illustrate the spatial distribution of all insurgent attacks, including small arms fire and roadside improvised explosive devices, and all airstrikes from February 23, 2007 to July 5, 2008, respectively.}
\label{fig:event_spatial}
\end{figure}

Figure~\ref{fig:event_spatial} presents the spatial distributions of the two main event types in the application. Panel~(a) illustrates the locations of all insurgent attacks, including small arms fire and roadside improvised explosive devices, from February 23, 2007 to July 5, 2008. Panel~(b) shows the locations of all US airstrikes over the same period.

\begin{figure}[!t]
\centering
\subfloat[Expected airstrike difference]{
\includegraphics[width = 0.33\textwidth,trim = 5 5 5 5, clip]{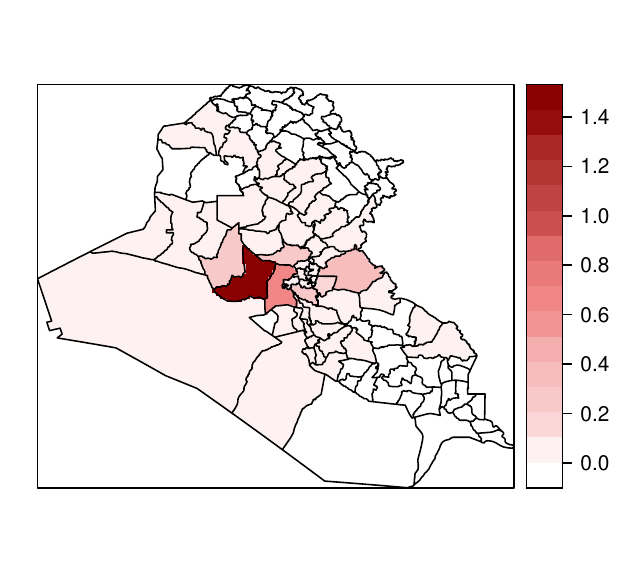}
\label{fig:airstrike_diff}
}\hspace{2mm}
\subfloat[Histogram of continuous aid]{
\includegraphics[width = 0.37\textwidth,trim = 5 5 5 5, clip]{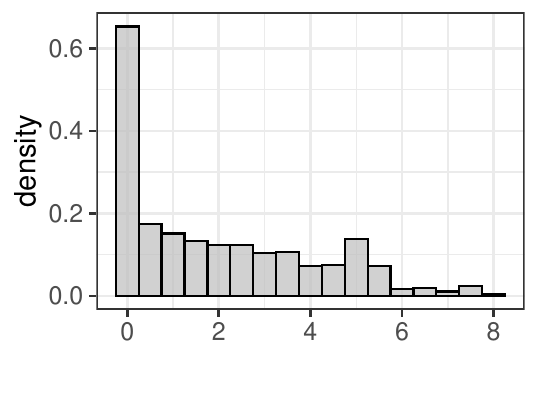}
\label{fig:hist_aid}
}
\caption{Panel (a) shows the expected change in the number of airstrikes under intervention \(F^M_{h_2}\) relative to intervention \(F^M_{h_1}\) for each district. Panel (b) shows the histogram of continuous aid on the log scale.}
\label{fig:airstrike_aid}
\end{figure}

Figure~\ref{fig:airstrike_aid} presents two additional quantities related to the empirical analysis. Panel~(a) illustrates the expected change in airstrike intensity under intervention \(F^M_{h_2}\) relative to intervention \(F^M_{h_1}\), computed for each district. Panel~(b) presents the distribution of continuous aid, measured on the log scale, which serves as the moderator in the empirical analysis.

\section{Additional simulation results}\label{a: additional_sim}

In this section, we first discuss how we use Monte Carlo apprximation to the true CATE values.  We then present additional simulation results for the H\'ajek and IPW estimators.  We also compare those two estimators and examine the efficiency gain resulting from the use of the estimated propensity score rather than the true propensity score.

\subsection{Monte Carlo approximation procedure}
\label{a:montecarlo}

As a concrete example, we outline the procedure for approximating the true CATE value $\tau_{\bm {h}_1,\bm{h}_2}(r)$ that compares two intervention distributions $F_{\bm h_1}$ and $F_{\bm h_2}$. At each time period $t$ and a given Monte Carlo draw $k= 1,\dots,K$, we sample treatment point patterns $w_{t-M+1}^{(jk)},\dots, w_{t}^{(jk)}$ from the intervention distribution $F_{\bm{h}_j}$ separately for $j = 1,2$. Then, we generate outcome point patterns $y_{t-M+1}^{(jk)},\dots, y_{t}^{(jk)}$ based on the treatment path $(\lWt[-M], \allowbreak w_{t-M+1}^{(jk)},\dots, w_{t}^{(jk)}),$ and compute the difference of outcomes in each pixel $S_i$ at time $t$, 
$D_{it}^{(k)} = N_{S_i}(y_t^{2k})-N_{S_i}(y_t^{1k})$. Let $r_{it}$ be the  value of the moderator in pixel $S_i$ at time $t$. We compute  $\tau_{t,\bm {h}_1,\bm{h}_2}^{(k)}(r)$ by regressing $D_{it}^{(k)}$ on $\basef[_1](r_{it}),\dots,\basef[_L](r_{it})$ as shown in Equation \cref{eq:model}.  Averaging $\tau_{t,\bm {h}_1,\bm{h}_2}^{(k)}(r)$ over all the iterations and time periods gives an approximation of $\tautr[\bm h_1][\bm h_2]$. 

\subsection{The H\'ajek estimator}\label{a: additional_sim_Hajek}

\begin{figure}[!t]
\centering
\includegraphics[width = 0.93\textwidth,trim = 0 20 0 25, clip]{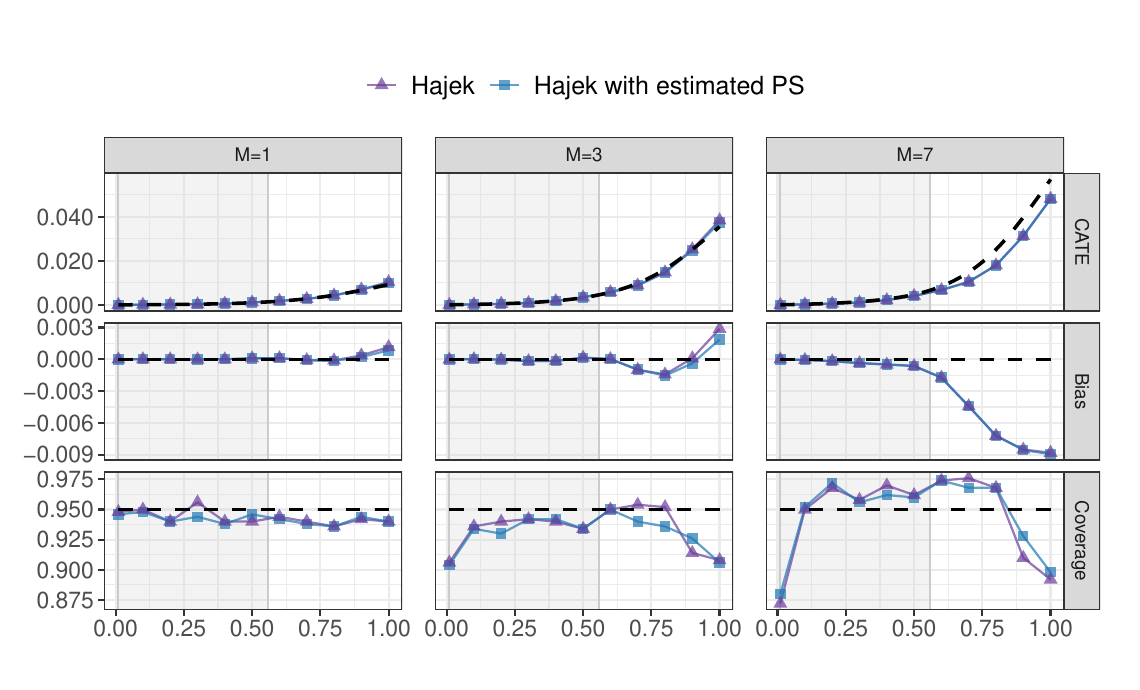}
\caption{The average of estimated CATE (first row), bias (second row), and coverage (third row) based on H\'ajek estimators for $F^M_{h_1}$ versus $F^M_{h_3}$ across 500 simulations for the spatial moderator. Dashed lines in the first, second, and third rows represent the true CATE, bias of zero, and the nominal coverage of 95\%, respectively. The purple line with triangles denotes the estimator based on the true propensity score, while the blue line with squares represent the estimators based on the estimated propensity scores. The shaded region indicates the 0.025 and 0.975 quantile range of the moderator values.}
\label{fig:spatial_est_13_appendix}
\end{figure}

\begin{figure}[p]
\centering
\subfloat[Spatial moderator]{
\includegraphics[width = \textwidth,trim = 0 10 0 60, clip]{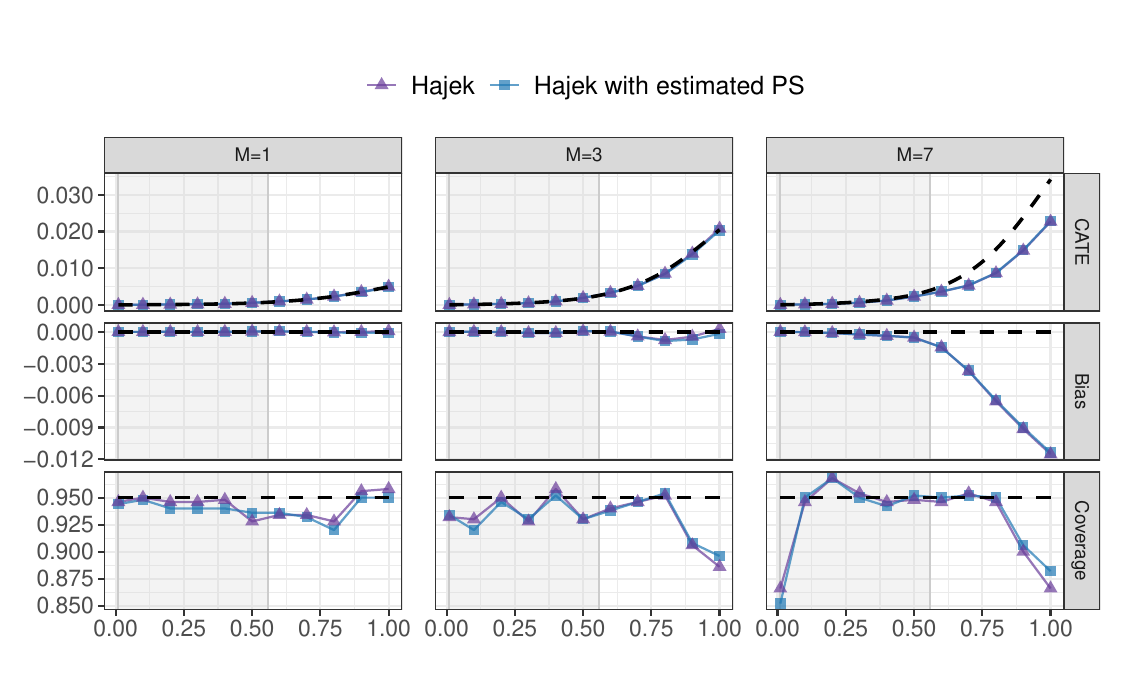}
\label{fig:spatial_est_12}
}  \\
\subfloat[Spatio-temporal moderator]{
\includegraphics[width = \textwidth,trim = 0 10 0 30, clip]{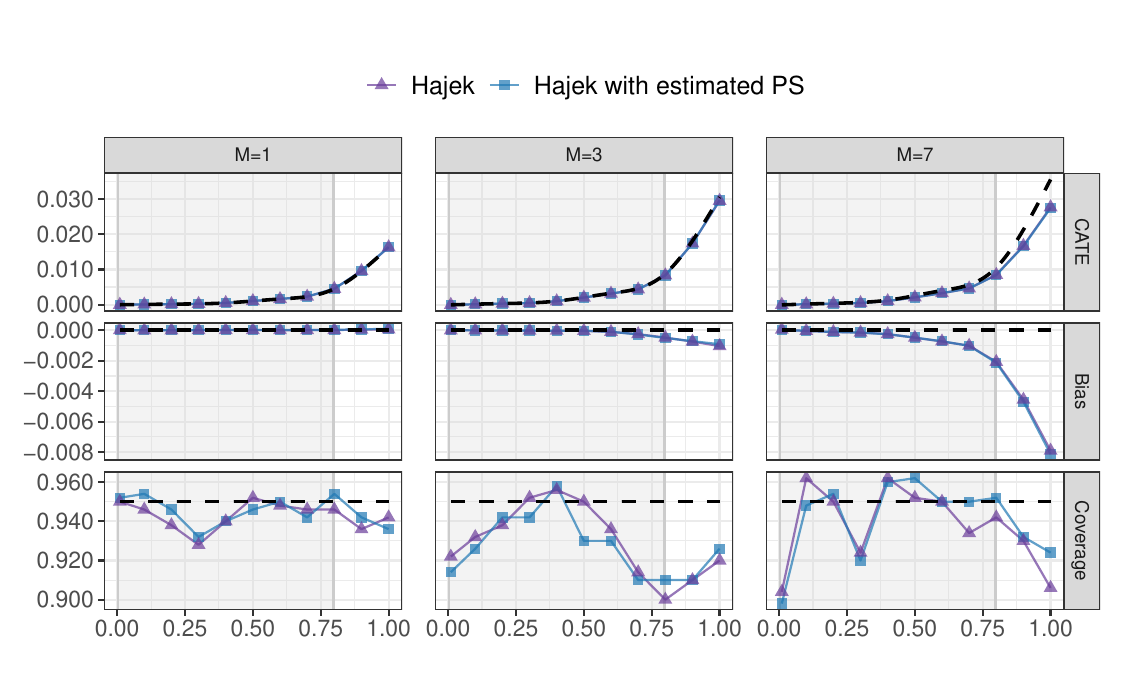}
\label{fig:st_est_12}
}
\caption{The average of estimated CATE (first row), bias (second row) and coverage (third row) based on the H\'ajek estimator for $F^M_{h_1}$ versus $F^M_{h_2}$ across 500 simulations. Dashed lines in the first, second, and third rows represent the true CATE, zero bias, and the theoretical minimum coverage, respectively. Plot~(a) corresponds to the spatial moderator, while plot~(b) pertains to the spatio-temporal moderator, both for various values of M = 1, 3, 7. The purple line with triangles denotes the estimator based on the true propensity score, while the blue line with squares represent the estimators based on the estimated propensity scores. The shaded region indicates the 0.025 and 0.975 quantile range of the moderator values.}
\label{fig:sim_est_12}
\end{figure}

\begin{figure}[p]
\centering
\subfloat[Spatial moderator]{
\includegraphics[width = \textwidth,trim = 0 10 0 60, clip]{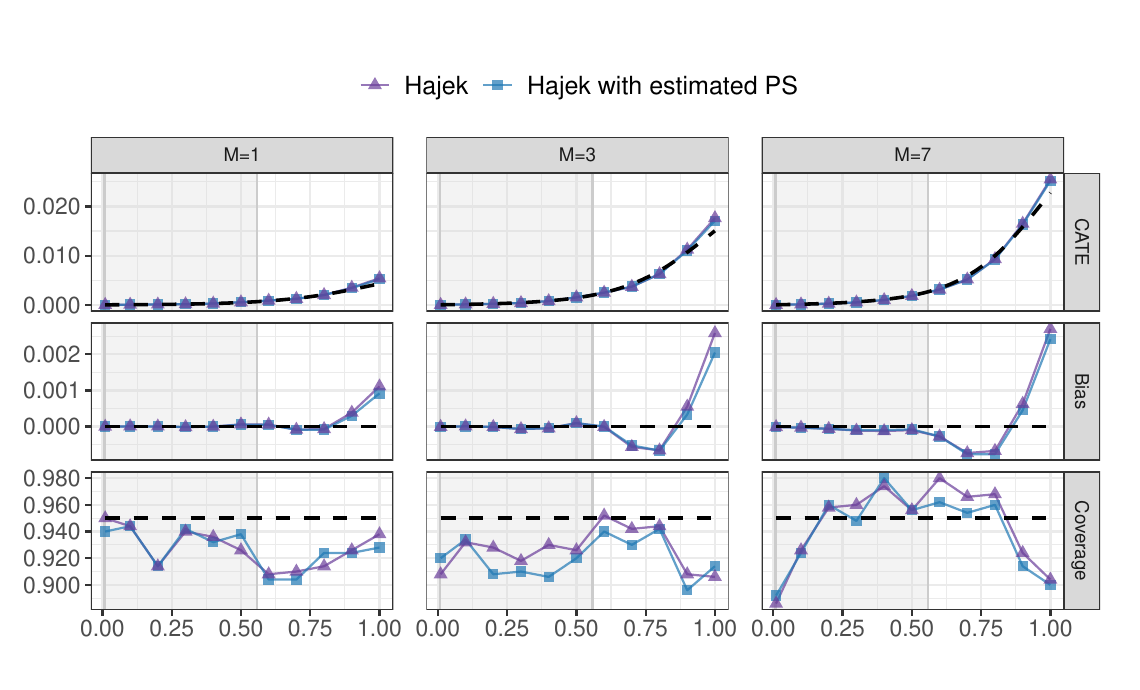}
\label{fig:spatial_est_23}
}  \\
\subfloat[Spatio-temporal moderator]{
\includegraphics[width = \textwidth,trim = 0 10 0 30, clip]{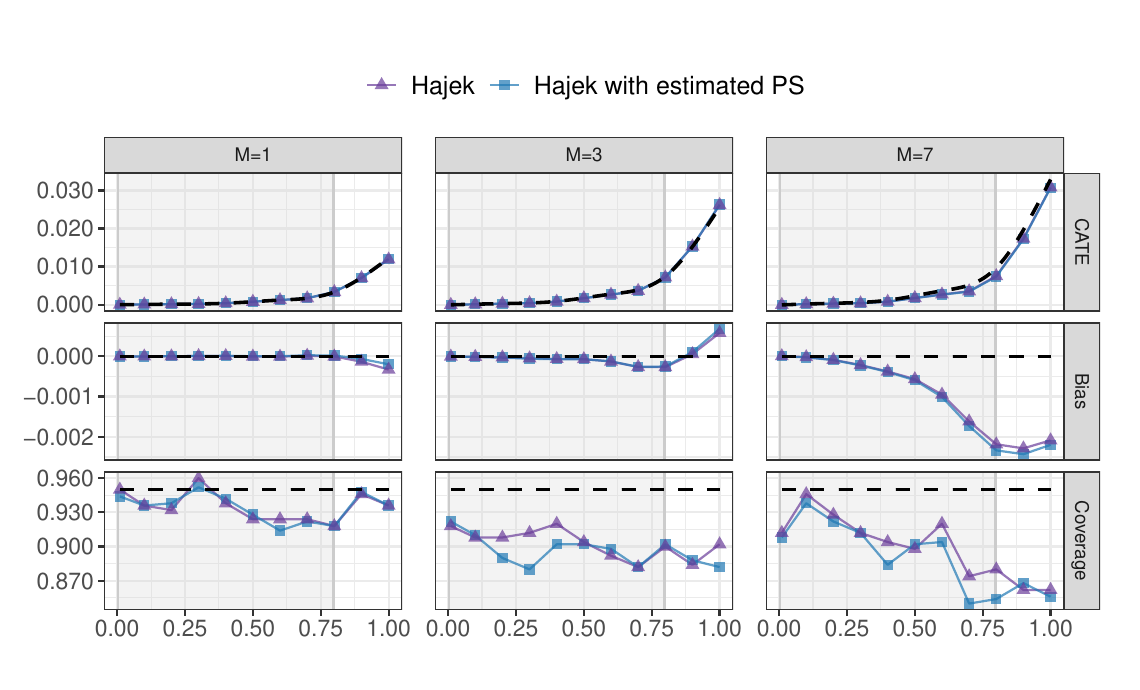}
\label{fig:st_est_23}
}
\caption{The average of estimated CATE (first row), bias (second row) and coverage (third row) based on the H\'ajek estimator for $F^M_{h_2}$ versus $F^M_{h_3}$ across 500 simulations. Dashed lines in the first, second, and third rows represent the true CATE, zero bias, and the theoretical minimum coverage, respectively. Plot~(a) corresponds to the spatial moderator, while plot~(b) pertains to
the spatio-temporal moderator, both for various values of M = 1, 3, 7. The purple line with triangles denotes the estimator based on the true propensity score, while the blue line with squares represent the estimators based on the estimated propensity scores. The shaded region indicates the 0.025 and 0.975 quantile range of the moderator values.}
\label{fig:sim_est_23}
\end{figure}

We present additional simulation results for the H\'ajek estimator. \cref{fig:spatial_est_13_appendix} shows the results for spatial effect moderator in the scenario assessed in \cref{subsec: simulation_res}. \cref{fig:sim_est_12} shows the results for the comparison between the two interventions, $F^M_{h_1}$ and $F^M_{h_2}$, while \cref{fig:sim_est_23} compares $F^M_{h_2}$ with $F^M_{h_3}$.

\cref{fig:spatial_est_13_appendix} presents the results for the spatial moderator in the comparison of $F^M_{h_1}$ and $F^M_{h_3}$. The H\'ajek estimators perform well for the spatial moderator, closely tracking the true CATEs with minimal bias and maintaining empirical coverage near the nominal 95\% level. The estimator performs similarly regardless of whether true or estimated propensity scores are used. Even for $M = 7$, the average bias remains below 0.001. As the number of intervention time periods increases, empirical coverage decreases slightly but remains above 0.87 across all scenarios. In regions with limited data, particularly near the upper end of the moderator range, small biases emerge, but coverage remains stable, indicating that the variance estimates effectively account for data sparsity.

In \cref{fig:sim_est_12}, we present results for the comparison between $F^M_{h_1}$ and $F^M_{h_2}$.
For the spatial moderator (\cref{fig:spatial_est_12}), we observe that the H\'ajek estimator performs well for $M = 1$, with estimated CATEs closely tracking the true CATEs and minimal bias. The empirical coverage of the 95\% confidence intervals remains close to the nominal level within the shaded region, highlighting the 0.025 and 0.975 quantiles of the moderator values. However, as $M$ increases to 3 and 7, the bias begins to increase slightly near the upper end of the moderator range (values close to 1). This indicates that longer time series  may be required in these regions to achieve better performance. For the spatio-temporal moderator (\cref{fig:st_est_12}), similar trends emerge. The H\'ajek estimator performs well for small $M$, with coverage rates remaining stable and close to 0.95. However, for $M = 7$, the bias increases for larger moderator values, although it remains small overall.

In \cref{fig:sim_est_23}, we present results for the comparison between $F^M_{h_2}$ and $F^M_{h_3}$. For the spatial moderator (\cref{fig:spatial_est_23}), the H\'ajek estimator again performs well when $M = 1$, with minimal bias and empirical coverage rates close to the nominal 0.95 level. As $M$ increases, deviations from the true CATE become more apparent near the upper range of the moderator values. In the spatio-temporal moderator case (\cref{fig:st_est_23}), the performance of the H\'ajek estimator remains robust for $M = 1$, with small bias and stable coverage rates. For larger $M$, the estimated CATEs continue to closely follow the true CATEs within the shaded region, although small biases appear near the large values of the moderator.

Overall, these additional results demonstrate that the H\'ajek estimator performs consistently well in the different scenarios. For both spatial and spatio-temporal moderators, the empirical coverage rates remain close to the nominal 95\% level, especially for small $M$. However, as $M$ increases, deviations from the true CATEs become noticeable at higher moderator values, suggesting that longer time series are needed to maintain performance in sparse regions.

\subsection{The IPW estimator}\label{a: additional_sim_IPW}

\begin{figure}[p]

\centering
\subfloat[Spatial moderator]{
\includegraphics[width = \textwidth,trim = 0 10 0 60, clip]{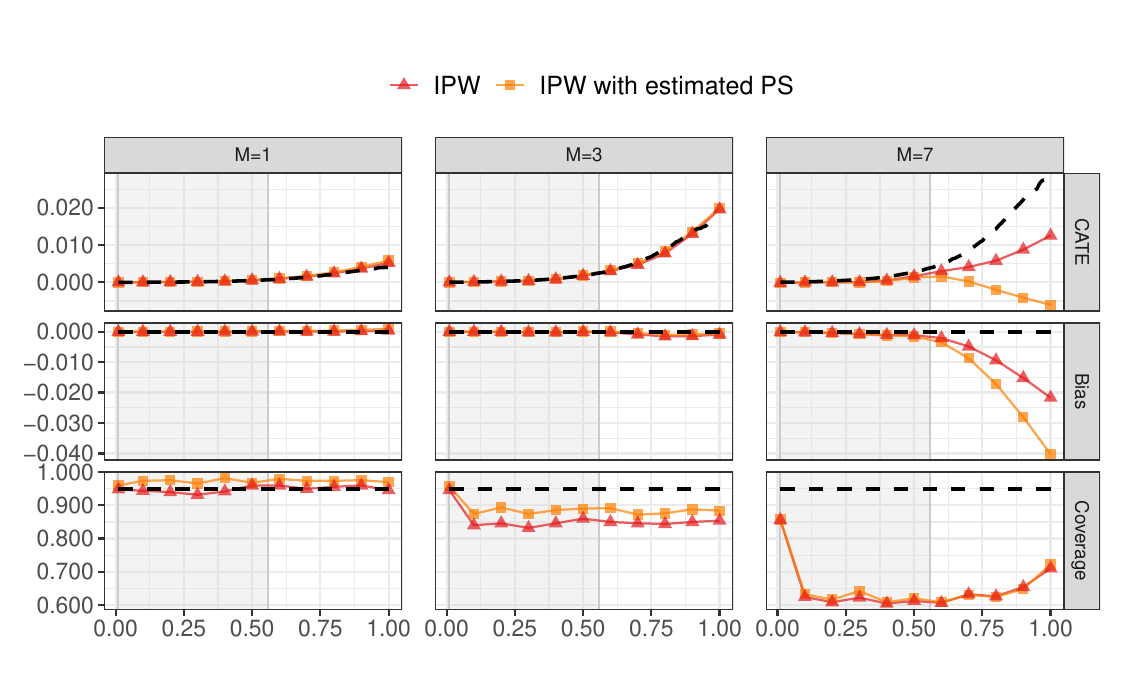}
\label{fig:IPW_spatial_est_12}
}  \\
\subfloat[Spatio-temporal moderator]{
\includegraphics[width = \textwidth,trim = 0 10 0 30, clip]{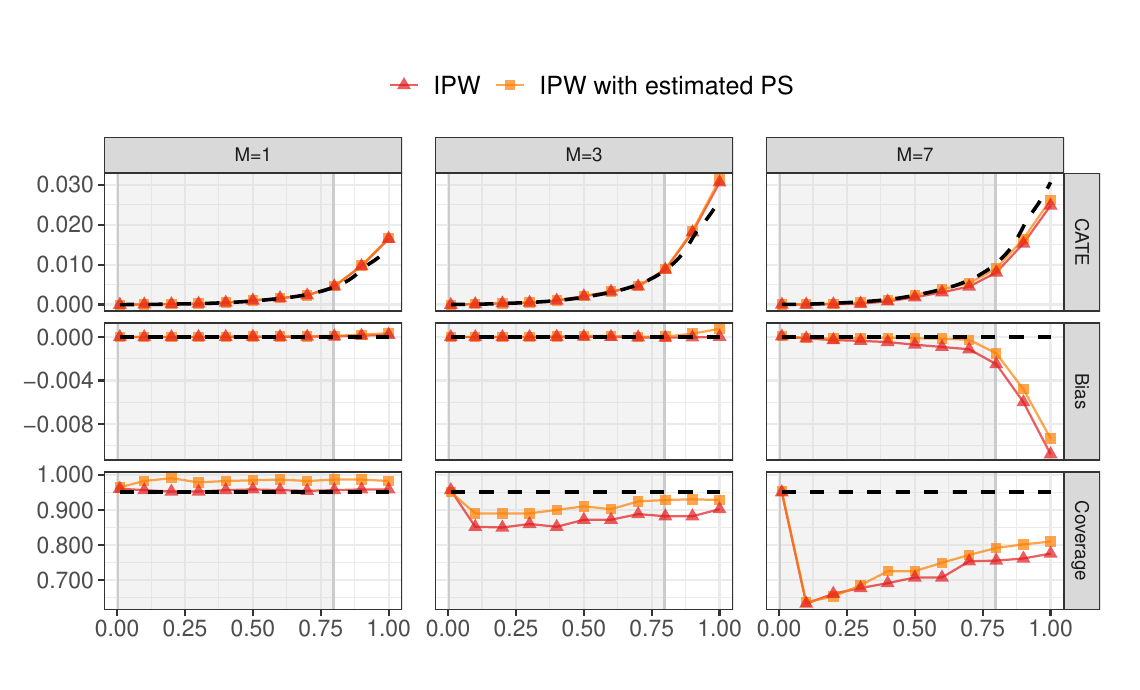}
\label{fig:IPW_st_est_12}
}

\caption{The average of estimated CATE (first row), bias (second row) and coverage (third row) based on IPW estimators for $F^M_{h_1}$ versus $F^M_{h_2}$ across 500 simulations. Dashed lines in the first, second, and third rows represent the true CATE, zero bias, and the theoretical minimum coverage, respectively. Plot~(a) corresponds to the spatial moderator, while plot~(b) pertains to
the spatio-temporal moderator, both for various values of M = 1, 3, 7. The purple line with triangles denotes the estimator based on the true propensity score, while the blue line with squares represent the estimators based on the estimated propensity scores. The shaded region indicates the 0.025 and 0.975 quantile range of the moderator values.}
\label{fig:IPW_sim_est_12}

\end{figure}

\begin{figure}[p]

\centering
\subfloat[Spatial moderator]{
\includegraphics[width = \textwidth,trim = 0 10 0 60, clip]{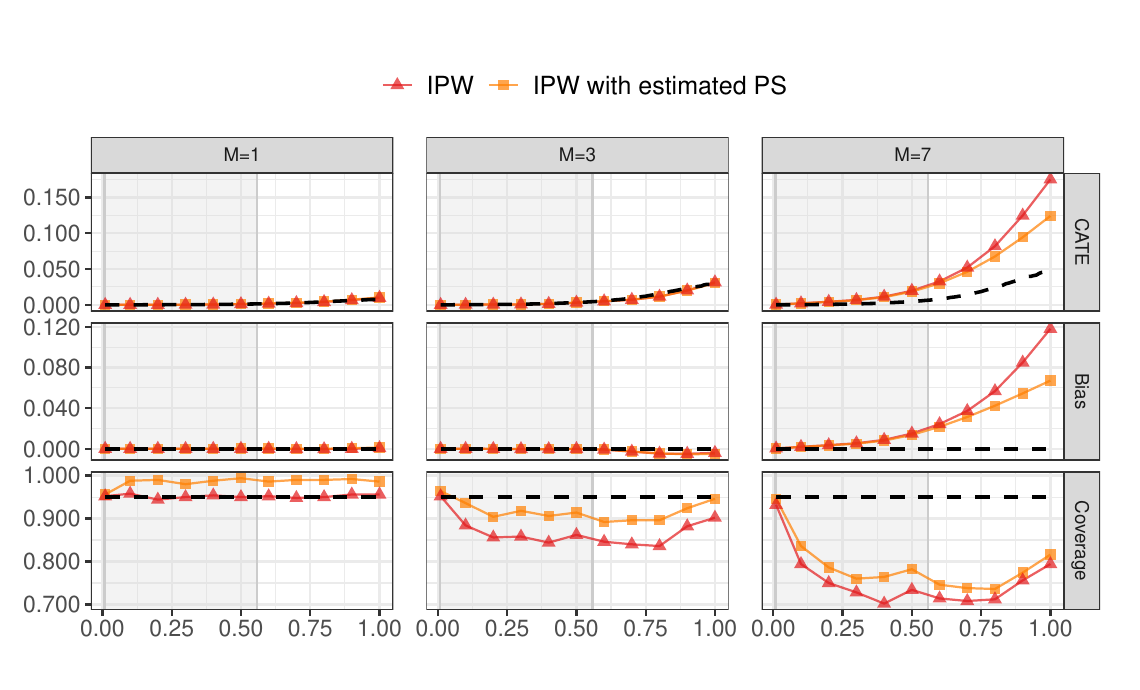}
\label{fig:IPW_spatial_est_13}
}  \\
\subfloat[Spatio-temporal moderator]{
\includegraphics[width = \textwidth,trim = 0 10 0 30, clip]{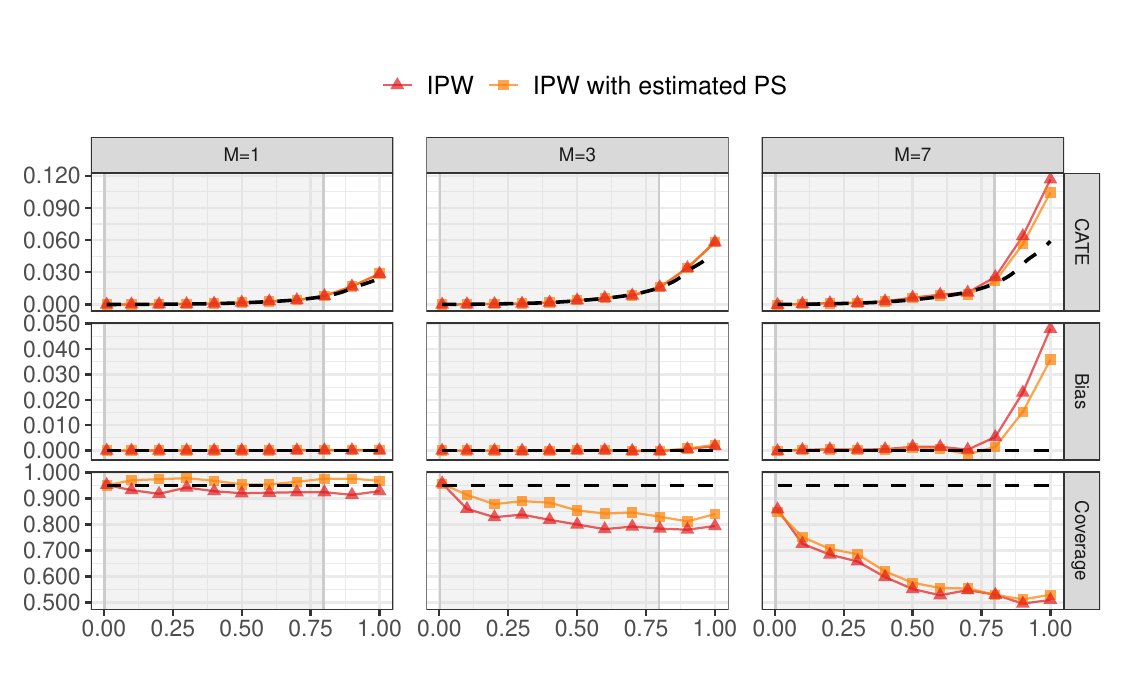}
\label{fig:IPW_st_est_13}
}

\caption{The average of estimated CATE (first row), bias (second row) and coverage (third row) based on IPW estimators for $F^M_{h_1}$ versus $F^M_{h_3}$ across 500 simulations. Dashed lines in the first, second, and third rows represent the true CATE, zero bias, and the theoretical minimum coverage, respectively. Plot~(a) corresponds to the spatial moderator, while plot~(b) pertains to
the spatio-temporal moderator, both for various values of M = 1, 3, 7. The purple line with triangles denotes the estimator based on the true propensity score, while the blue line with squares represent the estimators based on the estimated propensity scores. The shaded region indicates the 0.025 and 0.975 quantile range of the moderator values.}
\label{fig:IPW_sim_est_13}

\end{figure}

\begin{figure}[p]

\centering
\subfloat[Spatial moderator]{
\includegraphics[width = \textwidth,trim = 0 10 0 60, clip]{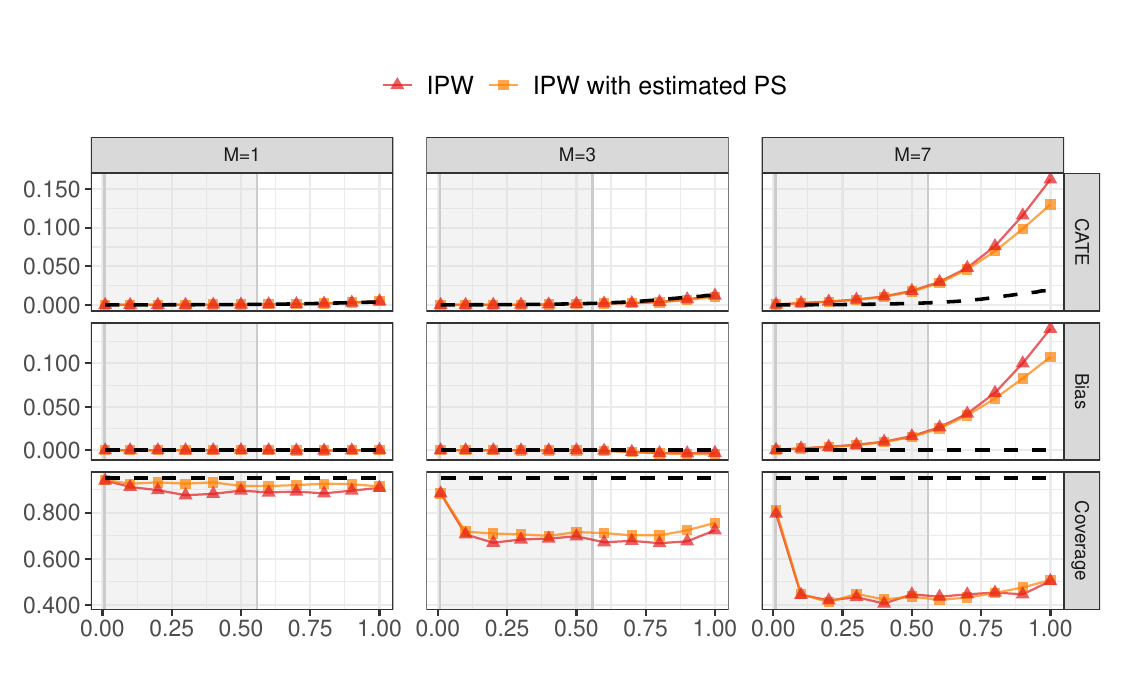}
\label{fig:IPW_spatial_est_23}
}  \\
\subfloat[Spatio-temporal moderator]{
\includegraphics[width = \textwidth,trim = 0 10 0 30, clip]{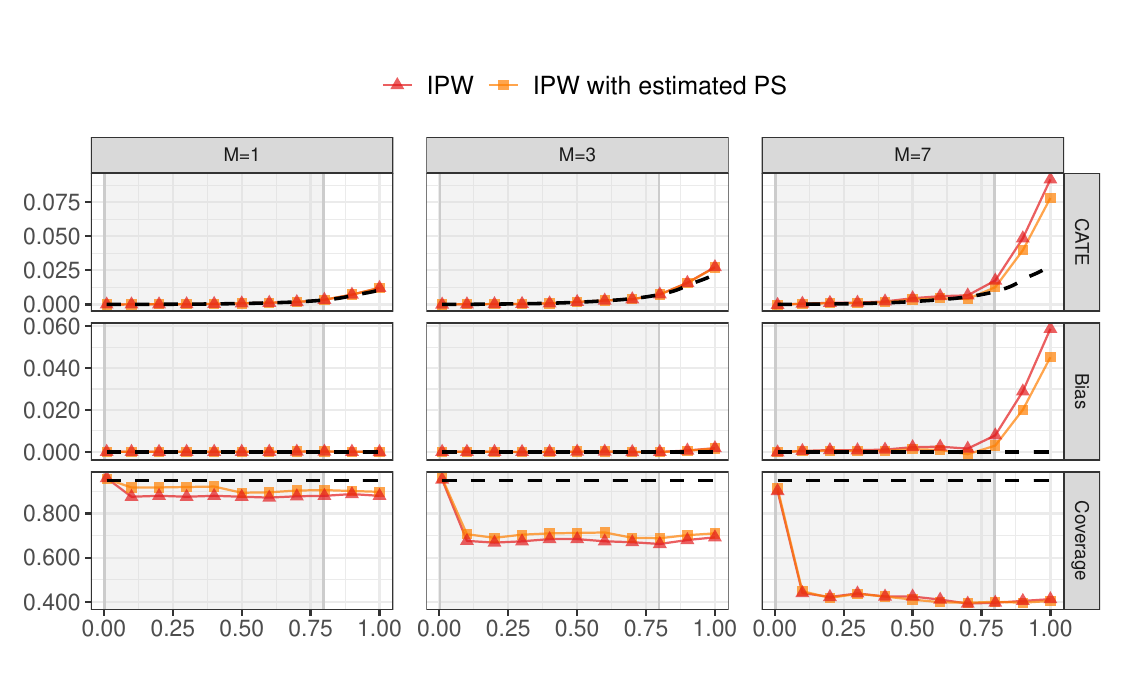}
\label{fig:IPW_st_est_23}
}

\caption{The average of estimated CATE (first row), bias (second row) and coverage (third row) based on IPW estimators for $F^M_{h_2}$ versus $F^M_{h_3}$ across 500 simulations. Dashed lines in the first, second, and third rows represent the true CATE, zero bias, and the theoretical minimum coverage, respectively. Plot~(a) corresponds to the spatial moderator, while plot~(b) pertains to
the spatio-temporal moderator, both for various values of M = 1, 3, 7. The purple line with triangles denotes the estimator based on the true propensity score, while the blue line with squares represent the estimators based on the estimated propensity scores. The shaded region indicates the 0.025 and 0.975 quantile range of the moderator values.}
\label{fig:IPW_sim_est_23}

\end{figure}

We present the performance of IPW estimators in the simulation studies described in \cref{sec:simulation}. Figures~\ref{fig:IPW_sim_est_12},~\ref{fig:IPW_sim_est_13},~and~\ref{fig:IPW_sim_est_23} show the results of the comparison of $F^M_{h_1}$ v.s $F^M_{h_2}$, $F^M_{h_1}$ v.s $F^M_{h_3}$ and $F^M_{h_2}$ v.s $F^M_{h_3}$ respectively. 

In \cref{fig:IPW_sim_est_12}, which compares $F^M_{h_1}$ and $F^M_{h_2}$, the IPW estimators perform well in the spatial moderator case when $M = 1$, where the estimated CATEs align closely with the true CATEs and exhibit minimal bias. The empirical coverage remains near the nominal 95\% level. As $M$ increases to 3 and 7, unstabilized weights result in larger biases, particularly near the upper range of the moderator values, and the empirical coverage deteriorate as $M$ increases.  In contrast, stabilized weights produce smaller biases and maintain more reliable coverage. For the spatio-temporal moderator, a similar pattern holds: stabilized weights consistently outperform their unstabilized counterparts, delivering more accurate estimates and better coverage properties. Additionally, estimators using the estimated propensity scores generally achieve comparable or even superior performance relative to those based on the true propensity scores.

\Cref{fig:IPW_sim_est_13} displays the results for $F^M_{h_1}$ versus $F^M_{h_3}$. In the spatial moderator case, the IPW estimators perform well when $M = 1$, but as $M$ increases, bias becomes more pronounced for unstabilized weights, particularly near higher moderator values. Empirical coverage remains mostly stable across the range but drops for larger $M$. Similar trends are observed for the spatio-temporal moderator. 

For $F^M_{h_2}$ versus $F^M_{h_3}$, the results are shown in \cref{fig:IPW_sim_est_23}. In the spatial moderator case, the IPW estimators again perform well for $M = 1$, with estimated CATEs exhibiting minimal bias and empirical coverage close to 0.95. As $M$ increases, unstabilized weights result in noticeable bias at higher moderator values and the coverage drops. In particular, for $M=7$, the empirical coverage falls below 50\% for all IPW-type estimators. The spatio-temporal results are consistent with earlier findings.

Overall, the IPW estimators perform well in low-variance settings, such as $M = 1$. However, for larger values of $M$, stabilized weights substantially enhance performance, reducing bias and improving empirical coverage. Across all comparisons and moderator types, H\'ajek estimators often outperform IPW estimators. These findings align with broader results in causal inference, where stabilized weights are known to improve finite-sample properties, especially in higher-variance settings.

\subsection{Efficiency due to the use of the estimated propensity score}

We examine the relative efficiency of the IPW and H\'ajek estimators when using the estimated propensity score rather than the true propensity score.  In particular, we computed the Monte Carlo standard deviation of the estimators across 500 simulations and present the ratio of standard deviation for the proposed estimator based on the true propensity score over that based on the estimated propensity score. If the ratio is greater than 1, the estimated propensity score makes the H\'ajek and IPW estimators more efficient than the true propensity score. According to \cref{thm: Hajek efficiency} and \cref{thm: IPW efficiency}, these ratios should be no less than 1 asymptotically.

\begin{figure}[p]
\centering
\subfloat[Spatial moderator]{
\includegraphics[width = \textwidth,trim = 0 0 0 60, clip]{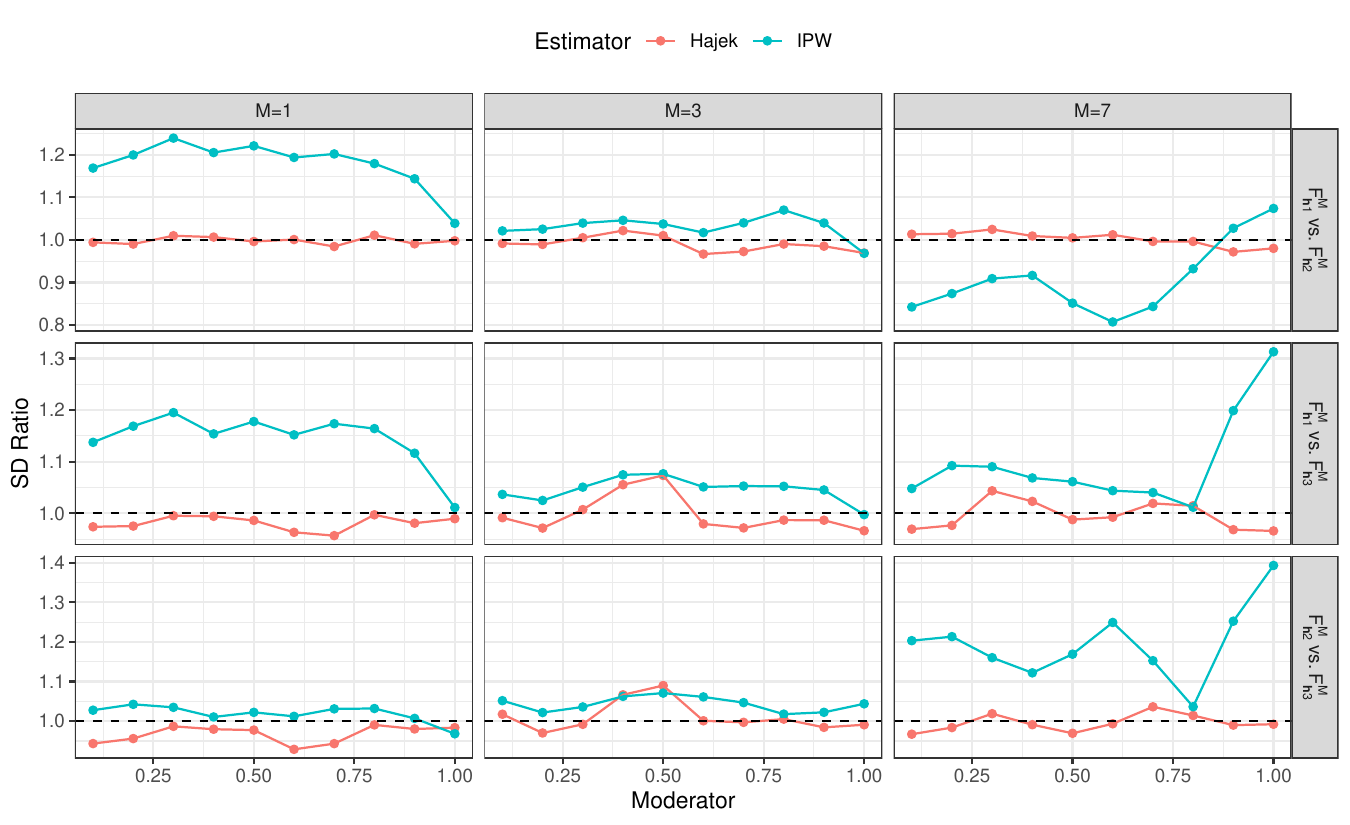}
\label{fig:variance_ratio_s}
}  \\
\subfloat[Spatio-temporal moderator]{
\includegraphics[width = \textwidth,trim = 0 0 0 0, clip]{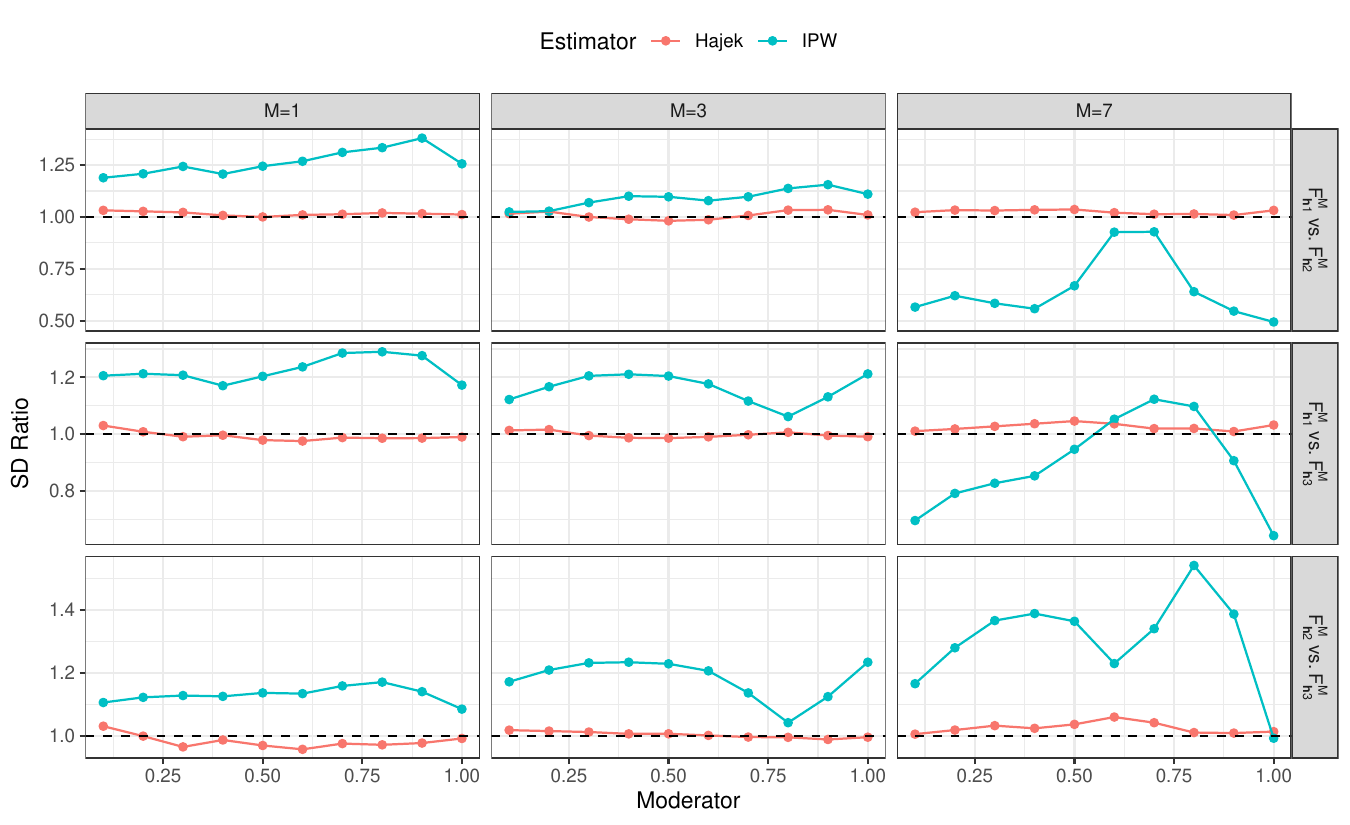}
\label{fig:variance_ratio_st}
}
\caption{Standard deviation ratio of the proposed estimator based on the true propensity score over the proposed
estimator based on the estimated propensity score. The results are based on Monte Carlo approximation
with T = 1000 and 500 simulations. The estimated propensity score is obtained from the correctly specified model.  We compare interventions that are over different time periods: M = 1,3,7. In one time period, the expected number of outcome events yielded by $F^M_{h_1}$, $F^M_{h_2}$ and $F^M_{h_3}$ are 3,5,7 respectively.} 
\label{fig:variance_ratio}
\end{figure}

The results are shown in \cref{fig:variance_ratio}.  We find that the estimated propensity score yields a significant efficiency gain for the IPW estimator for $M=1$ and $M=3$, while the variance ratio for the H\'ajek estimator remains close to 1 in all scenarios. Occasionally, the variance ratio falls below 1, especially in high-variability scenarios with $M=7$. This may be because the Monte Carlo variance does not sufficiently approximate the true asymptotic variances for the sample sizes considered. 

\subsection{Efficiency comparison of IPW and H\'ajek estimators}\label{a: subsec: Hajek_IPW}

\begin{figure}[p]
\centering
\subfloat[Spatial moderator]{
\includegraphics[width = \textwidth,trim = 0 0 0 30, clip]{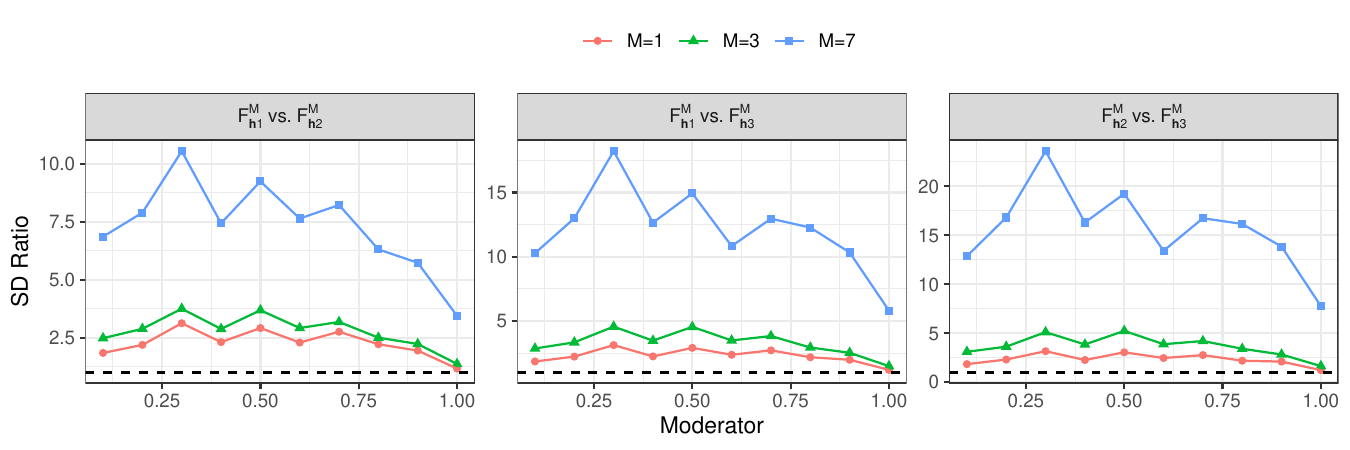}
\label{fig:s_Hajek_IPW}
}  \\
\subfloat[Spatio-temporal moderator]{
\includegraphics[width = \textwidth,trim = 0 0 0 0, clip]{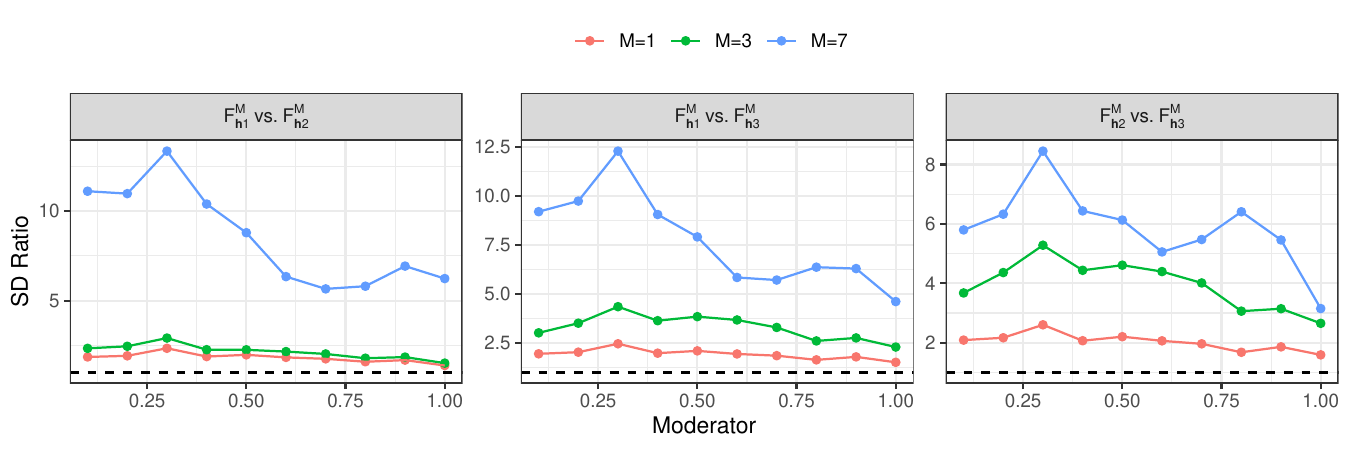}
\label{fig:st_Hajek_IPW}
}
\caption{Standard deviation ratio of the H\'ajek estimator based on the true propensity score over the IPW
estimator based on the true propensity score. The results are based on Monte Carlo approximation
with $T = 500$ and 500 simulations.  We compare interventions that are over different time periods: M = 1,3,7. In one time period, the expected number of outcome events yielded by $F_{h_1}^M$, $F_{h_2}^M$ and $F_{ h_3}^M$ are 3,5,7 respectively.} 
\label{fig:Hajek_IPW}
\end{figure}

Finally, we compare the variance of the IPW estimator with that of the H\'ajek estimator. We compute the Monte Carlo standard deviation of the IPW and H\'ajek estimators based on the estimated propensity score across 500 simulations with $T=500$ under several scenarios. \cref{fig:Hajek_IPW} presents the ratio of these estimated variances. We find that the H\'ajek estimator is consistently more efficient than the IPW estimator in all scenarios considered here.

\section{Additional results for the empirical application} \label{a:sec:add_application}

In this section, we present the results of an additional empirical analysis. We first evaluate the quality of the propensity score model using out-of-sample prediction.  We then show additional application result that complement \cref{sec:application} and conduct additional analyses to assess the sensitivity of our results to alternative specifications, including different lag definitions of the moderator, alternative truncation rules for the estimated propensity scores, varying pixel resolutions, and adjustments to the level of airstrike intensity.

\subsection{Out-of-sample prediction for propensity scores}
\label{a:sec:ps_prediction}

To evaluate the appropriateness of the propensity score model, we compare the predicted number of airstrikes with the observed number of airstrikes in the following four governorates in Iraq: Baghdad, Diyala, Salah al-Din, and Anbar. We assess both out-of-sample predictions, where the model is trained on the first 80\% of the observations, and in-sample predictions, where the model is trained using the entire dataset.

\begin{figure}[!t]
\centering
\includegraphics[width = \textwidth,trim = 0 0 0 0, clip]{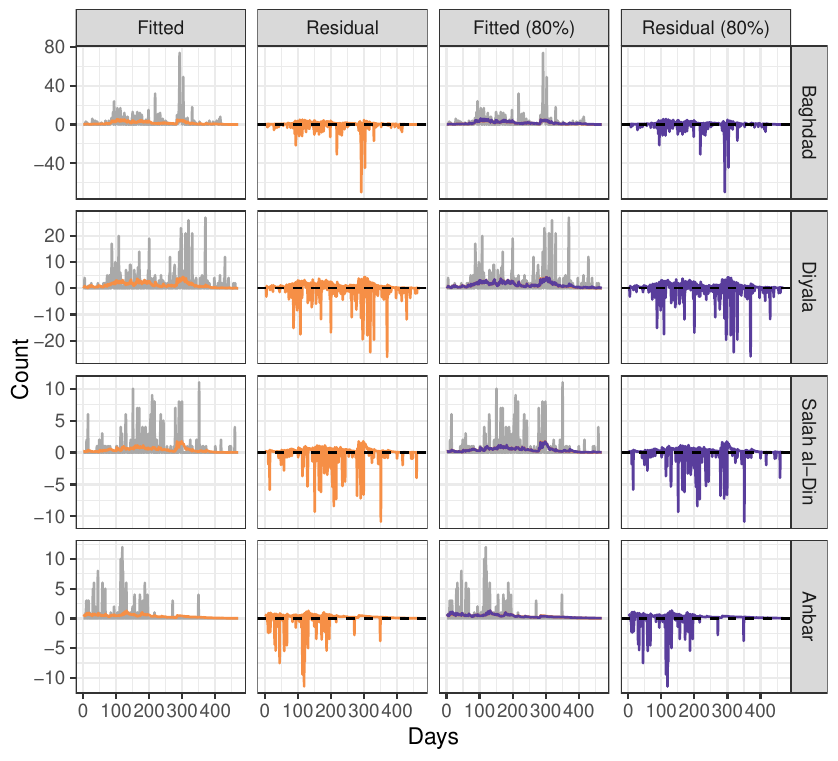}
\caption{Actual and predicted counts of airstrikes for four governorate in Iraq (rows). In the first column, the orange and gray lines indicate the fitted and actual counts, respectively. The second column shows the residual plot. In the third and fourth columns, the blue line indicates the results of out-of-sample prediction employing the first $80$\% of observations.} 
\label{fig:ps_prediction}
\end{figure}

\cref{fig:ps_prediction} presents the results. The estimated propensity score model captures the general trend of airstrikes in these governorates, although it misses some spikes. The discrepancies are likely due to the large variance in the outcome that is inherent in any point process. Furthermore, the model trained on the first 80\% of the data performs similarly to the estimated model using all observations, suggesting that the propensity score model is not overfit.

\subsection{Estimated CATE differences for the binary aid moderator}
\label{a:subsec:binary_cate_difference}

\begin{figure}[!t]
\centering
\includegraphics[width = 0.75\textwidth,trim = 0 10 0 0, clip]{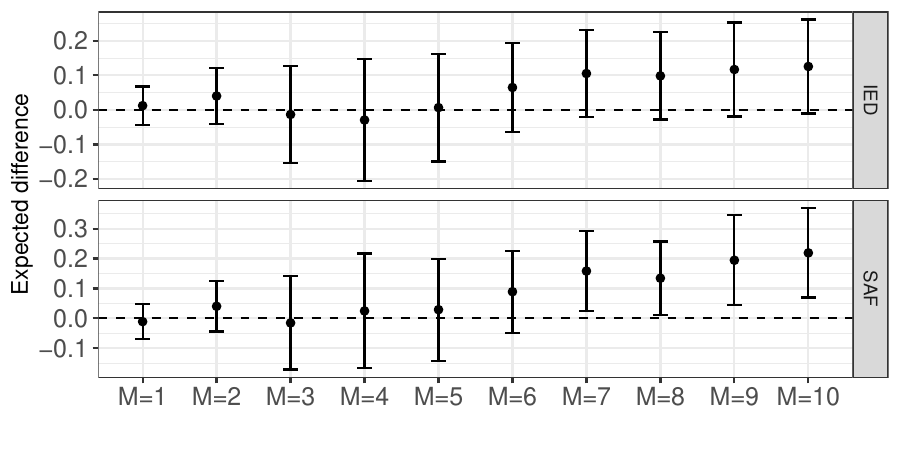}
\caption{Estimated differences in CATEs for the binary aid moderator. The figure shows the estimated difference in CATEs between districts that received aid in the previous month and those that did not. The corresponding 95\% confidence intervals are also shown.}
\label{fig:app_binary_beta_aid}
\end{figure}

In this section, we report the estimated differences in CATEs between districts that received aid in the previous month and those that did not. These results correspond to the binary aid moderator analysis presented in \cref{fig:app_binary_cate_aid} in the main text. \Cref{fig:app_binary_beta_aid} shows the estimated difference in CATEs, with 95\% confidence intervals.

\subsection{Different moderator definitions}
\label{a:subsec:aid_lag}

\begin{figure}[!t]

\centering
\subfloat[Estimated CATE]{
\includegraphics[width = 0.8\textwidth,trim = 0 10 0 0, clip]{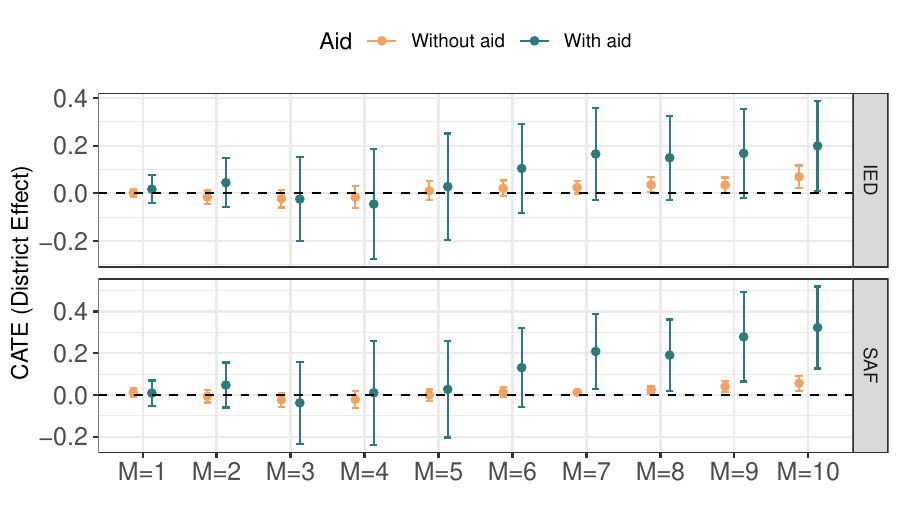}
\label{fig:app_bianry_cate_aid_twoweeks}
}  \\
\subfloat[Expected difference]{
\includegraphics[width = 0.8\textwidth,trim = 0 10 0 0, clip]{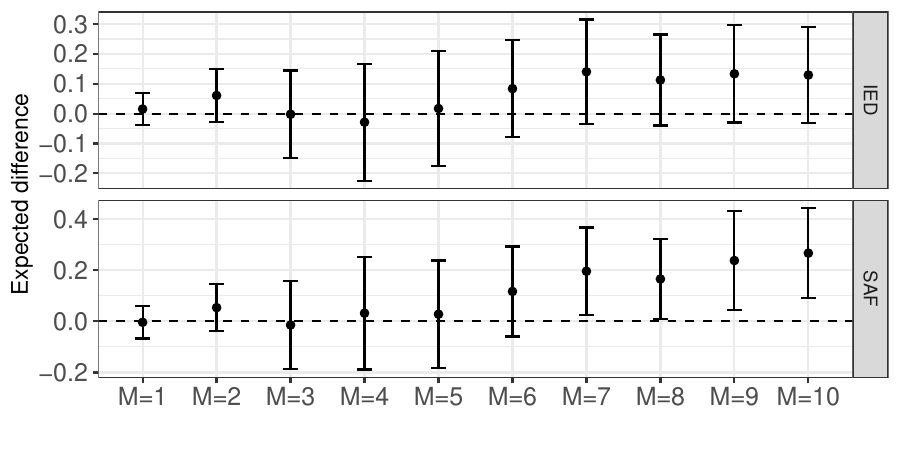}
\label{fig:app_bianry_beta_aid_twoweeks}
}
\caption{Results for the binary aid moderator. Plot~(a) shows the estimated CATEs of increasing airstrike intensity on the number of insurgency attacks in a district with or without aid in the previous \emph{two weeks}. Plot~(b) shows the estimated difference in the CATEs between a district that received aid in the previous \emph{two weeks} and those that did not. The corresponding 95\%  confidence intervals are also shown for all estimators. } 
\label{fig:app_twoweeks_binary}

\end{figure}

\begin{figure}[!t]

\centering
\subfloat[Estimated CATE]{
\includegraphics[width = 0.8\textwidth,trim = 0 10 0 0, clip]{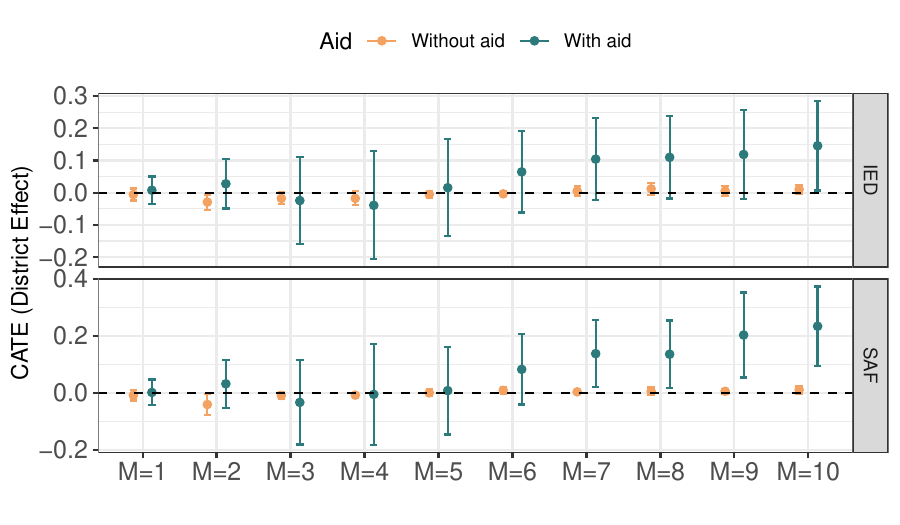}
\label{fig:app_bianry_cate_aid_twomonths}
}  \\
\subfloat[Expected difference]{
\includegraphics[width = 0.8\textwidth,trim = 0 10 0 0, clip]{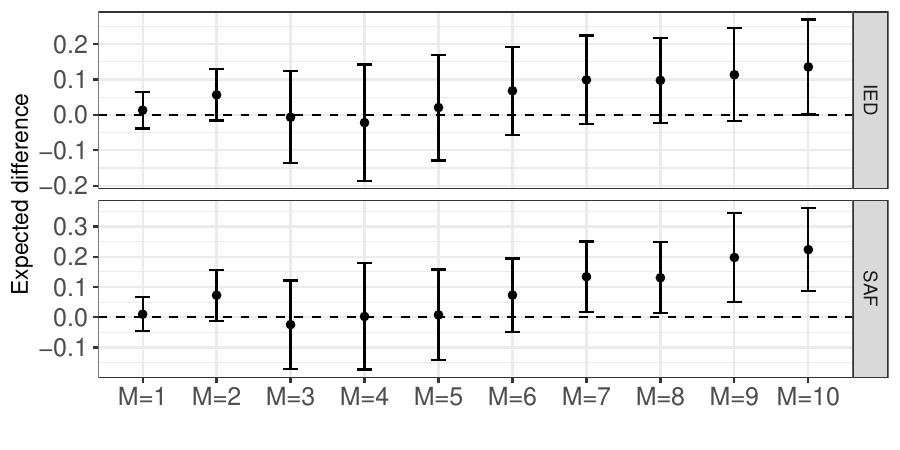}
\label{fig:app_bianry_beta_aid_twomonths}
}
\caption{Results for the binary aid moderator. Plot~(a) shows the estimated CATEs of increasing airstrike intensity on the number of insurgency attacks in a district with or without aid in the previous \emph{two months}. Plot~(b) shows the estimated difference in the CATEs between a district that received aid in the previous \emph{two months} and those that did not. The corresponding 95\%  confidence intervals are also shown for all estimators. } 
\label{fig:app_twomonths_binary}

\end{figure}

\begin{figure}[!t]
\centering
\includegraphics[width = 0.85\textwidth,trim = 0 0 0 0, clip]{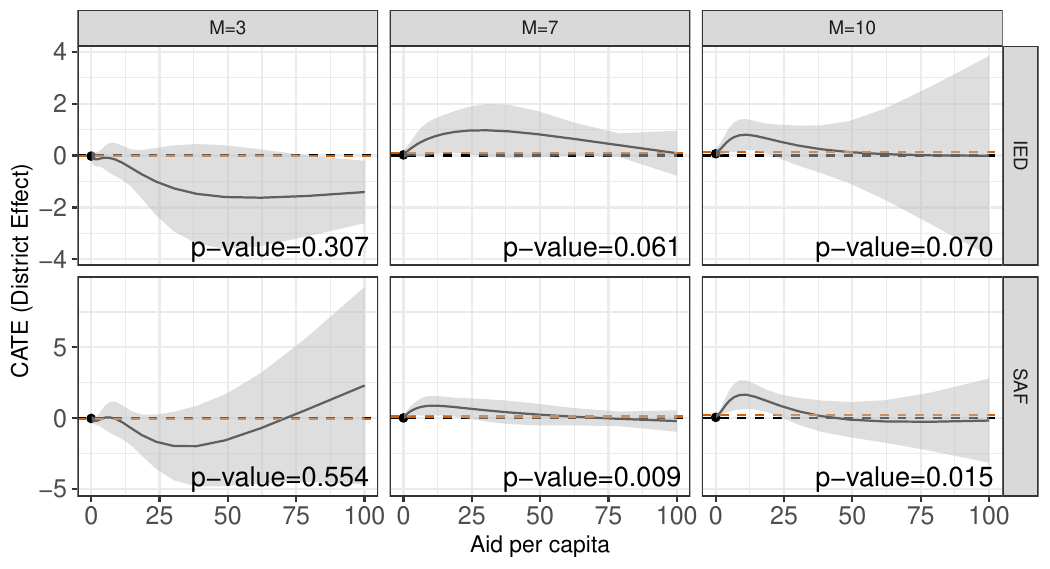}
\caption{Estimated CATEs of increasing airstrike intensity on the number of insurgency attacks in for different values of aid per capita received in the previous \emph{two weeks} with the shaded region indicating the 95\% confidence intervals. The red lines represent the estimated average treatment effects.The p-values correspond to tests for the overall heterogeneity effect.} 
\label{fig:app_continuous_aid_twoweeks}
\end{figure}

\begin{figure}[!t]
\centering
\includegraphics[width = 0.85\textwidth,trim = 0 0 0 0, clip]{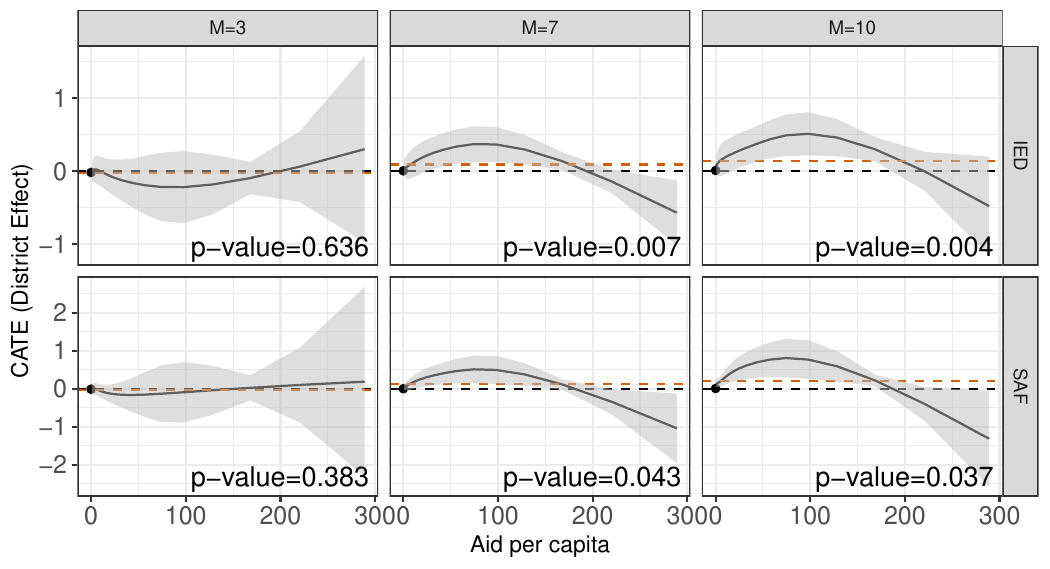}
\caption{Estimated CATEs of increasing airstrike intensity on the number of insurgency attacks in for different values of aid per capita received in the previous \emph{two months} with the shaded region indicating the 95\% confidence intervals. The red lines represent the estimated average treatment effects. The p-values correspond to tests for the overall heterogeneity effect.} 
\label{fig:app_continuous_aid_twomonths}
\end{figure}

We conduct an analysis similar to the one presented in \cref{sec:application} but define the aid moderator using different lag lengths. Specifically, for the binary moderator, we consider the indicators of aid provision during the previous two weeks and two months. For the continuous moderator, we use the aid per capita received in the previous two weeks and two months. The results for the binary aid moderator are presented in Figures~\ref{fig:app_twoweeks_binary} and~\ref{fig:app_twomonths_binary}, while Figures~\ref{fig:app_continuous_aid_twoweeks} and~\ref{fig:app_continuous_aid_twomonths} show the corresponding results for the continuous aid moderator.

For the binary moderator, the results are largely consistent across the two lag lengths. As seen in Figures~\ref{fig:app_twoweeks_binary} and~\ref{fig:app_twomonths_binary}, the estimated CATEs remain stable across all intervention lengths $M = 1, \dots, 10$, with confidence intervals overlapping between the two-week and two-month definitions of aid. For SAF, the CATEs become statistically significant when the intervention lasts at least seven days, regardless of the lag length. Similarly, for IED, the results remain inconclusive, consistent with the findings in the main analysis. Although the overall patterns remain similar across the lag lengths, the magnitude of the estimated CATEs tends to be larger when the moderator is the provision of aid over the previous two weeks compared to the previous two months. For example, for IED attacks with $M = 10$, the estimated CATE is 0.20 for districts that received aid during the previous two weeks, compared to 0.15 for those who received aid during the previous two months. This difference reflects the more immediate effect of the recent aid provision.

For the continuous moderator, shown in Figures~\ref{fig:app_continuous_aid_twoweeks} and~\ref{fig:app_continuous_aid_twomonths}, the results show some sensitivity to the lag length, particularly for longer intervention periods. Across all durations of the intervention $M$, confidence intervals tend to be wider when using aid in the previous two weeks, particularly for SAF. This increased uncertainty is likely due to greater variability in the short-term aid measure, which may capture more noise relative to the more stable two-month lag. For shorter intervention lengths ($M = 3$ and $M = 7$), the differences in the estimated CATEs between the two lag lengths are minor, and overall trends remain consistent. Specifically, CATEs exhibit a flat pattern for small $M$, increase for moderate $M$, and decrease as $M$ becomes large, regardless of the lag length. 

In summary, while the choice of lag length affects the magnitude and precision of the estimated CATEs, the overall patterns observed in the main analysis are preserved. Aid provision during the previous two weeks tends to produce slightly larger estimated effects and wider confidence intervals compared to aid measured over the previous two months, particularly for longer intervention.

\subsection{Sensitivity to truncation levels of the estimated propensity scores}
\label{a:subsec:trunc_level}

\begin{figure}[!t]

\centering
\subfloat[Estimated CATE]{
\includegraphics[width = 0.8\textwidth,trim = 0 10 0 0, clip]{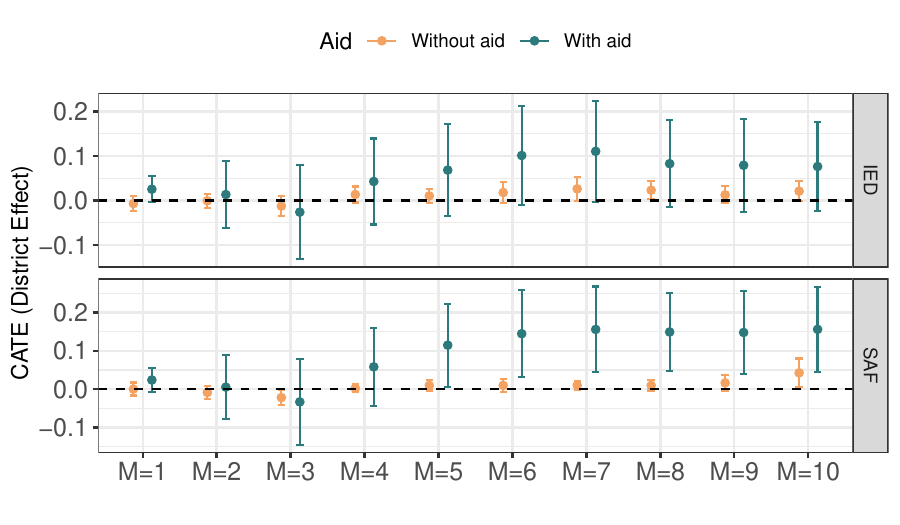}
\label{fig:app_bianry_cate_aid_onemonth_90}
}  \\
\subfloat[Expected difference]{
\includegraphics[width = 0.8\textwidth,trim = 0 10 0 0, clip]{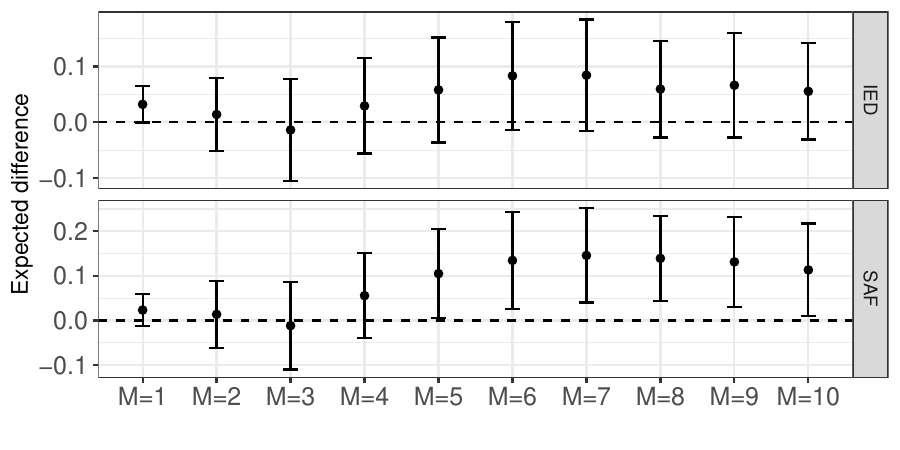}
\label{fig:app_bianry_beta_aid_onemonth_90}
}
\caption{Results for the binary aid moderator based on propensity scores truncated at the \emph{$90^{th}$ percentile}. Plot~(a) shows the estimated CATEs of increasing airstrike intensity on the number of insurgency attacks in a district with or without aid in the previous one months. Plot~(b) shows the estimated difference in the CATEs between a district that received aid in the previous one months and those that did not. The corresponding 95\%  confidence intervals are also shown for all estimators. } 
\label{fig:app_onemonths_binary_90}

\end{figure}

\begin{figure}[!t]

\centering
\subfloat[Estimated CATE]{
\includegraphics[width = 0.8\textwidth,trim = 0 10 0 0, clip]{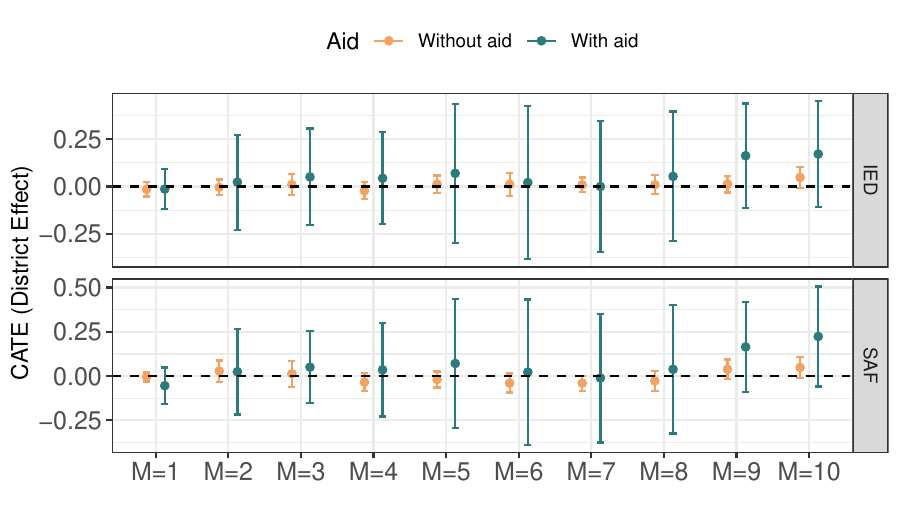}
\label{fig:app_bianry_cate_aid_onemonth_98}
}  \\
\subfloat[Expected difference]{
\includegraphics[width = 0.8\textwidth,trim = 0 10 0 0, clip]{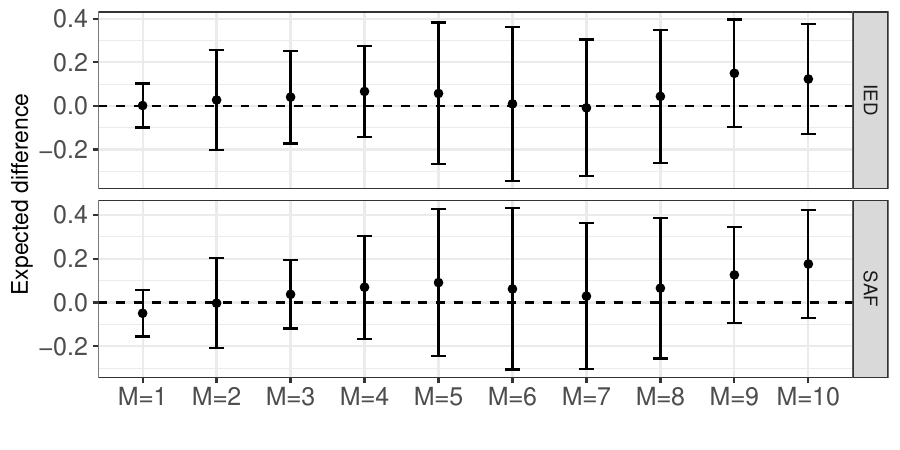}
\label{fig:app_bianry_beta_aid_onemonth_98}
}
\caption{Results for the binary aid moderator based on propensity scores truncated at the \emph{$98^{th}$ percentile}. Plot~(a) shows the estimated CATEs of increasing airstrike intensity on the number of insurgency attacks in a district with or without aid in the previous one months. Plot~(b) shows the estimated difference in the CATEs between a district that received aid in the previous one months and those that did not. The corresponding 95\%  confidence intervals are also shown for all estimators. } 
\label{fig:app_onemonths_binary_98}

\end{figure}

\begin{figure}[!t]
\centering
\includegraphics[width = 0.85\textwidth,trim = 0 0 0 0, clip]{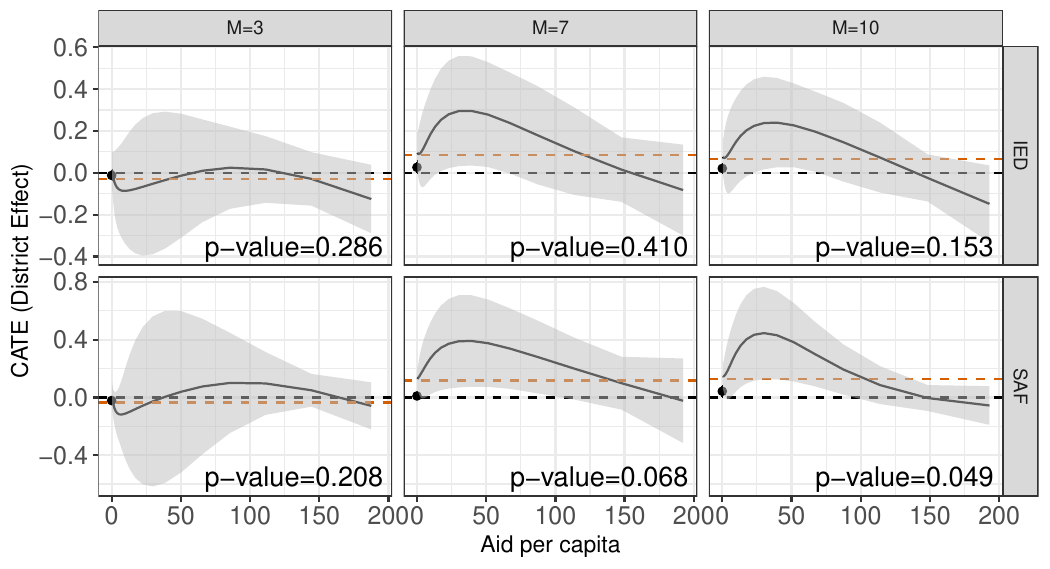}
\caption{Estimated CATEs of increasing airstrike intensity on the number of insurgency attacks in for different values of aid per capita received in the previous two weeks with the shaded region indicating the 95\% confidence intervals. The results are using propensity scores truncated at the $90^{th}$ percentile.} 
\label{fig:app_continuous_aid_onemonth_90}
\end{figure}

\begin{figure}[!t]
\centering
\includegraphics[width = 0.85\textwidth,trim = 0 0 0 0, clip]{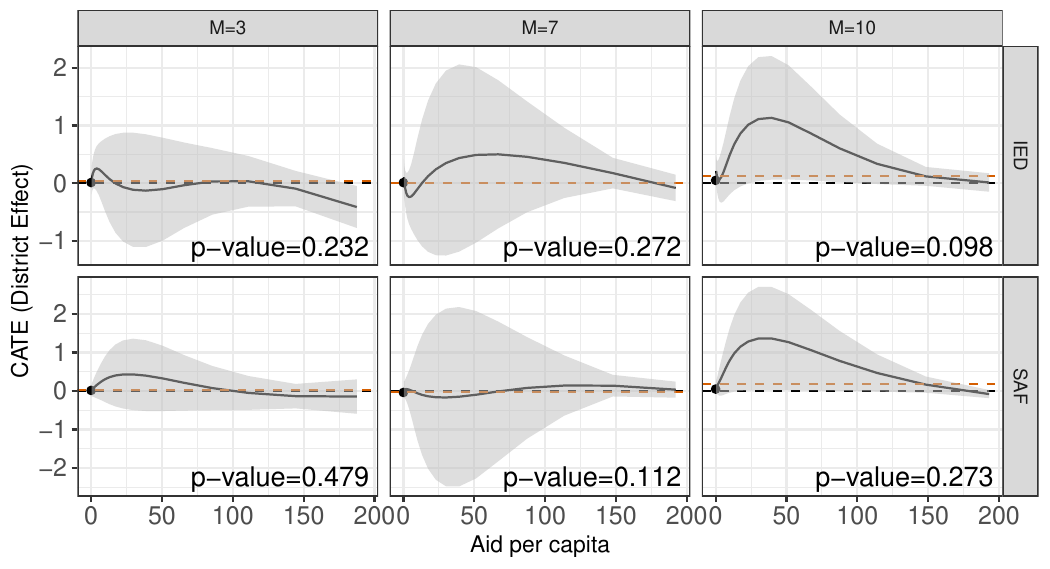}
\caption{Estimated CATEs of increasing airstrike intensity on the number of insurgency attacks in for different values of aid per capita received in the previous two months with the shaded region indicating the 95\% confidence intervals. The red lines represent the estimated average treatment effects. The results are using propensity scores truncated at the $98^{th}$ percentile. The p-values correspond to tests for the overall heterogeneity effect.} 
\label{fig:app_continuous_aid_onemonth_98}
\end{figure}

To evaluate the sensitivity of our results to the level of truncation of the estimated propensity scores, we compare the results using truncation at the 90\% and 98\% quantiles. Figures~\ref{fig:app_onemonths_binary_90} and~\ref{fig:app_onemonths_binary_98} show the results for the binary aid moderator, while Figures~\ref{fig:app_continuous_aid_onemonth_90} and~\ref{fig:app_continuous_aid_onemonth_98} present the corresponding results for the continuous aid moderator.

For the binary aid moderator, truncating the propensity scores at higher quantiles, such as 98\%, tends to produce larger estimates of the CATEs with wider confidence intervals compared to truncation at the 90\% quantile. As shown in \cref{fig:app_onemonths_binary_90} and \cref{fig:app_onemonths_binary_98}, general patterns remain consistent across truncation levels. For SAF, the estimated difference in CATEs between districts with and without aid becomes positive and statistically significant for intervention periods of at least 7 days. However, for IED, the estimation uncertainty is substantial, and the results remain inconclusive regardless of the truncation level.

For the continuous aid moderator, results are shown in \cref{fig:app_continuous_aid_onemonth_90} and \cref{fig:app_continuous_aid_onemonth_98}. Here, we observe greater sensitivity to the choice of truncation level. Higher truncation levels, such as the 98\% quantile, lead to larger estimated CATEs and wider confidence intervals, reflecting the increased variability of the estimates. For instance, in Figure~A.16, the estimated CATE for SAF at $M = 10$ decreases substantially when moving from the 90\% to the 98\% truncation level. Although the magnitude of the estimates changes, the overall patterns are preserved at all truncation levels, with CATE estimates exhibiting little change for small values of $M$, increasing for moderate values, and decreasing as $M$ becomes large.

The p-values corresponding to tests of no heterogeneity effect are also reported in the figures. For the continuous aid moderator, the p-values tend to decrease as $M$ increases, suggesting stronger evidence of heterogeneity for longer intervention periods. However, there are cases where the significance depends on the truncation level. For example, in \cref{fig:app_continuous_aid_onemonth_90}, the p-values for SAF at $M = 10$ are significant when truncated at the level 90\%, while the same results are not significant under the 98\% truncation level (\cref{fig:app_continuous_aid_onemonth_98}).

Overall, while the choice of truncation level affects the magnitude of the estimated effects, it does not alter the general trends observed in the main analysis. For the binary aid moderator, the estimated CATEs for SAF remain positive and significant for intervention periods of at least seven days, while the results for IED remain inconclusive. For the continuous aid moderator, the results are more sensitive to truncation, but the observed patterns are robust, with evidence of heterogeneity increasing as the intervention period grows longer.

\subsection{Sensitivity to pixel resolution}
\label{a:subsec:pixel}

In this section, we examine the sensitivity of our results to the choice of spatial resolution used in the second-step regression. In addition to the baseline $128 \times 128$ grid used in \cref{sec:application}, we consider both finer and coarser discretizations, as well as an alternative specification based on administrative districts. Specifically, we conduct analyses using a $256 \times 256$ grid, $64 \times 64$ and $32 \times 32$ grids, and a specification in which each district is treated as a single unit. In all cases, we analyze both binary and continuous aid moderators for SAF and IED events. To facilitate comparison, all pixel-level estimates are rescaled to the district level. For the continuous aid moderator, when aid varies within a pixel, we use the average aid value within that pixel.

\begin{figure}[!t]

\centering
\subfloat[Estimated CATE]{
\includegraphics[width = 0.8\textwidth,trim = 0 10 0 0, clip]{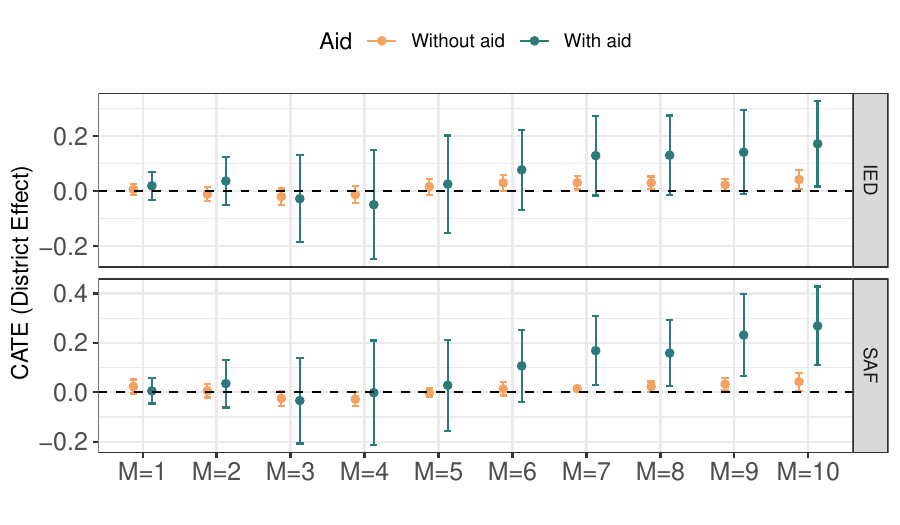}
\label{fig:app_bianry_cate_aid_onemonth_95_256}
}  \\
\subfloat[Expected difference]{
\includegraphics[width = 0.8\textwidth,trim = 0 10 0 0, clip]{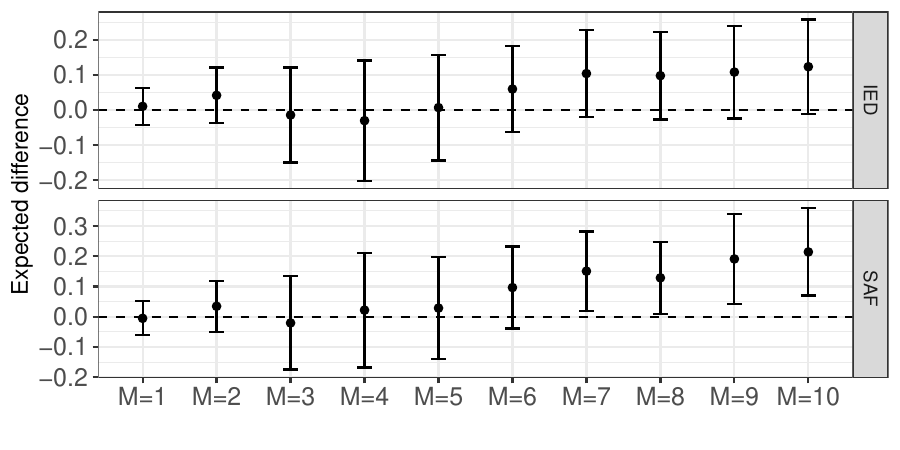}
\label{fig:app_bianry_beta_aid_onemonth_95_256}
}
\caption{Results for the binary aid moderator based on $256\times256$ pixel grid. Plot~(a) shows the estimated CATEs of increasing airstrike intensity on the number of insurgency attacks in a district with or without aid in the previous one months. Plot~(b) shows the estimated difference in the CATEs between a district that received aid in the previous one months and those that did not. The corresponding 95\%  confidence intervals are also shown for all estimators. } 
\label{fig:app_onemonths_binary_95_256}

\end{figure}

\begin{figure}[!t]
\centering
\includegraphics[width = 0.85\textwidth,trim = 0 0 0 0, clip]{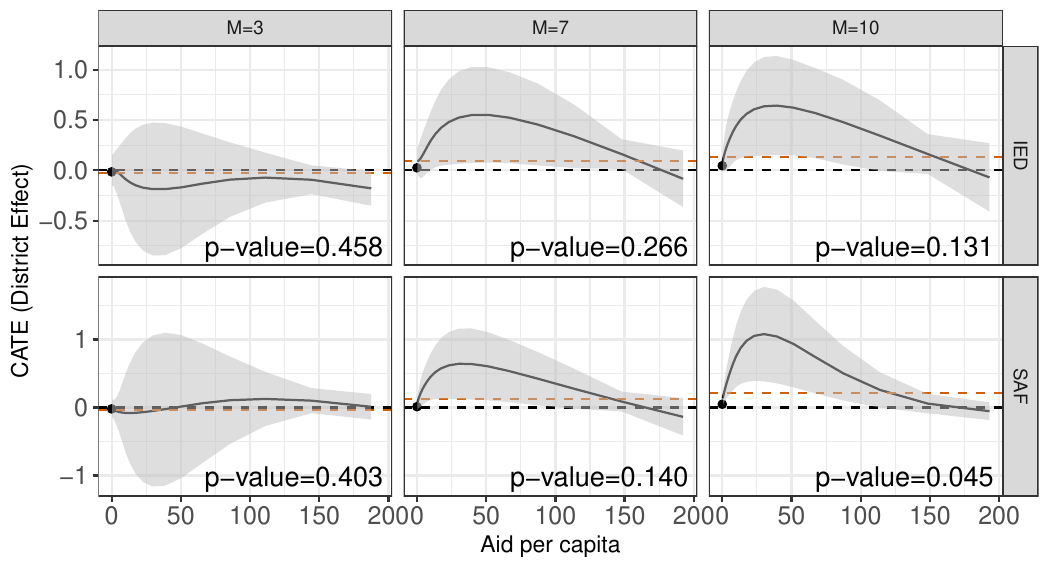}
\caption{Estimated CATEs of increasing airstrike intensity on the number of insurgency attacks in for different values of aid per capita received in the previous two months with the shaded region indicating the 95\% confidence intervals. The red lines represent the estimated average treatment effects. The results are based on $256\times256$ pixel grid. The p-values correspond to tests for the overall heterogeneity effect.} 
\label{fig:app_continuous_aid_onemonth_95_256}
\end{figure}

\begin{figure}[!t]

\centering
\subfloat[Estimated CATE]{
\includegraphics[width = 0.8\textwidth,trim = 0 10 0 0, clip]{plot_binary_cate_onemonth_95_64.pdf}
\label{fig:app_bianry_cate_aid_onemonth_95_64}
}  \\
\subfloat[Expected difference]{
\includegraphics[width = 0.8\textwidth,trim = 0 10 0 0, clip]{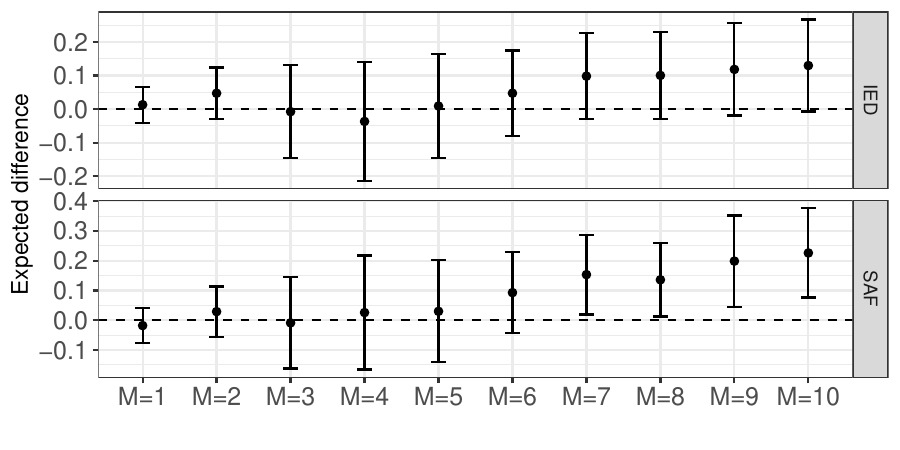}
\label{fig:app_bianry_beta_aid_onemonth_95_64}
}
\caption{Results for the binary aid moderator based on $64\times64$ pixel grid. Plot~(a) shows the estimated CATEs of increasing airstrike intensity on the number of insurgency attacks in a district with or without aid in the previous one months. Plot~(b) shows the estimated difference in the CATEs between a district that received aid in the previous one months and those that did not. The corresponding 95\%  confidence intervals are also shown for all estimators. } 
\label{fig:app_onemonths_binary_95_64}

\end{figure}

\begin{figure}[!t]
\centering
\includegraphics[width = 0.85\textwidth,trim = 0 0 0 0, clip]{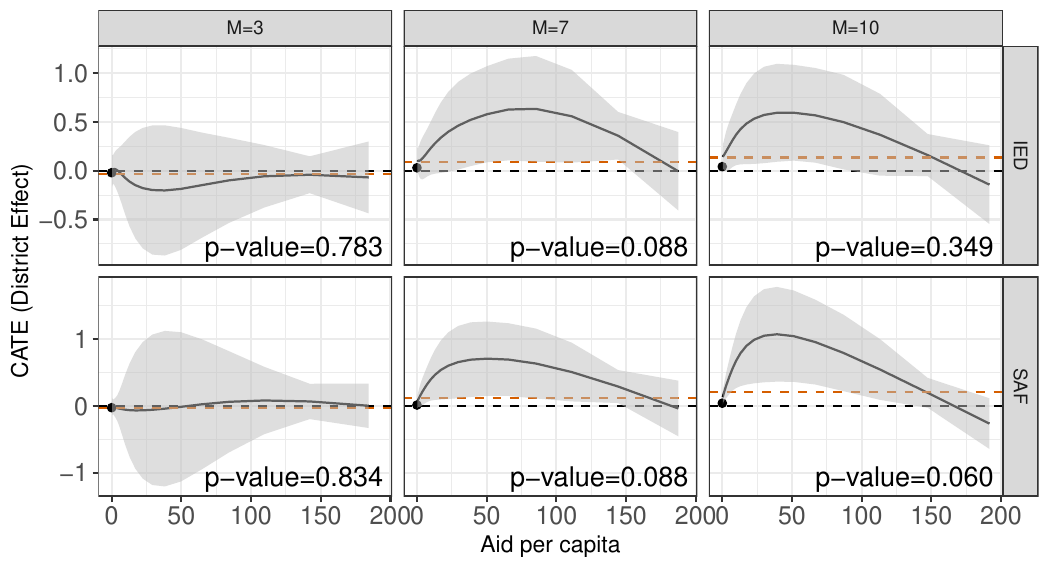}
\caption{Estimated CATEs of increasing airstrike intensity on the number of insurgency attacks in for different values of aid per capita received in the previous two months with the shaded region indicating the 95\% confidence intervals. The red lines represent the estimated average treatment effects. The results are based on $64\times64$ pixel grid. The p-values correspond to tests for the overall heterogeneity effect.} 
\label{fig:app_continuous_aid_onemonth_95_64}
\end{figure}

\begin{figure}[!t]

\centering
\subfloat[Estimated CATE]{
\includegraphics[width = 0.8\textwidth,trim = 0 10 0 0, clip]{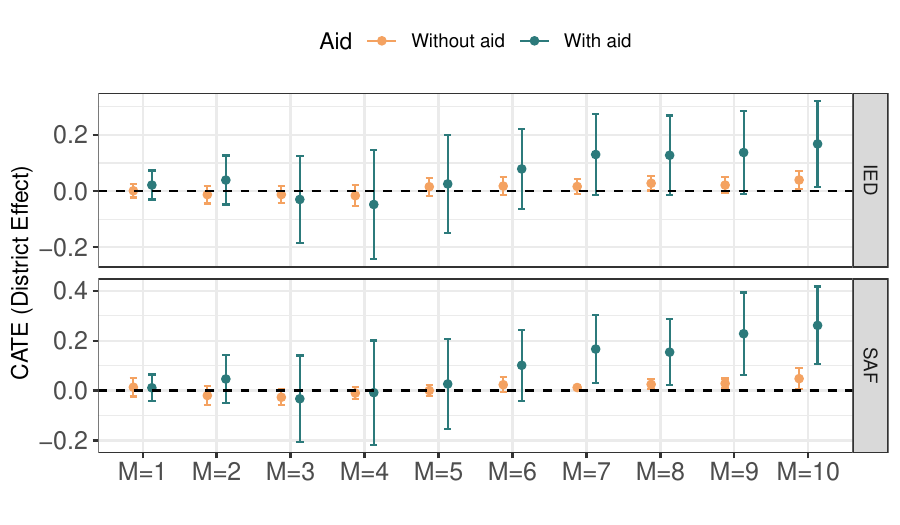}
\label{fig:app_bianry_cate_aid_onemonth_95_32}
}  \\
\subfloat[Expected difference]{
\includegraphics[width = 0.8\textwidth,trim = 0 10 0 0, clip]{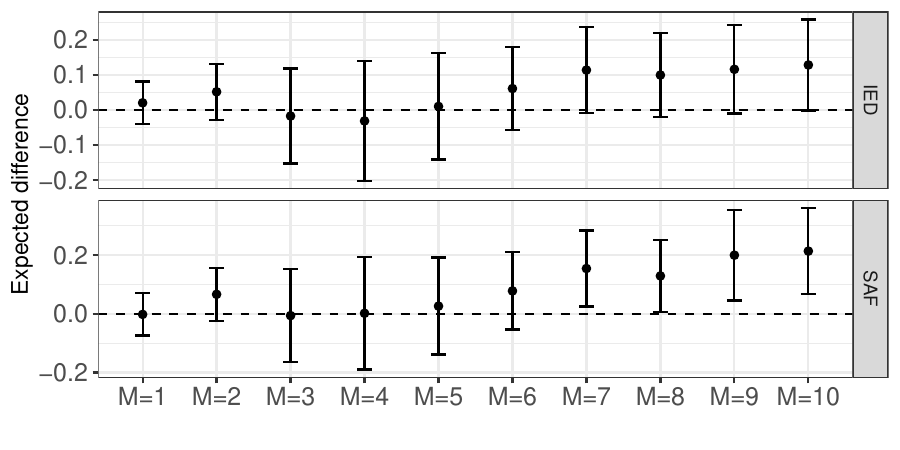}
\label{fig:app_bianry_beta_aid_onemonth_95_32}
}
\caption{Results for the binary aid moderator based on $64\times64$ pixel grid. Plot~(a) shows the estimated CATEs of increasing airstrike intensity on the number of insurgency attacks in a district with or without aid in the previous one months. Plot~(b) shows the estimated difference in the CATEs between a district that received aid in the previous one months and those that did not. The corresponding 95\%  confidence intervals are also shown for all estimators. } 
\label{fig:app_onemonths_binary_95_32}

\end{figure}

\begin{figure}[!t]
\centering
\includegraphics[width = 0.85\textwidth,trim = 0 0 0 0, clip]{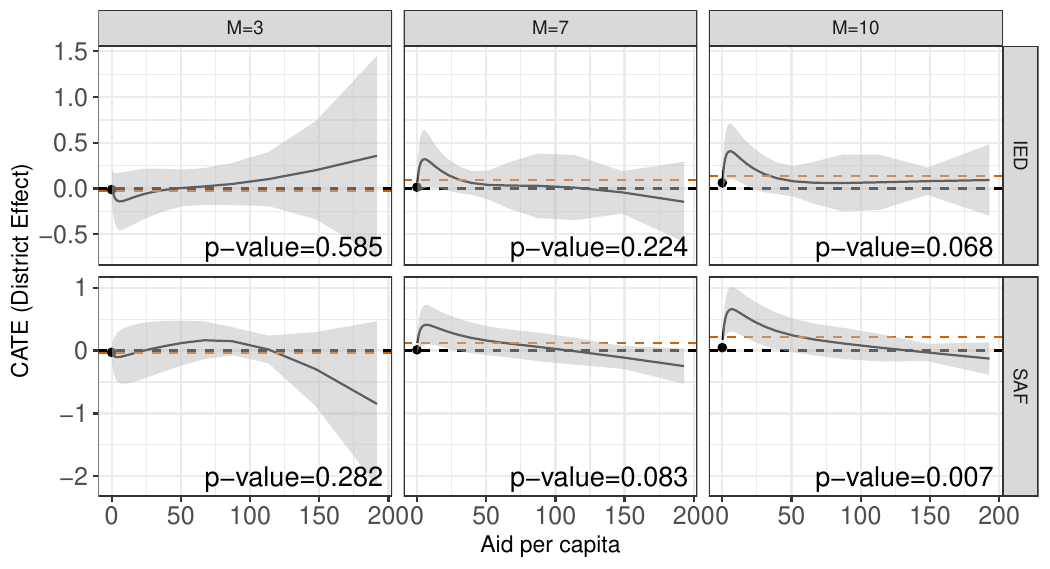}
\caption{Estimated CATEs of increasing airstrike intensity on the number of insurgency attacks in for different values of aid per capita received in the previous two months with the shaded region indicating the 95\% confidence intervals. The red lines represent the estimated average treatment effects. The results are based on $64\times64$ pixel grid. The p-values correspond to tests for the overall heterogeneity effect.} 
\label{fig:app_continuous_aid_onemonth_95_32}
\end{figure}

\begin{figure}[!t]

\centering
\subfloat[Estimated CATE]{
\includegraphics[width = 0.8\textwidth,trim = 0 10 0 0, clip]{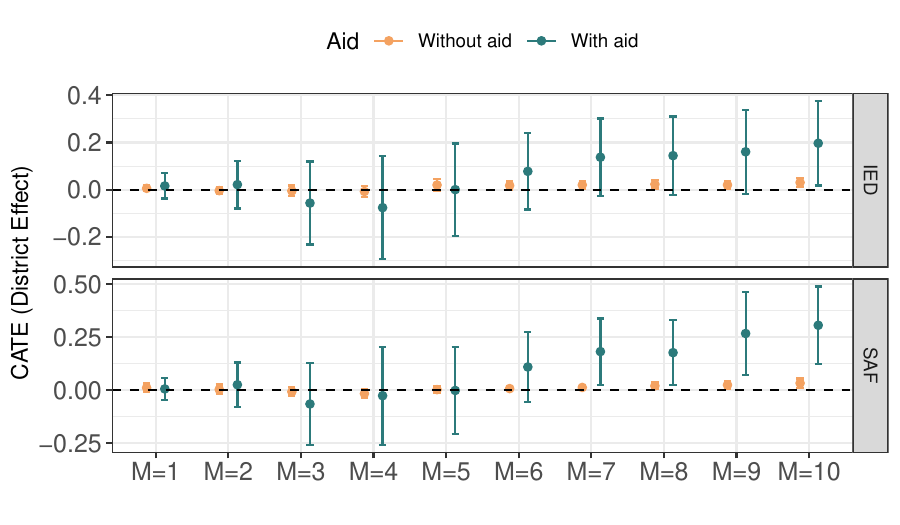}
\label{fig:app_bianry_cate_aid_onemonth_95_district}
}  \\
\subfloat[Expected difference]{
\includegraphics[width = 0.8\textwidth,trim = 0 10 0 0, clip]{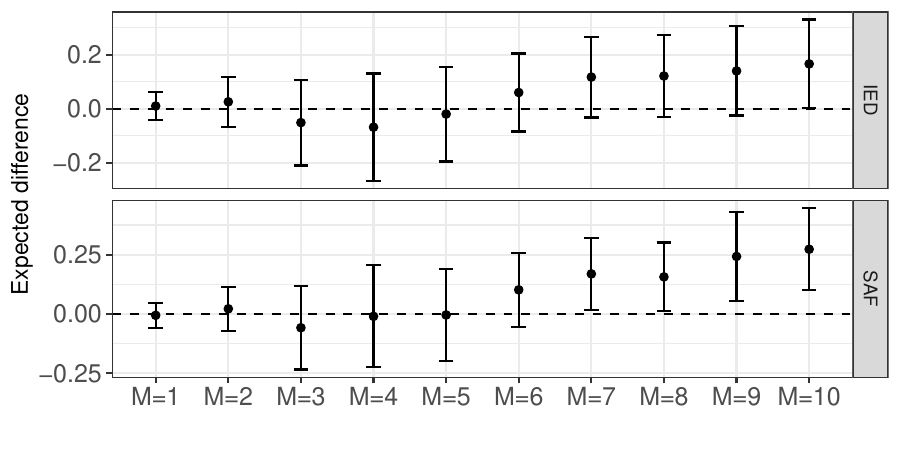}
\label{fig:app_bianry_beta_aid_onemonth_95_district}
}
\caption{Results for the binary aid moderator based on $64\times64$ pixel grid. Plot~(a) shows the estimated CATEs of increasing airstrike intensity on the number of insurgency attacks in a district with or without aid in the previous one months. Plot~(b) shows the estimated difference in the CATEs between a district that received aid in the previous one months and those that did not. The corresponding 95\%  confidence intervals are also shown for all estimators. } 
\label{fig:app_onemonths_binary_95_district}

\end{figure}

\begin{figure}[!t]
\centering
\includegraphics[width = 0.85\textwidth,trim = 0 0 0 0, clip]{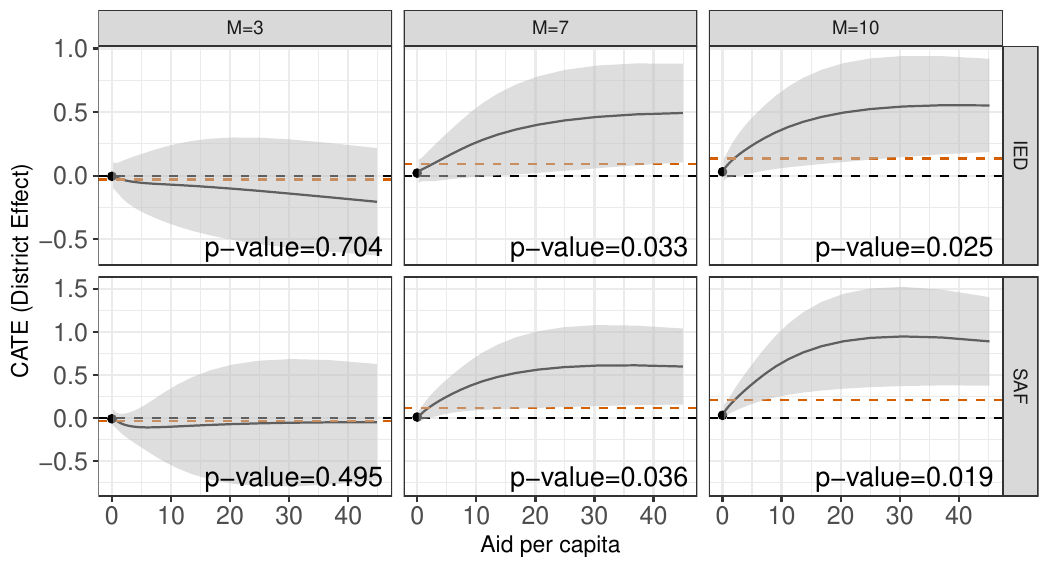}
\caption{Estimated CATEs of increasing airstrike intensity on the number of insurgency attacks in for different values of aid per capita received in the previous two months with the shaded region indicating the 95\% confidence intervals. The red lines represent the estimated average treatment effects. The results are based on $64\times64$ pixel grid. The p-values correspond to tests for the overall heterogeneity effect.} 
\label{fig:app_continuous_aid_onemonth_95_district}
\end{figure}

The results for the $256 \times 256$ grid are presented in \cref{fig:app_onemonths_binary_95_256,fig:app_continuous_aid_onemonth_95_256}, the $64 \times 64$ and $32 \times 32$ grids in \cref{fig:app_onemonths_binary_95_64,fig:app_continuous_aid_onemonth_95_64,fig:app_onemonths_binary_95_32,fig:app_continuous_aid_onemonth_95_32}, and the district-based analysis in \cref{fig:app_onemonths_binary_95_district,fig:app_continuous_aid_onemonth_95_district}.

For the binary aid moderator, the estimated effects are nearly identical across all specifications, including the finer $256 \times 256$ grid and the district-level analysis. This indicates that the estimated CATE is highly robust to the choice of spatial resolution in the second-step regression.

For the continuous aid moderator, the results remain broadly consistent across specifications. The estimates obtained using the $256 \times 256$, $128 \times 128$, and $64 \times 64$ grids are nearly indistinguishable, and the $32 \times 32$ grid yields similar patterns. The small differences observed under coarser resolutions are primarily attributable to how the district-level aid values are aggregated to the pixel level. In particular, when moving to coarser grids, pixels that overlap district boundaries require averaging aid values across districts, which introduces minor discrepancies relative to finer discretizations.

The district-based specification can be interpreted as a reweighted version of the $128 \times 128$ pixel-level regression. Specifically, it corresponds to performing the pixel-level regression with weights proportional to the inverse of the number of pixels contained in each district, so that each district contributes equally to the estimation. Under this specification, the results for the binary moderator remain very similar to those from the pixel-based analyses. For the continuous moderator, the pattern differs slightly: while the estimated CATE still increases with aid at lower levels, the decline at higher levels is less pronounced, resulting in a flatter relationship. This difference is likely due to the change in weighting, as the $128 \times 128$ specification implicitly assigns greater weight to larger districts, whereas the district-level analysis weights all districts equally.

\subsection{Additional analysis for the binary moderator}\label{a: application_inten}

As discussed in \cref{sec:application}, the baseline distribution of airstrikes is not uniform across space, which means that increasing airstrike intensity relative to baseline causes some regions to experience a greater increase in airstrikes than others. To account for this variation, we study how the treatment effect varies with the binary moderator, defined as the indicator of receiving aid during the previous month, while controlling for expected changes in the number of airstrikes. 

Specifically, we include the expected change in airstrike numbers and its interaction with the aid indicator in the regression model. The model is given by:  
$$
\tau^{\mathrm{Proj.}}_{t,\bm{h}_1,\bm{h}_2}(r; \bbeta_t) = \beta_{t,0} + \beta_{t,1} r + \beta_{t,2} q + \beta_{t,3} r \times q,
$$
where $r$ is the binary aid moderator, and $q$ is the expected change in the number of airstrikes in a district. After estimating the coefficients, we plug in the median value of $q$ to represent the expected airstrike intensity and compare the CATEs for areas with and without aid.

\begin{figure}[!t]

\centering
\subfloat[Estimated CATE]{
\includegraphics[width = 0.8\textwidth,trim = 0 10 0 0, clip]{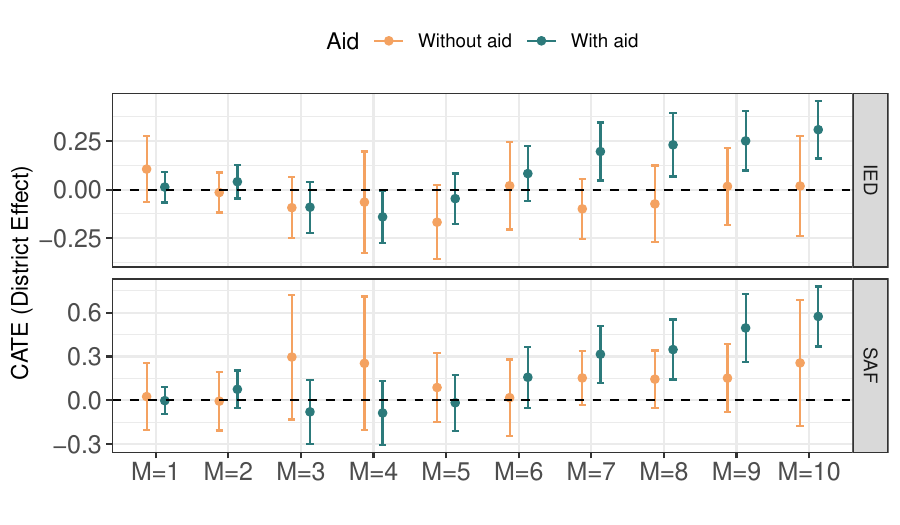}
\label{fig:app_bianry_cate_aid_inten}
}  \\
\subfloat[Expected difference]{
\includegraphics[width = 0.8\textwidth,trim = 0 10 0 0, clip]{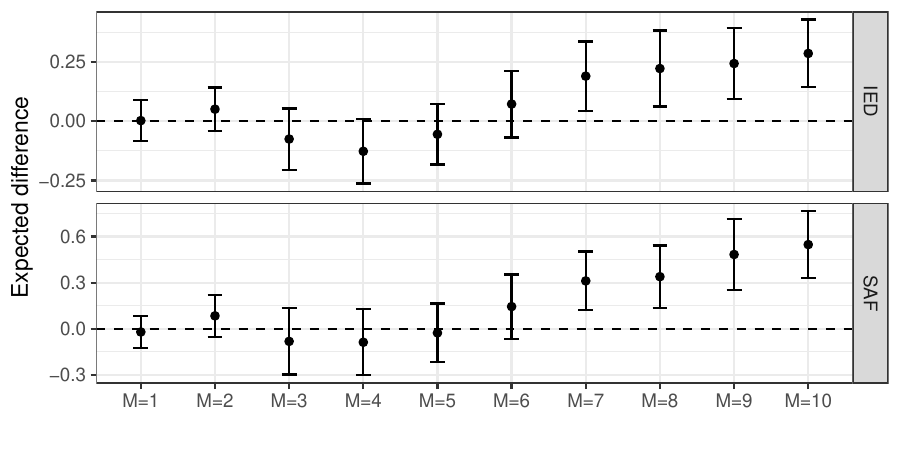}
\label{fig:app_bianry_beta_aid_inten}
}
\caption{Results for the binary aid moderator for regions with \emph{median level} of increased airstrike intensity. Plot~(a) shows the estimated CATEs of increasing airstrike intensity on the number of insurgency attacks in a district with or without aid in the previous one months controlling the expected changes in the airstrike intensity. Plot~(b) shows the estimated difference in the CATEs between a district that received aid in the previous one months and those that did not. The corresponding 95\%  confidence intervals are also shown for all estimators. } 
\label{fig:app_binary_inten}

\end{figure}

\Cref{fig:app_binary_inten} presents the results for both IED and SAF attacks. Plot~(a) shows the estimated CATEs for intervention windows ranging from $M = 1$ to $M = 10$, while Plot~(b) displays the estimated difference in the CATEs between districts with and without aid. The findings generally align with the results in \cref{sec:application}, where the expected intensity of airstrikes was not included in the model. For SAF attacks, districts that received aid in the previous month continue to show a stronger reaction to airstrikes, particularly when the intervention spans at least seven days. Importantly, for IED attacks, the estimated differences in CATEs now become statistically significant, suggesting a stronger effect in regions with prior aid.

Overall, controlling for expected airstrike intensity does not alter the main conclusion that regions receiving aid in the previous month exhibit a stronger response to increased airstrike intensity. These results indicate that the observed heterogeneity in treatment effects is not driven by differences in the intensity of airstrikes across regions with and without aid.

\subsection{Additional analysis for adaptive interventions}\label{a:sec:add_adaptive}

This section extends the main application from static interventions, where the counterfactual treatment distribution remains fixed across $M$ consecutive days and across time, to adaptive interventions, in which the counterfactual treatment distribution may depend on the observed history. 

In all other respects, the analysis follows the same structure as in \cref{sec:application}. The treatment variable (airstrikes), the outcome variables (SAF and IED), the spatial resolution, the moderators (whether a district received any aid and the amount of aid per capita in the previous month) and the propensity score model remain unchanged. The main difference lies in the construction of the baseline distribution of airstrikes and the definition of the counterfactual interventions. To obtain the baseline intensity, we fit a non-homogeneous Poisson point process using historical data from 2006. The specification of this model is similar to the propensity score model described in \cref{subsec:design}, except that the covariate vector $\bm X_t$ excludes time splines, the surge indicator, and prior airstrikes. The exclusion of prior airstrikes is intentional. If the current airstrike intensity were modeled as explicitly dependent on past airstrikes, then under an adaptive intervention spanning multiple periods, such dependence could artificially amplify or attenuate changes in the expected airstrike intensity at later time points, making the interpretation of the estimated effects more complicated.

We define two counterfactual treatment distributions by halving and doubling the baseline intensity, denoted $h_t'(\omega)=0.5\,\lambda_t^{\text{Base}}(\omega)$ and $h_t''(\omega)=2\,\lambda_t^{\text{Base}}(\omega)$, where $\lambda_t^{\text{Base}}(\omega)$ is the estimated baseline intensity. The corresponding counterfactual interventions over $M$ time periods are defined as $F_{\bm h_t'} = \left(F_{h_t'},F_{h_{t-1}'},\dots,F_{h_{t-M+1}'}\right)$ and $F_{\bm h_t''} = \left(F_{h_t''},F_{h_{t-1}''},\dots,F_{h_{t-M+1}''}\right)$.

For an adaptive intervention lasting one period ($M=1$), the observed history remains unchanged, and the contrast between $F_{h_t'}$ and $F_{h_t''}$ represents how the expected outcomes would differ if the expected number of airstrikes were shifted from one-half to twice the estimated baseline intensity. When the adaptive intervention extends over multiple periods ($M>1$), the intensity is halved or doubled at each period conditional on the observed history up to that time. For later periods, however, the observed history itself is modified because the distribution of airstrikes in earlier periods differs under the two intervention rules. Consequently, when $M>1$, the comparison should be interpreted as contrasting two adaptive assignment strategies that update sequentially according to past realizations, rather than comparing two fixed sequences of airstrike intensities based on the estimated baseline intensity.

\begin{figure}[!t]
\centering
\includegraphics[width = 0.75\textwidth,trim = 0 10 0 0, clip]{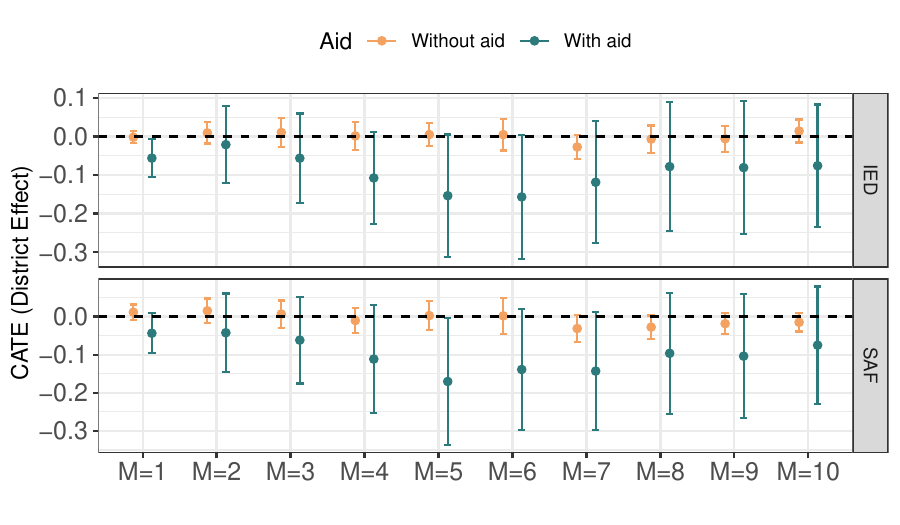}
\caption{Estimated district-level CATEs of changing the adaptive intervention rule from halving to doubling the baseline airstrike intensity, by intervention length ($M=1,\dots,10$). The top and bottom panels show IED and SAF outcomes, respectively. Error bars represent 95\% confidence intervals.}
\label{fig:adp_binary_cate}
\end{figure}

\begin{figure}[!t]
\centering
\includegraphics[width = 0.75\textwidth,trim = 0 10 0 0, clip]{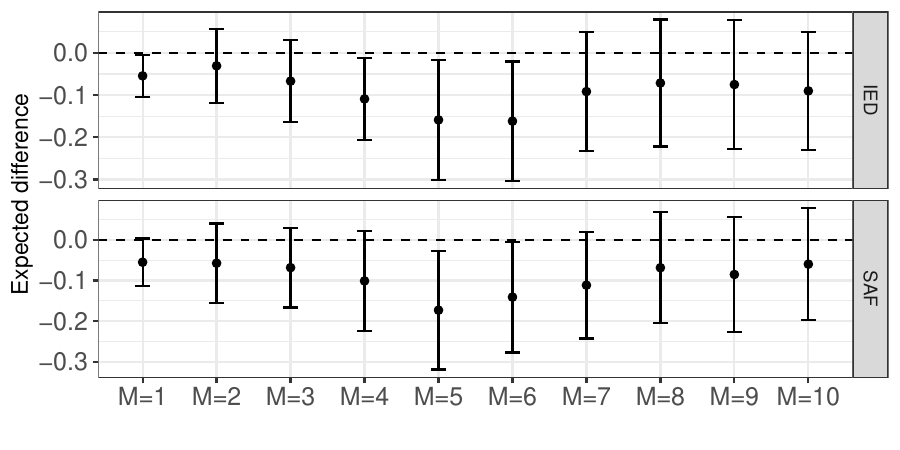}
\caption{Estimated difference in district-level CATEs between districts with and without aid in the previous month ($r=1$ minus $r=0$) for adaptive interventions of different durations ($M=1,\dots,10$). Error bars represent 95\% confidence intervals.}
\label{fig:adp_binary_diff}
\end{figure}

We first examine the binary moderator indicating whether a district received any U.S.\ aid spending in the previous month. For each $M$ ranging from one to ten, we estimate district-level conditional average treatment effects (CATEs) for districts with and without aid, using the same model specification as in the main text. The estimated CATEs with 95\% confidence intervals are presented in \Cref{fig:adp_binary_cate}, and the corresponding differences are shown in \Cref{fig:adp_binary_diff}.

For SAF, districts that received aid consistently experienced fewer attacks than those without aid. The estimated reduction is largest when the airstrike strategy shifts from $F_{\bm h_t'}$ to $F_{\bm h_t''}$ over a five-day period and gradually diminishes as the duration of the intervention increases. The decrease in SAF attacks is statistically significant when the adaptive intervention is maintained for approximately five to six days. For other values of $M$, the estimation uncertainty is substantial, making it difficult to draw firm conclusions about whether districts with and without aid respond differently to the change in airstrike strategy from $F_{\bm h'}$ to $F_{\bm h''}$. The results for IED display similar patterns to those observed for SAF.

\begin{figure}[!t]
\centering
\includegraphics[width=0.75\textwidth,trim = 0 0 0 0, clip]{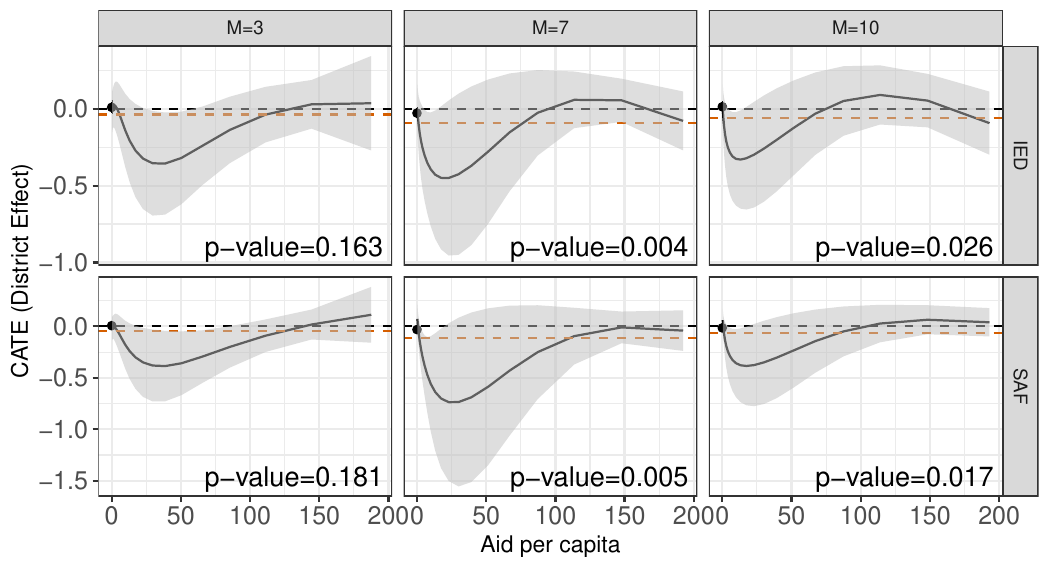}
\caption{Estimated district-level CATEs of changing the adaptive intervention rule from halving to doubling the baseline airstrike intensity, as a function of aid per capita in the previous month. The top and bottom panels correspond to IED and SAF outcomes, respectively. Shaded regions represent 95\% confidence intervals.The red lines represent the estimated average treatment effects. The p-values correspond to tests for the overall heterogeneity effect.}
\label{fig:adp_continuous}
\end{figure}

We next examine heterogeneity in the adaptive intervention effects as a function of aid per capita in the previous month. The working model is identical to the one in \cref{subsec: findings}, using four natural cubic spline basis functions for strictly positive aid and an indicator for zero aid. \Cref{fig:adp_continuous} presents the estimated CATEs for varying levels of prior aid $r$, along with the average treatment effect when the treatment strategy shifts from $F_{\bm h_t'}$ to $F_{\bm h_t''}$. The p-values shown in the figure correspond to tests for the overall heterogeneity effect described in \cref{subsec: test}.

The patterns for SAF and IED are broadly similar. For IED attacks, the estimated effects display consistent behavior across intervention horizons $M=3,7,10$, with the evidence for heterogeneity becoming stronger as $M$ increases. The estimated CATEs show an initial decrease as prior aid rises from \$0 to approximately \$25 per capita, followed by an increase between \$25 and \$125, after which the effect gradually declines as aid becomes larger. These results indicate that districts respond differently to the change in airstrike strategy from $F_{\bm h_t'}$ to $F_{\bm h_t''}$ depending on the amount of aid previously received, and that such heterogeneity becomes more pronounced when the adaptive intervention spans multiple time periods.

\end{document}